\documentclass[epj,nopacs]{svjour}
\usepackage{graphicx}           
\usepackage{amsmath}
\usepackage{amssymb}            

\usepackage{braket}             
\usepackage{pdflscape}          
\usepackage{hyperref}
\usepackage{caption}            
\usepackage{dcolumn}            

\usepackage{multirow}
\usepackage{xspace} 
\usepackage{tablefootnote}

\usepackage[table,usenames,dvipsnames]{xcolor}
\usepackage{arydshln} 

\usepackage{feynmp}             
\usepackage{color}              
\usepackage[normalem]{ulem}
\usepackage[title,titletoc,toc]{appendix}
\usepackage{cuted}
\usepackage{widetext}

\newcolumntype{f}[1]{D{.}{.}{#1}}

\newcommand\cnt[2]{\multicolumn{#1}{c}{#2}}
\newcommand\lft[2]{\multicolumn{#1}{l}{#2}}

\newcommand\cntl[2]{\multicolumn{#1}{c|}{#2}}

\definecolor{darkgreen}{rgb}{0,0.7,0.2}
\definecolor{lightgreen}{rgb}{0,0.85,0}

\newcommand\AwayA[1]{{\color{OliveGreen} \sout{#1}}} 
\newcommand\AwayB[1]{{\color{RedViolet} \sout{#1}}}     

\renewcommand\AwayA[1]{}
\renewcommand\AwayB[1]{}

\newcommand{\mup}{\ensuremath{\mu}{\rm p}\xspace}
\newcommand{\mud}{\ensuremath{\mu}{\rm d}\xspace}
\newcommand{\muHet}{\ensuremath{\mu^3}{\rm He}\ensuremath{^+}\xspace}
\newcommand{\muHef}{\ensuremath{\mu^4}{\rm He}\ensuremath{^+}\xspace}

\newcommand{\TwoSOne}{\ensuremath{2\textrm{S}_{1/2}}\xspace}
\newcommand{\TwoPOne}{\ensuremath{2\textrm{P}_{1/2}}\xspace}
\newcommand{\TwoPThree}{\ensuremath{2\textrm{P}_{3/2}}\xspace}
\newcommand{\TwoSOneFCero}{\ensuremath{2\textrm{S}_{1/2}^{\textrm{F}=0}}\xspace}
\newcommand{\TwoSOneFOne}{\ensuremath{2\textrm{S}_{1/2}^{\textrm{F}=1}}\xspace}
\newcommand{\TwoPOneFCero}{\ensuremath{2\textrm{P}_{1/2}^{\textrm{F}=0}}\xspace}
\newcommand{\TwoPOneFOne}{\ensuremath{2\textrm{P}_{1/2}^{\textrm{F}=1}}\xspace}
\newcommand{\TwoPThreeFOne}{\ensuremath{2\textrm{P}_{3/2}^{\textrm{F}=1}}\xspace}
\newcommand{\TwoPThreeFTwo}{\ensuremath{2\textrm{P}_{3/2}^{\textrm{F}=2}}\xspace}

\newcommand{\rel}{\ensuremath{r_E} }
\newcommand{\rrel}{\ensuremath{{r_E}^2} }

\newcommand{\rze}{\ensuremath{r_Z} }

\newcommand{\rr}{\ensuremath{\langle r^2 \rangle}}

\newcommand{\rp}{\ensuremath{r_\mathrm{p}}}
\newcommand{\rd}{\ensuremath{r_\mathrm{d}}}
\newcommand{\rh}{\ensuremath{r_\mathrm{h}}}

\newcommand{\etal}{{\it et\,al.}}

\newcommand{\mev}{\,\ensuremath{\mathrm{meV}}\xspace}
\newcommand{\fm}{\,\ensuremath{\mathrm{fm}}\xspace}
\newcommand{\insqfm}{\,\ensuremath{\mathrm{fm}^{-2}}\xspace}

\newcommand{\ELSrad}{\ensuremath{\Delta E_\mathrm{r\mathrm{-dep.}}^\mathrm{LS}} }
\newcommand{\ELSradind}{\ensuremath{\Delta E_\mathrm{r\mathrm{-indep.}}^\mathrm{LS}} }

\newcommand{\ELSTPE}{\ensuremath{\Delta E_\mathrm{TPE}^\mathrm{LS}}}
\newcommand{\ELSFriar}{\ensuremath{\Delta E_\mathrm{Friar}^\mathrm{LS}}}

\newcommand{\ELSinelast}{\ensuremath{\Delta E_\mathrm{inelastic}^\mathrm{LS}}}

\newcommand{\EFS}{\ensuremath{\Delta E_\mathrm{FS}} }

\renewcommand{\baselinestretch}{1.2}

\def\LSVAL{\ensuremath{1644.3466}}
\def\LSERR{\ensuremath{  0.0146}}
\def\TABLSVAL{{\ensuremath{\bf 1644}}\bf .{\ensuremath{\bf 3466}}}
\def\TABLSERR{{\ensuremath{\bf   0}}\bf .{\ensuremath{\bf 0146}}}

\def\RADVAL{\ensuremath{103.5184}}

\def\RADVALERR{\RADVAL \ensuremath{(98)}}

\def\POLVALRND{\ensuremath{15.30}}
\def\POLVALFINAL{\ensuremath{15.30}}
\def\POLERRFINAL{\ensuremath{0.52}}  

\def\POLVALERR{\POLVALRND \ensuremath{(52)}}

\def\FSVAL{\ensuremath{144.7993}}
\def\FSERR{\ensuremath{0.0101}}
\def\TABFSVAL{{\ensuremath{\bf 144}}\bf .{\ensuremath{\bf 7993}}}
\def\TABFSERR{{\ensuremath{\bf 0}}\bf .{\ensuremath{\bf 0101}}}
\def\FSVALERR{\FSVAL \ensuremath{(101)}}

\begin{document}\sloppy

\title{Theory of the {\emph n}\,=\,2 levels in muonic helium-3 ions}

\author{
Beatrice Franke\inst{1,2}\,\thanks{authors contributed equally}\,\thanks{\emph{email:} bfranke@triumf.ca} \and 
Julian J.~Krauth\inst{1,3{\rm\,a}}\,\thanks{\emph{email:} jkrauth@uni-mainz.de} \and 
Aldo Antognini\inst{4,5} \and 
Marc~Diepold\inst{1} \and 
Franz~Kottmann\inst{4} \and 
Randolf~Pohl\inst{3,1}\,\thanks{\emph{email:} pohl@uni-mainz.de}
%
}                     
\authorrunning{B.~Franke, J.~J.~Krauth \textit{et al.}}
%
%
\institute{
Max--Planck--Institut f{\"u}r Quantenoptik, 85748 Garching, Germany \and 
TRIUMF, 4004 Wesbrook Mall, Vancouver, BC V6T 2A3, Canada \and
Johannes Gutenberg-Universit\"at Mainz, QUANTUM, Institut f\"ur Physik \& Exzellenzcluster PRISMA, 55099 Mainz, Germany \and
Institute for Particle Physics and Astrophysics, ETH Zurich, 8093 Zurich, Switzerland \and
Paul Scherrer Institute, 5232 Villigen, Switzerland
}
\date{(Dated: December 20, 2017)}
%


\abstract{
The present knowledge of Lamb shift, fine-, and hyperfine structure of the 2S and 2P states in muonic helium-3 ions is reviewed in anticipation of the results of a first measurement of several $\mathrm{2S\rightarrow2P}$ transition frequencies in the muonic helium-3 ion, \muHet. This ion is the bound state of a single negative muon $\mu^-$ and a bare helium-3 nucleus (helion), $\mathrm{^3He^{++}}$.\\
A term-by-term comparison of all available sources, including new, updated, and so far unpublished calculations, reveals reliable values and uncertainties of the QED and nuclear structure-dependent contributions to the Lamb shift and the hyperfine splitting. These values are essential for the determination of the helion rms charge radius and the nuclear structure effects to the hyperfine splitting in \muHet. With this review we continue our series of theory summaries in light muonic atoms [see Antognini \etal, Ann.~Phys.~\textbf{331}, 127 (2013); Krauth \etal, Ann.~Phys.~\textbf{366}, 168 (2016); and Diepold \etal, arXiv:1606.05231 (2016)].
%
} 

\maketitle

\section{Introduction}
\label{sec:Intro}
Laser spectroscopy of light muonic atoms and ions, where a single negative muon
orbits a bare nucleus, holds the promise for a vastly improved determination of
nuclear parameters, compared to the more traditional methods of elastic electron
scattering and precision laser spectroscopy of regular electronic atoms.

The CREMA collaboration has so far determined the charge radii of the proton and
the deuteron, by measuring several transitions in muonic hydrogen (\mup) \cite{Pohl:2010:Nature_mup1,Antognini:2013:Science_mup2,Antognini:2013:Annals} and muonic deuterium (\mud) \cite{Pohl:2016:mud,Krauth:2016:mud}. Interestingly, both values differ by as much as six standard deviations from the respective CODATA-2014 values \cite{Mohr:2016:CODATA14}, which contain data from laser spectroscopy in atomic hydrogen/deuterium and electron scattering. 
This discrepancy has been coined ``proton radius puzzle" \cite{Pohl:2013:ARNPS,Carlson:2015:Puzzle,Hill:2017:PRP}. 
However, the discrepancy exists for the deuteron, too. 
Interestingly, for the proton and the deuteron, the muonic isotope shift is compatible with the electronic one from the 1S-2S transition in H and D \cite{Parthey:2010:PRL_IsoShift,Jentschura:2011:IsoShift}. 
The respective radii are
\begin{align}
  \rp (\mup) =& ~0.84087(\hphantom{0}26)^\mathrm{exp}(29)^\mathrm{th} \nonumber \\ 
             =& ~0.84087(\hphantom{0}39)\fm &\text{\cite{Pohl:2010:Nature_mup1,Antognini:2013:Science_mup2}} \label{eq:mup}\\
  \rp (\rm CODATA'14) =& ~0.87510(610)\fm  &\text{\cite{Mohr:2016:CODATA14}}\\[10pt]
  \rd (\mud)            =& ~2.12562(\hphantom{0}13)^\mathrm{exp}(77)^\mathrm{th} \nonumber \\
                        =& ~2.12562(\hphantom{0}78)\fm &\text{\cite{Pohl:2016:mud}} \label{eq:mud}\\
  \rd (\rm CODATA'14) =& ~2.14130(250)\fm.  &\text{\cite{Mohr:2016:CODATA14}}
\end{align}

Very recently, the CREMA collaboration has measured a total of five transitions in muonic helium-3 and -4 ions \cite{Antognini:2011:Conf:PSAS2010}, which have been analyzed now. 


These measurements will help to improve our understanding of nuclear model theories \cite{Machleidt:2011:nuclforces,NevoDinur:2016:TPE} and shed more light on the proton radius puzzle. Several ideas exist to solve the puzzle \cite{Antognini:2016:PRP}, some within the standard model \cite{Miller:2013:pol,Jentschura:2015:virtPart} and others proposing muon specific forces beyond the standard model \cite{Tucker-Smith:2011,Batell:2011:PV_muonic_forces,Karshenboim:2014:darkForces,Carlson:2015:BSM}. 
These ideas lead to predictions which can be tested with precise charge radius determinations in muonic helium ions.

\begin{figure*}
  \centering
  \includegraphics[width=0.7\linewidth]{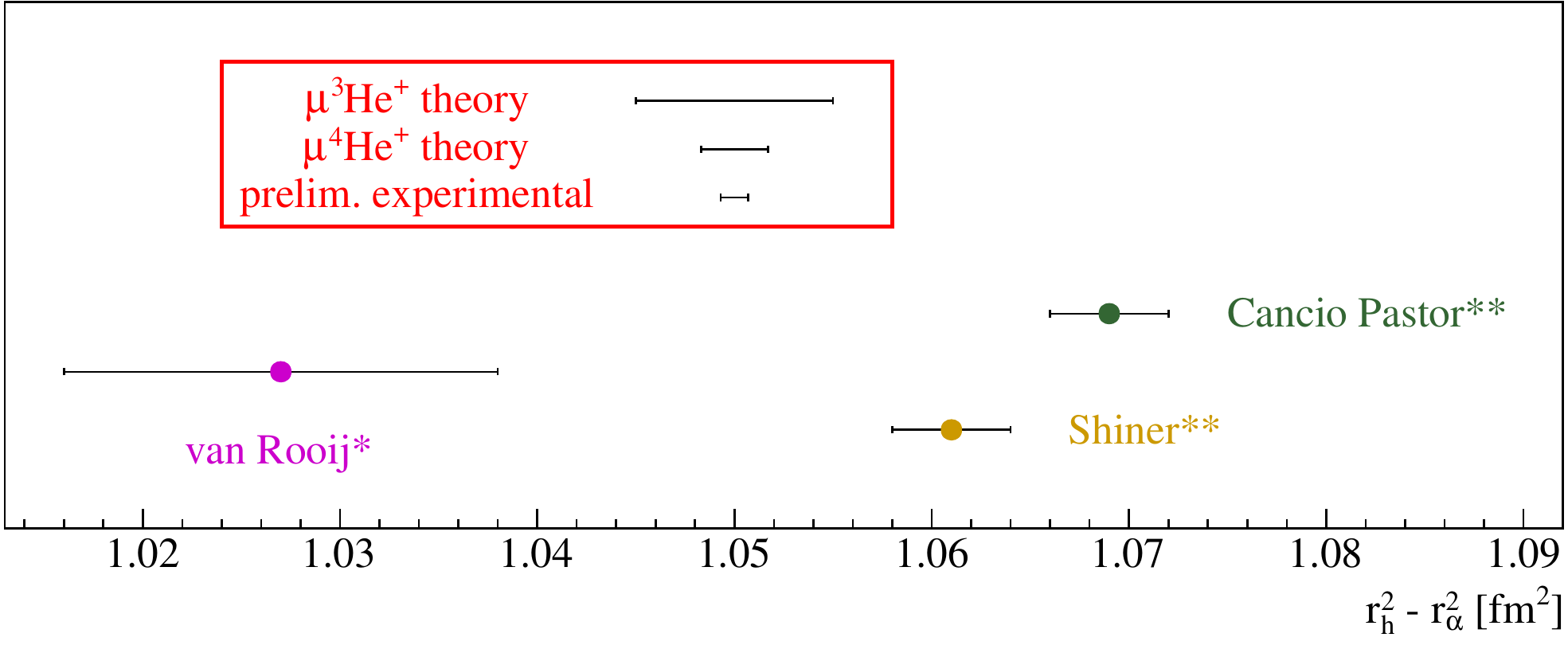}
  \caption{Difference of squared helion-to-alpha particle charge radii as obtained from laser spectroscopy of transitions in regular, electronic helium-3 and helium-4 atoms \cite{Shiner:1995:heliumSpec,Rooij:2011:HeSpectroscopy,CancioPastor:2012:PRL108} when combined with accurate theory (*\cite{Patkos:2017:HeIsoII}, **\cite{Patkos:2016:HeIso}). A $4\,\sigma$ discrepancy persists. 
Also shown are the individual theory uncertainties which enter $\rh^2-r_\alpha^2$ (\muHef: \cite{Diepold:2016:muHe4theo}, \muHet: this work), as well as the expected uncertainty from our laser spectroscopy of the Lamb shift in muonic helium ions.
Note that the combination of the two theoretical uncertainties should contain correlations which will partly cancel in the total uncertainty.
}
  \label{fig:iso_shift}
\end{figure*}

The measurement of the charge radius in both, helium-3 and helium-4 ions will in addition help understand the discrepancy between several measurements of the helium isotope shift in electronic helium  \cite{Shiner:1995:heliumSpec,Rooij:2011:HeSpectroscopy,CancioPastor:2012:PRL108,Patkos:2016:HeIso,Patkos:2017:HeIsoII} which yield the difference of the squared charge radii (see Fig.\,\ref{fig:iso_shift}). 

Several other experiments are on the way to contribute to the puzzle in the future \cite{Antognini:2016:PRP} by precision spectroscopy measurements in electronic hydrogen \cite{Vutha:2012:H2S2P,Beyer:2013:AdP_2S4P,Peters:2013:AdP} and He$^+$ \cite{Herrmann:2009:He1S2S,Kandula:2010:XUV_comb_metrology}, as well as by electron scattering at very low $Q^2$ \cite{Mihovilovic:2013:ISR_exp_MAMI,Gasparian:2014:PRad} and muon-scattering \cite{Gilman:2013:MUSE}. 
The He$^+$ spectroscopy, in combination with our measurement in muonic helium ions, will be able to determine the Rydberg constant independently from hydrogen and deuterium. This is particularly interesting as the proton charge radius and the Rydberg constant are highly correlated which means that a change in the Rydberg constant could also resolve the puzzle \cite{Beyer:2013:AdP_2S4P}.

The determination of the helion charge radius from muonic helium spectroscopy
requires accurate knowledge of the corresponding theory.
Similar to muonic hydrogen \cite{Antognini:2013:Annals}, deuterium \cite{Krauth:2016:mud}, and helium-4 ions \cite{Diepold:2016:muHe4theo}, we feel therefore obliged to summarize the current knowledge on the state of theory contributions to the Lamb shift, fine-, and hyperfine structure in muonic helium-3 ions.

The accuracy to be expected from the experiment will be on the order of 20\,GHz, which corresponds to $\sim 0.08\mev$~\footnote{$1\mev~\widehat=~241.799\,\mathrm{GHz}$}. In order to exploit the experimental precision, theory should, ideally, be accurate to a level of
\begin{equation}
\sigma_\mathrm{theory}\sim \mathcal{O}( 0.01\,\mathrm{meV}).
  \label{eq:uncertainty}
\end{equation}  
This would result in a nearly hundred-fold better accuracy in the helion rms charge radius \rh ~compared to the value from electron scattering of 
\begin{equation}
  \rh = 1.973(14)\fm,
\end{equation}  
deduced by Sick \cite{Sick:2014:HeZemach}.

A more precise value has been given by Angeli~\etal~\cite{Angeli:2013:radii}, which should be discarded. Their value is based on a charge radius extraction from \muHef by Carboni~\etal~\cite{Carboni:1978:LS_mu4he} and on the isotope shift measurement from Shiner~\etal~\cite{Shiner:1995:heliumSpec}. The Carboni measurement has however shown to be wrong \cite{Hauser:1992:LS_search}, and the more recent measurement of the electronic isotope shift by van Rooij \etal\ \cite{Rooij:2011:HeSpectroscopy} disagrees by $4\,\sigma$ from the Shiner one \cite{Shiner:1995:heliumSpec}, see Fig.\,\ref{fig:iso_shift}. 

We anticipate here that the total uncertainty in the theoretical calculation of the Lamb shift transition amounts to 0.52\mev (corresponding to a relative uncertainty of $\sim$0.03\%), neglecting the charge radius contribution to be extracted from the \muHet measurement. This value is completely dominated by the two-photon exchange contributions which are difficult to calculate but have seen wonderful progress in recent years \cite{NevoDinur:2016:TPE,Hernandez:2016:POLupdate,Carlson:2016:tpe}. The total uncertainty of the pure QED contributions (without the two-photon exchange) amounts to 0.04\mev and is thus in the desired order of magnitude. Note that while the theory uncertainty from the two-photon exchange in \rp\ is of similar size as the experimental uncertainty (Eq.\,(\ref{eq:mup})), already for \mud the theory uncertainty is vastly dominant (Eq.\,(\ref{eq:mud})). Experiments with muonic atoms are thus a sensitive tool to determine the two-photon exchange contributions.

\section{Overview}
\label{sec:overview}
\begin{figure*}[t]
\centering
 \includegraphics[width=0.75\linewidth]{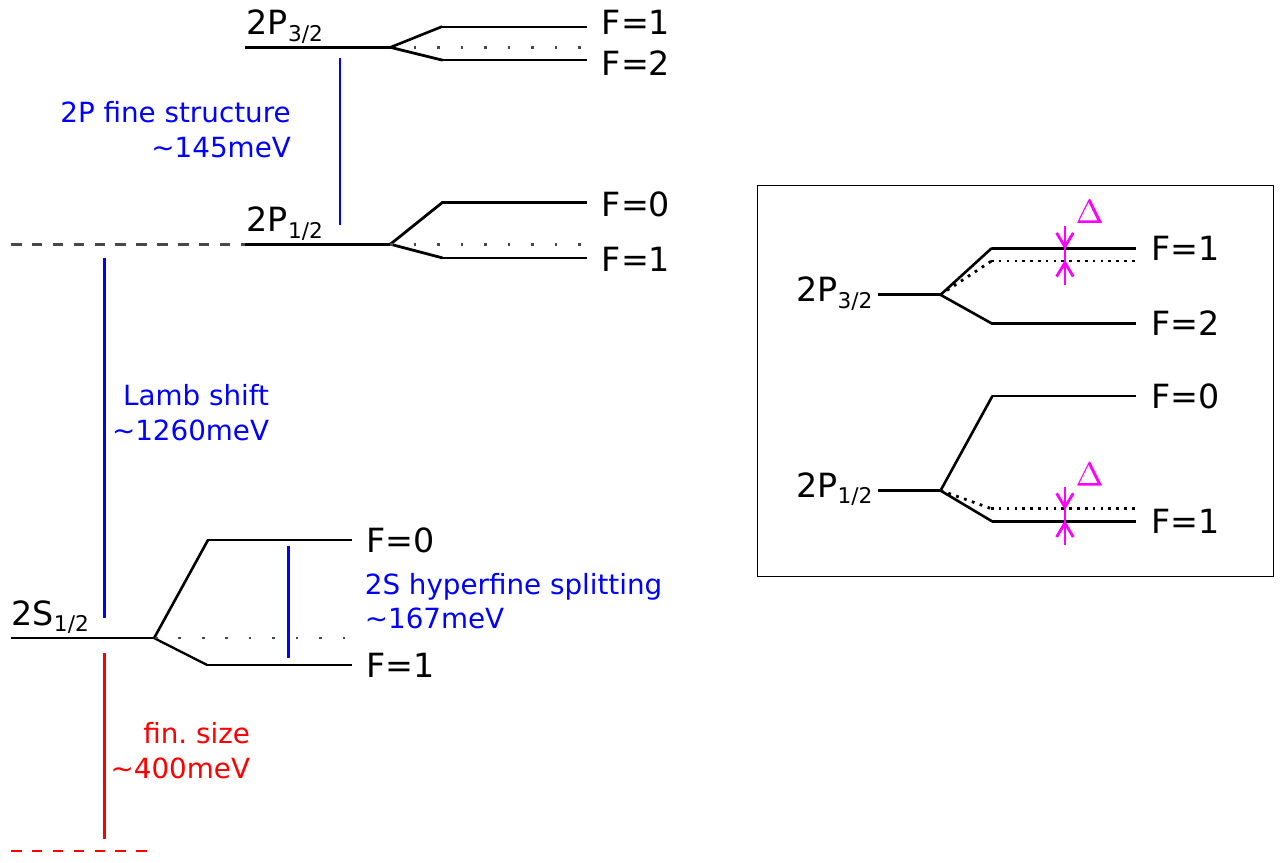}
 \caption{The 2S and 2P energy levels in the muonic helium-3 ion. 
  The inset on the right displays the shift $\Delta$ 
  of the 2P levels due to the mixing of levels with same quantum number $F$, as described in Sec.\,\ref{sec:2Plevels}.
  The figure is not to scale.}
  \label{fig:energy_level}
\end{figure*}

%
The $n=2$ energy levels of the muonic helium-3 ion are sketched in Fig.\,\ref{fig:energy_level}. 
The helion has nuclear spin $I=1/2$, just as the proton. Hence the level scheme is very similar to the one of muonic hydrogen. However, the helion magnetic moment $g=-2.127\,625\,308(25)$ \cite{Mohr:2016:CODATA14} (here given in units of the nuclear magneton) is negative, which swaps the ordering of the hyperfine levels.

A note on the sign convention of the Lamb shift contributions used in this article: The 2S level is shifted below the 2P levels due to the Lamb shift. This means that, fundamentally, the 2S Lamb shift should be given a \textit{negative} sign.\\
However, following long-established conventions we assign the {\em measured} $\TwoSOne \rightarrow \TwoPOne$ energy difference a \textit{positive} sign, i.e. E(2P) -- E(2S) $>$ 0. This is in accord with almost all publications we review here and we will mention explicitly when we have inverted the sign with respect to the original publications where the authors calculated level shifts.\\
Moreover, we obey the traditional definition of the Lamb shift as the terms beyond the Dirac equation and the leading order recoil corrections, i.e.\ excluding effects of the hyperfine structure. In particular, this means that the mixing of the hyperfine levels (Sec.\,\ref{sec:2Plevels}) does {\em not} influence the Lamb shift. 

The Lamb shift 
is dependent on the rms charge radius of the nucleus and is treated in Sec.\,\ref{sec:LS}. We split the Lamb shift contributions into \textit{nuclear structure-independent} contributions and \textit{nuclear structure-dependent} ones. The latter are composed out of one-photon exchange diagrams which represent the finite size effect and two-photon exchange diagrams which contain the polarizability contributions.  

In Sec.\,\ref{sec:HFS}, we treat the 2S hyperfine structure, which depends on the Zemach radius. It also has two-photon exchange contributions. However, these have not been calculated yet and can only be estimated with a large uncertainty.

In Sec.\,\ref{sec:2Plevels}, we compile the 2P level structure which includes fine- and hyperfine splitting, and the mixing of the hyperfine levels \cite{Brodsky:1967:zeemanspectrum}. 


For the theory compilation presented here, we use the calculations from many sources mentioned in the following. The names of the authors of the respective groups are ordered alphabetically.

The first source is E.~Borie who was one of the first to publish detailed calculations of many terms involved in the Lamb shift of muonic atoms. Her most recent calculations for \mup, \mud, \muHef, and \muHet are all found in her Ref.\,\cite{Borie:2012:LS_revisited_AoP}. Several updated versions of this paper are available on the arXiv. In this work we always refer to \cite{Borie:2014:arxiv_v7} which is version-7, the most recent one at the time of this writing.

The second source is the group of Elekina, Faustov, Krutov, and Martynenko \etal~(termed ``Martynenko group" in here for simplicity). The calculations we use in here are found in Krutov \etal~\cite{Krutov:2014:JETP120_73} for the Lamb shift, in Martynenko \etal~\cite{Martynenko:2010:2SHFS_muHe,Martynenko:2008:muheHFS} and Faustov \etal~\cite{Faustov:2014:radrec} for the 2S hyperfine structure, and Elekina \etal~\cite{Elekina:2010:2Pmu3He} for the 2P fine- and hyperfine structure.

Jentschura and Wundt calculated some Lamb shift contributions in their Refs.\,\cite{Jentschura:2011:SemiAnalytic,Jentschura:2011:PRA84_012505}. They are referred to as ``Jentschura'' for simplicity.

The group of Ivanov, Karshenboim, Korzinin, and Shelyuto is referred to ``Karshenboim group'' for simplicity. Their calculations are found in Korzinin \etal~\cite{Korzinin:2013:PRD88_125019} and in Karshenboim \etal~\cite{Karshenboim:2012:PRA85_032509} for Lamb shift and fine structure contributions. 

The group of Bacca, Barnea, Hernandez, Ji, and Nevo Dinur, situated at TRIUMF and Hebrew University, has performed \textit{ab initio} calculations on two-photon exchange contributions of the Lamb shift. Their calculations are found in Nevo Dinur \etal~\cite{NevoDinur:2016:TPE} and Hernandez \etal~\cite{Hernandez:2016:POLupdate}. For simplicity we refer to them as ``TRIUMF-Hebrew group''.

A recent calculation of the two-photon exchange using scattering data and dispersion relations has been performed by Carlson, Gorchtein, and Vanderhaeghen \cite{Carlson:2016:tpe}.

Item numbers \# in our tables
follow the nomenclature in Refs.~\cite{Antognini:2013:Annals,Krauth:2016:mud}.
In the tables, we usually identify the ``source'' of all values
entering ``our choice'' by the first letter of the (group of) authors
given in adjacent columns (e.g.\ ``B'' for Borie).
We denote as average ``avg.'' in the tables the center of the band covered by 
all values $v_i$ under consideration, 
with an uncertainty of half the spread, i.e.\
\begin{equation}
  \label{eq:avg}
  \begin{aligned}
  \mathrm{avg.} ~ = & ~ &
  \frac{1}{2}\,\big[ {\rm MAX}(v_i) + {\rm MIN}(v_i) \big] \\[1ex]
  & \pm &
  \frac{1}{2}\,\big[ {\rm MAX}(v_i) - {\rm MIN}(v_i) \big].
  \end{aligned}
\end{equation}
If individual uncertainties are provided by the authors we add these in quadrature. 
We would like to point out that uncertainties due to uncalculated higher order terms are 
often not indicated explicitly by the authors. In the case some number is given, we include it in our sum. But in general our method can not account for uncertainty estimates of uncalculated higher order terms.

Throughout the paper, 
$Z$ denotes the nuclear charge with $Z=2$ for the helion and alpha particle,
$\alpha$ is the fine structure constant,
$m_r = 199\,m_e$ is the reduced mass of the muon-nucleon system.
``VP'' is short for ``vacuum polarization'',
``SE'' is ``self-energy'',
``RC'' is ``recoil correction''.
``Perturbation theory'' is abbreviated as ``PT'', and SOPT and TOPT denote
$2^{\rm nd}$ and $3^{\rm rd}$ order perturbation theory, respectively.

\section{Lamb shift in muonic helium-3}
\label{sec:LS}
\subsection{Nuclear structure-independent contributions}
\label{sec:LS:QED}

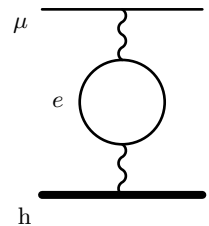
\begin{figure}
  \begin{center}
    \begin{fmffile}{uehling}
      \begin{fmfgraph*}(60,70)
        \fmfstraight
        \fmftopn{t}{3}
        \fmfbottomn{b}{3}
        \fmf{plain,tension=1.0}{t1,t3}
        \fmf{plain,width=3        }{b1,b3}
        \fmf{photon}{t2,c1}
        \fmf{photon}{c3,b2}
        \fmfpoly{smooth, pull=?, tension=0.3}{c0,c1,c2,c3}
        \fmffreeze
        \fmfv{label=h}{b1}
        \fmfv{label.angle=-150,label=$\mu$}{t1}
        \fmfv{label.angle=180,label=$e$}{c2}
      \end{fmfgraph*}
    \end{fmffile}
  \end{center}
  \caption{Item \#1, the leading order 1-loop electron vacuum polarization
    (eVP), also called Uehling term.}    
  \label{fig:uehling}
\end{figure}

Nuclear structure-independent contributions have been calculated by Borie, Martynenko group, Karshenboim group, and Jentschura. The contributions are listed in Tab.\,\ref{tab:LS:QED}, labeled with \#$i$.
The leading contribution is the one-loop electron vacuum polarization (eVP) of order $\alpha(Z\alpha)^2$, the so-called Uehling term (see Fig.\,\ref{fig:uehling}). It accounts for 99.5\% of the radius-independent part of the Lamb shift, so it is very important that this contribution is well understood. There are two different approaches to calculate this term. 

%

Borie \cite{Borie:2014:arxiv_v7} (p.\,4, Tab.) and the Karshenboim group \cite{Korzinin:2013:PRD88_125019} (Tab.\,I) use relativistic Dirac wavefunctions to calculate a relativistic Uehling term (item \#3). 
A relativistic recoil correction (item \#19) has to be added to allow comparison to nonrelativistic calculations (see below).
Borie provides the value of this correction explicitly in \cite{Borie:2014:arxiv_v7} Tab.\,6, whereas the Karshenboim group only gives the total value which includes the correction, thus corresponding to  ($\#3+\#19$).

Nonrelativistic calculations of the Uehling term (item \#1) exist from the Martynenko group \cite{Krutov:2014:JETP120_73} (No.\,1, Tab.\,1) and Jentschura \cite{Jentschura:2011:PRA84_012505}, which are in very good agreement. Additionally, a relativistic correction (item \#2) has to be applied. This relativistic correction already accounts for relativistic recoil effects (item \#19).
Item \#2 has been calculated by the Martynenko group \cite{Krutov:2014:JETP120_73} (No.\,7+10, Tab.\,1), Borie \cite{Borie:2014:arxiv_v7} (Tab.\,1), Jentschura \cite{Jentschura:2011:PRA84_012505,Jentschura:2011:SemiAnalytic} (Eq.\,17), and Karshenboim \etal~\cite{Karshenboim:2012:PRA85_032509}, which agree well within all four groups, however do not have to be included in Borie's and Korzinin \etal's value because their relativistic Dirac wavefunction approach already accounts for relativistic recoil effects.

Both approaches agree well within the required uncertainty. As \textit{our choice} for the Uehling term with relativistic correction ($\#1+\#2$) or ($\#3+\#19$) we take the average
\begin{equation}
\Delta E (\mathrm{Uehling\,+\,rel.~corr.}) = 1642.3962\pm0.0018\mev.
\end{equation}

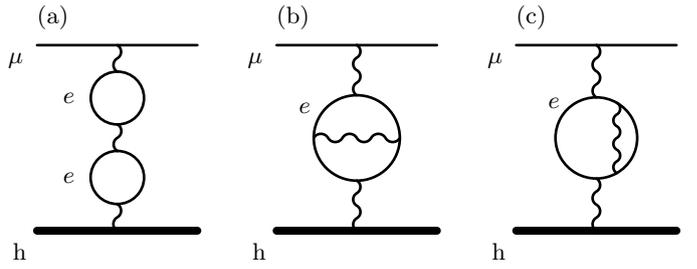
\begin{figure}
  \begin{center}
    \begin{minipage}{0.22\columnwidth}
      (a)\hfill\hspace{5px}
      \vspace{5px}\\
      \begin{fmffile}{item_4}
        \begin{fmfgraph*}(60,70)
          \fmfstraight
          \fmftopn{t}{7}
          \fmfbottomn{b}{7}
          \fmf{plain,tension=1.0}{t1,t7}
          \fmf{plain,width=3        }{b1,b7}
          \fmf{phantom,tension=0.0001}{t4,c1,x1,c2,c4,x2,c5,b4}
          \fmf{photon,tension=0}{t4,c1}
          \fmf{photon,tension=0}{c2,c4}
          \fmf{photon,tension=0}{c5,b4}
          \fmf{plain,left,tension=0}{c1,c2,c1}
          \fmf{plain,left,tension=0}{c4,c5,c4}
          \fmffreeze
          \fmfv{label=h}{b1}
          \fmfv{label.angle=-150,label=$\mu$}{t1}
          \fmfpoly{phantom}{c1,l1,c2,l2}
          \fmfv{label.angle=180,label=$e$}{l1}
          \fmfpoly{phantom}{c4,l3,c5,l4}
          \fmfv{label.angle=180,label=$e$}{l3}
        \end{fmfgraph*}
      \end{fmffile}
    \end{minipage}
    \hspace{1cm}
    \begin{minipage}{0.22\columnwidth}
      (b)\hfill\mbox{~}
      \vspace{5px}\\
      \begin{fmffile}{item_4b}
        \begin{fmfgraph*}(60,70)
          \fmfstraight
          \fmftopn{t}{3}
          \fmfbottomn{b}{3}
          \fmf{plain}{t1,t2,t3}
          \fmf{plain,width=3}{b1,b2,b3}
          \fmf{photon}{t2,c0}
          \fmf{photon}{c4,b2}
          \fmfpoly{smooth,pull=?,tension=0.5}{c0,c1,c2,c3,c4,c5,c6,c7}
          \fmffreeze
          \fmf{photon}{c2,c6}
          \fmfv{label=h}{b1}
          \fmfv{label.angle=-150,label=$\mu$}{t1}
          \fmfv{label.angle=180,label=$e$}{c1}
        \end{fmfgraph*}
      \end{fmffile}
    \end{minipage}
    \hspace{1cm}
    \begin{minipage}{0.22\columnwidth}
      (c)\hfill\mbox{~}
      \vspace{5px}\\
      \begin{fmffile}{item_4c}
        \begin{fmfgraph*}(60,70)
          \fmfstraight
          \fmftopn{t}{3}
          \fmfbottomn{b}{3}
          \fmf{plain}{t1,t2,t3}
          \fmf{plain,width=3}{b1,b2,b3}
          \fmf{photon}{t2,c0}
          \fmf{photon}{c6,b2}
          \fmfpoly{smooth,pull=?,tension=0.8}{c0,c1,c2,c3,c4,c5,c6,c7,c8,c9,c10,c11}
          \fmffreeze
          \fmf{photon}{c7,c11}
          \fmfv{label=h}{b1}
          \fmfv{label.angle=-150,label=$\mu$}{t1}
          \fmfv{label.angle=180,label=$e$}{c1}
        \end{fmfgraph*}
      \end{fmffile}
    \end{minipage}
  \end{center}
  \vspace{-10px}
  \caption{Item \#4, the two-loop eVP 
    (K\"allen-Sabry) contribution. 
    This is Fig.\,1 (b,c,d) from the Martynenko group~\cite{Krutov:2014:JETP120_73}.
  }
  \label{fig:item_4}
\end{figure}

Item \#4, the second largest contribution in this section, is the two-loop eVP of order $\alpha^2(Z\alpha)^2$, the so-called K\"all\'en-Sabry term \cite{KallenSabry:1955} (see Fig.\,\ref{fig:item_4}). It has been calculated by Borie \cite{Borie:2014:arxiv_v7} (p.\,4, Tab.) and the Martynenko group \cite{Krutov:2014:JETP120_73} (No.\,2, Tab.\,1) which agree within 0.0037\mev. As \textit{our choice} we take the average.

\begin{figure}
  \begin{center}
      \mbox{~}\vfill
      \begin{fmffile}{item_5}
        \begin{fmfgraph*}(100,70)
          \fmfstraight
          \fmftopn{t}{8}
          \fmfbottomn{b}{8}
          \fmf{plain,tension=1.0}{t1,t8}
          \fmf{plain,width=3        }{b1,b8}
          \fmf{phantom,tension=0.0001}{t3,c1,c2,b3}
          \fmf{photon,tension=0}{t3,c1}
          \fmf{photon,tension=0}{c2,b3}
          \fmf{plain,left,tension=0}{c1,c2,c1}
          \fmf{phantom,tension=0.0001}{t6,c3,c4,b6}
          \fmf{photon,tension=0}{t6,c3}
          \fmf{photon,tension=0}{c4,b6}
          \fmf{plain,left,tension=0}{c3,c4,c3}
          \fmffreeze
          \fmfv{label=h}{b1}
          \fmfv{label.angle=-150,label=$\mu$}{t1}
          \fmfpoly{phantom}{c1,l1,c2,l2}
          \fmfv{label.angle=110,label.dist=10,label=$e$}{l1}
          \fmfpoly{phantom}{c3,l3,c4,l4}
          \fmfv{label.angle=100,label.dist=10,label=$e$}{l3}
        \end{fmfgraph*}
      \end{fmffile}
  \end{center}
  \caption{Item \#5, the one-loop eVP in 2-Coulomb lines.}
  \label{fig:item_5}
\end{figure}
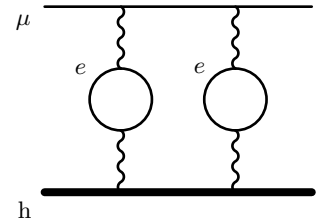

Item \#5 is the one-loop eVP in two Coulomb lines of order $\alpha^2(Z\alpha)^2$ (see Fig.\,\ref{fig:item_5}). It has been calculated by Borie \cite{Borie:2014:arxiv_v7} (Tab.\,6), the Martynenko group \cite{Krutov:2014:JETP120_73} (No.\,9, Tab.\,1), and Jentschura \cite{Jentschura:2011:SemiAnalytic} (Eq.\,13) of whom the latter two obtain the same result, which differs from Borie by 0.0033\mev. As \textit{our choice} we adopt the 
average.

The Karshenboim group \cite{Korzinin:2013:PRD88_125019} (Tab.\,I) has calculated the sum of item \#4 and \#5, the two-loop eVP (K\"all\'en-Sabry) and one-loop eVP in two Coulomb lines (Fig.\,\ref{fig:item_4} and \ref{fig:item_5}).
Good agreement between all groups is observed.
%
%

Item \#6+7 is the third order eVP of order $\alpha^3(Z\alpha)^2$. It has been calculated by the Martynenko group \cite{Krutov:2014:JETP120_73} (No.\,$4+11+12$, Tab.\,1) and the Karshenboim group \cite{Korzinin:2013:PRD88_125019} (Tab.\,I). Borie \cite{Borie:2014:arxiv_v7} (p.\,4) adopts the value from Karshenboim \etal. Martynenko \etal\ and Karshenboim \etal\ differ by $0.004\mev$, which is in agreement considering the uncertainty of $0.003\mev$ given by the Martynenko group. As \textit{our choice} we adopt the average and obtain an uncertainty of 0.0036\mev 
via Gaussian propagation of uncertainty.

Item \#29 is the second order eVP of order $\alpha^2(Z\alpha)^4$. It has been calculated by the Martynenko group \cite{Krutov:2014:JETP120_73} (No.\,$8+13$, Tab.\,1) and the Karshenboim group \cite{Korzinin:2013:PRD88_125019} (Tab.\,VIII). Their values did agree in the case of \mud, however for \muHet they differ by 0.004\mev. 
This difference is twice as large as the value from Martynenko \etal\ but this contribution is small, so the uncertainty is not at all dominating. We reflect the difference by adopting the average as \textit{our choice}.

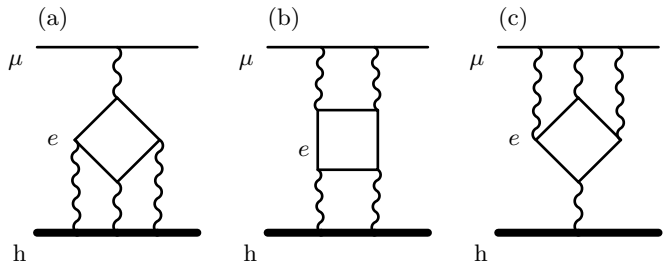
\begin{figure}
  \begin{center}
    \begin{minipage}{0.22\columnwidth}
      (a)\hfill\mbox{~}
      \vspace{5px}\\
      \begin{fmffile}{lblone}
        \begin{fmfgraph*}(60,70)
          \fmfstraight
          \fmftop{i1,v1,o1}
          \fmfbottom{i2,v2,v3,v4,o2}
          \fmf{plain}{i1,v1,o1}
          \fmf{plain,width=3}{i2,v2,v3,v4,o2}

          \fmf{photon}{v1,c0}
          \fmf{photon}{c2,v3}

          \fmfpoly{default, pull=?, tension=0.3}{c0,c1,c2,c3}
          \fmffreeze

          \fmf{photon}{c1,v2}
          \fmf{photon}{c3,v4}
          \fmfv{label.angle=180,label=$e$}{c1}
          \fmfv{label=h}{i2}
          \fmfv{label.angle=-150,label=$\mu$}{i1}
        \end{fmfgraph*}
      \end{fmffile}
    \end{minipage}
    \hspace{25px}
    \begin{minipage}{0.22\columnwidth}
      (b)\hfill\mbox{~}
      \vspace{5px}\\
      \begin{fmffile}{lbltwo}
        \begin{fmfgraph*}(60,70)
          \fmfstraight
          \fmftop{i1,v1,v2,o1}
          \fmfbottom{i2,v3,v4,o2}
          \fmf{plain}{i1,v1,v2,o1}
          \fmf{plain,width=3}{i2,v3,v4,o2}

          \fmf{photon}{v1,c0}
          \fmf{photon}{v2,c3}
          \fmf{photon}{v3,c1}
          \fmf{photon}{v4,c2}

          \fmfpoly{default, pull=?, tension=0.5}{c0,c1,c2,c3}

          \fmfv{label.angle=120,label=$e$}{c1}
          \fmfv{label=h}{i2}
          \fmfv{label.angle=-150,label=$\mu$}{i1}
        \end{fmfgraph*}
      \end{fmffile}
      \end{minipage}
    \hspace{25px}
    \begin{minipage}{0.22\columnwidth}
      (c)\hfill\mbox{~}
      \vspace{5px}\\
      \begin{fmffile}{lblthree}
        \begin{fmfgraph*}(60,70)
          \fmfstraight
          \fmftop{i1,v1,v3,v4,o1}
          \fmfbottom{i2,v2,o2}
          \fmf{plain}{i1,v1,v3,v4,o1}
          \fmf{plain,width=3}{i2,v2,o2}

          \fmf{photon}{v3,c0}
          \fmf{photon}{c2,v2}

          \fmfpoly{default, pull=?, tension=0.3}{c0,c1,c2,c3}
          \fmffreeze

          \fmf{photon}{v1,c1}
          \fmf{photon}{v4,c3}
          \fmfv{label.angle=180,label=$e$}{c1}
          \fmfv{label=h}{i2}
          \fmfv{label.angle=-150,label=$\mu$}{i1}
        \end{fmfgraph*}
      \end{fmffile}
      \end{minipage}
  \end{center}
  \vspace{-10px}
  \caption{The three contributions to Light-by-light scattering:
    (a) Wichmann-Kroll or ``1:3'' term, item~\#9,
    (b) Virtual Delbr\"uck or ``2:2'' term, item~\#10, and 
    (c) inverted Wichmann-Kroll or ``3:1'' term, item~\#9a$^\dagger$.}
  \label{fig:lbl}
\end{figure}

Items \#9, \#10, and \#9a are the terms of the Light-by-light (LbL) scattering contribution (see Fig.\,\ref{fig:lbl}). The sum of the LbL terms is calculated by the Karshenboim group \cite{Korzinin:2013:PRD88_125019} (Tab.\,I). Borie \cite{Borie:2014:arxiv_v7} also lists the value from Karshenboim \etal. 
Item \#9 is the \textit{Wichmann-Kroll} term, or ``1:3'' LbL, which is of order $\alpha(Z\alpha)^4$. This item has also been calculated by Borie \cite{Borie:2014:arxiv_v7} (p.\,4) and the Martynenko group \cite{Krutov:2014:JETP120_73} (No.\,5, Tab.\,1) who obtain the same result. 
Item \#10 is the \textit{virtual Delbr\"uck} or ``2:2'' LbL, which is of order $\alpha^2(Z\alpha)^3$. 
Item \#9a is the \textit{inverted Wichmann-Kroll} term, or ``3:1'' LbL, which is of order $\alpha^3(Z\alpha)^2$. The sum of the latter two is also given by the Martynenko group \cite{Krutov:2014:JETP120_73} (No.\,6, Tab.\,1). 
As \textit{our choice} we use the one from Karshenboim \etal, who are the first and only group to calculate all three LbL contributions. The groups are in agreement when taking into account the uncertainty of 0.0006\mev given by Karshenboim \etal.

\begin{figure}
  \begin{center}
     \hfill~(a)\hspace{60px}~\hspace{25px}~(b)\hspace{60px}~\hfill\vspace{5px}\\
      \begin{fmffile}{item_20}
        \begin{fmfgraph*}(70,60)
          \fmfstraight
          \fmftopn{t}{7}
          \fmfbottomn{b}{7}
          \fmf{plain}{t1,t7}
          \fmf{plain,width=3}{b1,b7}
          \fmf{photon,tension=0.1,left=1.0}{t3,t5}
          \fmf{photon}{t4,b4}
          \fmfv{label=h}{b1}
          \fmfv{label.angle=-150,label=$\mu$}{t1}
        \end{fmfgraph*}
        \hspace{40px}
        \begin{fmfgraph*}(70,60)
          \fmfstraight
          \fmftopn{t}{7}
          \fmfbottomn{b}{7}
          \fmf{plain}{t1,t7}
          \fmf{plain,width=3}{b1,b7}
          \fmf{photon}{t4,c1}
          \fmf{photon}{c2,b4}
          \fmfpoly{smooth, pull=?, tension=0.3}{c1,l1,c2,l2}
          \fmfv{label.angle=180,label=$\mu$}{l1}
          \fmfv{label=h}{b1}
          \fmfv{label.angle=-150,label=$\mu$}{t1}
        \end{fmfgraph*}
    \end{fmffile}
  \end{center}
  \caption{Item \#20, the muon-self energy (a) and the muon vacuum polarization (b), $\alpha(Z\alpha)^4$.   }
  \label{fig:onephotonSE}
\end{figure}
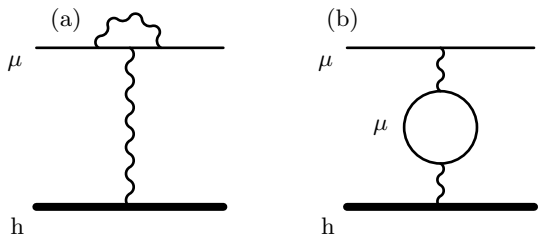

Item \#20 is the contribution from muon self-energy ($\mu$SE) and muon vacuum polarization ($\mu$VP) of order 
$\alpha(Z\alpha)^4$ (see Fig.\,\ref{fig:onephotonSE}). This item constitutes the third largest term in this section~\footnote{In ordinary hydrogen-like atoms this term is the leading order Lamb shift contribution: The leptons in the loop are the same as the orbiting lepton. This term can thus be rescaled from well-known results in hydrogen.}. This item has been calculated by Borie \cite{Borie:2014:arxiv_v7} (Tab.\,2, Tab.\,6) and the Martynenko group \cite{Krutov:2014:JETP120_73} (No.\,24, Tab.\,1). They differ by 0.001\mev. As \textit{our choice} we adopt the average. 

Items \#11, \#12, \#30, \#13, and \#31 are all corrections to VP or $\mu$SE and of order $\alpha^2(Z\alpha)^4$.

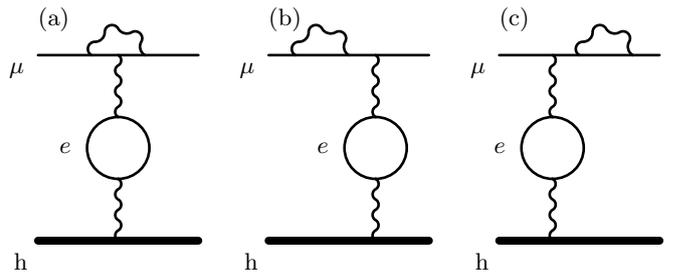
\begin{figure}
  \begin{center}
    \begin{minipage}{0.22\columnwidth}
      (a)\hfill\mbox{~}
      \vspace{8px}\\
      \begin{fmffile}{item_11a}
        \begin{fmfgraph*}(60,70)
          \fmfstraight
          \fmftopn{t}{7}
          \fmfbottomn{b}{7}
          \fmf{plain,tension=1.0}{t1,t7}
          \fmf{plain,width=3        }{b1,b7}
          \fmf{photon,tension=0.1,left=1.0}{t3,t5}
          \fmf{phantom,tension=0.0001}{t4,c1,c2,b4}
          \fmf{photon,tension=0}{t4,c1}
          \fmf{photon,tension=0}{c2,b4}
          \fmf{plain,left,tension=0}{c1,c2,c1}
          \fmffreeze
          \fmfpoly{phantom}{c1,l1,c2,l2}
          \fmfv{label.angle=180,label=$e$}{l1}
          \fmfv{label=h}{b1}
          \fmfv{label.angle=-150,label=$\mu$}{t1}
        \end{fmfgraph*}
      \end{fmffile}
    \end{minipage}
    \hspace{25px}
    \begin{minipage}{0.22\columnwidth}
      (b)\hfill\mbox{~}
      \vspace{8px}\\
      \begin{fmffile}{item_11b}
        \begin{fmfgraph*}(60,70)
          \fmfstraight
          \fmftopn{t}{7}
          \fmfbottomn{b}{7}
          \fmf{plain,tension=1.0}{t1,t7}
          \fmf{plain,width=3        }{b1,b7}
          \fmf{photon,tension=0.1,left=1.0}{t2,t4}
          \fmf{phantom,tension=0.0001}{t5,c1,c2,b5}
          \fmf{photon,tension=0}{t5,c1}
          \fmf{photon,tension=0}{c2,b5}
          \fmf{plain,left,tension=0}{c1,c2,c1}
          \fmffreeze
          \fmfpoly{phantom}{c1,l1,c2,l2}
          \fmfv{label.angle=180,label=$e$}{l1}
          \fmfv{label=h}{b1}
          \fmfv{label.angle=-150,label=$\mu$}{t1}
       \end{fmfgraph*}
      \end{fmffile}
    \end{minipage}
    \hspace{25px}
    \begin{minipage}{0.22\columnwidth}
      (c)\hfill\mbox{~}
      \vspace{8px}\\
      \begin{fmffile}{item_11c}
        \begin{fmfgraph*}(60,70)
          \fmfstraight
          \fmftopn{t}{7}
          \fmfbottomn{b}{7}
          \fmf{plain,tension=1.0}{t1,t7}
          \fmf{plain,width=3        }{b1,b7}
          \fmf{photon,tension=0.1,left=1.0}{t4,t6}
          \fmf{phantom,tension=0.0001}{t3,c1,c2,b3}
          \fmf{photon,tension=0}{t3,c1}
          \fmf{photon,tension=0}{c2,b3}
          \fmf{plain,left,tension=0}{c1,c2,c1}
          \fmffreeze
          \fmfpoly{phantom}{c1,l1,c2,l2}
          \fmfv{label.angle=180,label=$e$}{l1}
          \fmfv{label=h}{b1}
          \fmfv{label.angle=-150,label=$\mu$}{t1}
        \end{fmfgraph*}
      \end{fmffile}
    \end{minipage}
  \end{center}
  \vspace{-10px}
  \caption{Item~\#11, muon self-energy corrections to the electron vacuum polarization
    $\alpha^2 (Z\alpha)^4$.
    This figure is Fig.\,2 from Jentschura~\cite{Jentschura:2011:AnnPhys1}.
    It corresponds to Fig.~6(a) from Karshenboim~\cite{Korzinin:2013:PRD88_125019}.}
  \label{fig:item_11}
\end{figure}

Item \#11 is the $\mu$SE correction to eVP (see Fig.\,\ref{fig:item_11}). It has been calculated by all four groups. Martynenko \etal\ calculate this term (Eq.\,99) in \cite{Krutov:2014:JETP120_73}, however in their table (No.\,28) they use the more exact calculation from  Jentschura. Jentschura \cite{Jentschura:2011:SemiAnalytic} (Eq.\,29), and the Karshenboim group \cite{Korzinin:2013:PRD88_125019} (Tab.\,VIII a) are in excellent agreement. Borie \cite{Borie:2014:arxiv_v7} (Tab.\,16) differs significantly because she only calculates a part of this contribution in her App.\,C. This value does not enter her sum and thus is also not considered in here. On p.\,12 of \cite{Borie:2014:arxiv_v7} she states that this value should be considered as an uncertainty. As \textit{our choice} we adopt the number from Jentschura and Karshenboim \etal.

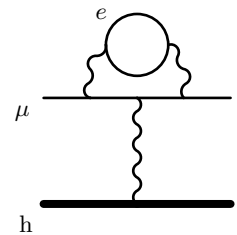
\begin{figure}

  \begin{center}
    %
    %
    \begin{minipage}{0.5\columnwidth}
      \mbox{~}\vfill
      \centering
      \begin{fmffile}{Mart_fig11b}
        \begin{fmfgraph*}(70,80)
          \fmfstraight
          \fmfleftn{i}{9}
          \fmfrightn{o}{9}
          \fmf{plain,tension=1.0}{i5,t1,t2,t3,o5}
          \fmf{plain,width=3}{i1,b2,o1}
          \fmf{photon,tension=0}{t2,b2}
          \fmf{phantom}{i7,c1}
          \fmf{phantom}{o7,c2}
          \fmf{photon,tension=0,left=0.3}{t1,c1}
          \fmf{photon,tension=0,left=0.3}{c2,t3}
          \fmf{plain,left,tension=0.5}{c1,c2,c1}
          \fmffreeze
          \fmfv{label.angle=100,label.dist=10,label=$e$}{c1}
          \fmfv{label=h}{i1}
          \fmfv{label.angle=-150,label=$\mu$}{i5}
        \end{fmfgraph*}
      \end{fmffile}
    \end{minipage}
  \end{center}
  \caption{Item \#12, {\em eVP loop in SE} are radiative corrections with VP effects.
    This is Fig.\,11(b) from a publication by the Martynenko group~\cite{Krutov:2014:JETP120_73} which
  is the same as Fig.\,4 in Pachucki~\cite{Pachucki:1996:LSmup}.
  It is Karshenboim's Fig.\,6(d) in Ref.~\cite{Korzinin:2013:PRD88_125019}.}
  \label{fig:item_12}
\end{figure}

Item \#12 is the eVP in $\mu$SE (see Fig.\,\ref{fig:item_12}). This item has been calculated by the Martynenko group \cite{Krutov:2014:JETP120_73} (No.\,27, Tab.\,1) and the Karshenboim group \cite{Korzinin:2013:PRD88_125019} (Tab.\,VIII d), which are in perfect agreement. On p.\,10 of \cite{Borie:2014:arxiv_v7} Borie mentions that she included the ``fourth order electron loops'' in ``muon Lamb shift, higher order'' term, which is our item \#21. As we include item \#21 from Borie, we will not on top include item \#12.

\begin{figure}
  \begin{center}
    \begin{minipage}{0.5\columnwidth}
      \mbox{~}\vfill
      \centering
      \begin{fmffile}{item_30}
        \begin{fmfgraph*}(70,80)
          \fmfstraight
          \fmfleftn{i}{9}
          \fmfrightn{o}{9}
          \fmf{plain,tension=1.0}{i5,t1,t2,t3,o5}
          \fmf{plain,width=3}{i1,b2,o1}
          \fmf{photon,tension=0}{t2,b2}
          \fmf{phantom}{i7,c1}
          \fmf{phantom}{o7,c2}
          \fmf{photon,tension=0,left=0.3}{t1,c1}
          \fmf{photon,tension=0,left=0.3}{c2,t3}
          \fmf{dashes,left,tension=0.5}{c1,c2,c1}
          \fmffreeze
          \fmfv{label.angle=100,label.dist=10,label=$h$}{c1}
          \fmfv{label=h}{i1}
          \fmfv{label.angle=-150,label=$\mu$}{i5}
        \end{fmfgraph*}
      \end{fmffile}
    \end{minipage}
  \end{center}
  \caption{Item \#30, hadronic VP in SE contribution,
  corresponds to Fig.~6(e) in Karshenboim \etal's~\cite{Korzinin:2013:PRD88_125019}.}
    \label{fig:item_30}
\end{figure}
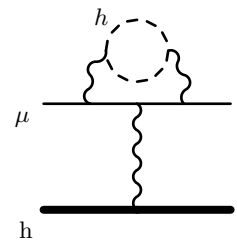

Item \#30 is the hadronic vacuum polarization (hVP) in $\mu$SE (see Fig.\,\ref{fig:item_30}). This item has only been calculated by the Karshenboim group \cite{Korzinin:2013:PRD88_125019} (Tab.\,VIII e) which we adopt as \textit{our choice}.

\begin{figure}
  \begin{center}
    \begin{minipage}{0.22\columnwidth}
      (a)\hfill\mbox{~}
      \vspace{5px}\\
      \begin{fmffile}{item_13a}
        \begin{fmfgraph*}(60,70)
          \fmfstraight
          \fmftopn{t}{7}
          \fmfbottomn{b}{7}
          \fmf{plain,tension=1.0}{t1,t7}
          \fmf{plain,width=3        }{b1,b7}
          \fmf{phantom,tension=0.0001}{t4,c1,x1,c2,c4,x2,c5,b4}
          \fmf{photon,tension=0}{t4,c1}
          \fmf{photon,tension=0}{c2,c4}
          \fmf{photon,tension=0}{c5,b4}
          \fmf{plain,left,tension=0}{c1,c2,c1}
          \fmf{plain,left,tension=0}{c4,c5,c4}
          \fmffreeze
          \fmfv{label=h}{b1}
          \fmfv{label.angle=-150,label=$\mu$}{t1}
          \fmfpoly{phantom}{c1,l1,c2,l2}
          \fmfv{label.angle=180,label=$e$}{l1}
          \fmfpoly{phantom}{c4,l3,c5,l4}
          \fmfv{label.angle=180,label=$\mu$}{l3}
        \end{fmfgraph*}
      \end{fmffile}
    \end{minipage}
    \hspace{40px}
    \begin{minipage}{0.35\columnwidth}
      (b)\hfill\mbox{~}
      \vspace{5px}\\
      \begin{fmffile}{item_13b}
        \begin{fmfgraph*}(100,70)
          \fmfstraight
          \fmftopn{t}{8}
          \fmfbottomn{b}{8}
          \fmf{plain,tension=1.0}{t1,t8}
          \fmf{plain,width=3        }{b1,b8}
          \fmf{phantom,tension=0.0001}{t3,c1,c2,b3}
          \fmf{photon,tension=0}{t3,c1}
          \fmf{photon,tension=0}{c2,b3}
          \fmf{plain,left,tension=0}{c1,c2,c1}
          \fmf{phantom,tension=0.0001}{t6,c3,c4,b6}
          \fmf{photon,tension=0}{t6,c3}
          \fmf{photon,tension=0}{c4,b6}
          \fmf{plain,left,tension=0}{c3,c4,c3}
          \fmffreeze
          \fmfv{label=h}{b1}
          \fmfv{label.angle=-150,label=$\mu$}{t1}
          \fmfpoly{phantom}{c1,l1,c2,l2}
          \fmfv{label.angle=110,label.dist=10,label=$e$}{l1}
          \fmfpoly{phantom}{c3,l3,c4,l4}
          \fmfv{label.angle=100,label.dist=10,label=$\mu$}{l3}
        \end{fmfgraph*}
      \end{fmffile}
    \end{minipage}
  \end{center}
  \vspace{-10px}
  \caption{Item \#13, the mixed eVP-$\mu$VP contribution.}    
  \label{fig:item_13}
\end{figure}
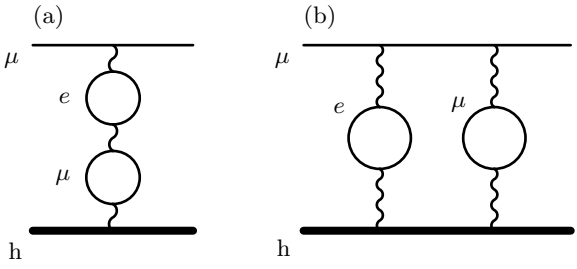

Item \#13 is the mixed $e$VP + $\mu$VP (see Fig.\,\ref{fig:item_13}). The calculations from Borie \cite{Borie:2014:arxiv_v7} (p.\,4) and the Martynenko group \cite{Krutov:2014:JETP120_73} (No.\,3, Tab.\,1) roughly agree, whereas the value from the Karshenboim group \cite{Korzinin:2013:PRD88_125019} (Tab.\,VIII b) is 0.002\mev larger. As \textit{our choice} we take the average.

\begin{figure}
  \begin{center}
    \begin{minipage}{0.22\columnwidth}
      (a)\hfill\mbox{~}
      \vspace{5px}\\
      \begin{fmffile}{item_31a}
        \begin{fmfgraph*}(60,70)
          \fmfstraight
          \fmftopn{t}{7}
          \fmfbottomn{b}{7}
          \fmf{plain,tension=1.0}{t1,t7}
          \fmf{plain,width=3        }{b1,b7}
          \fmf{phantom,tension=0.0001}{t4,c1,x1,c2,c4,x2,c5,b4}
          \fmf{photon,tension=0}{t4,c1}
          \fmf{photon,tension=0}{c2,c4}
          \fmf{photon,tension=0}{c5,b4}
          \fmf{plain,left,tension=0}{c1,c2,c1}
          \fmf{dashes,left,tension=0}{c4,c5,c4}
          \fmffreeze
          \fmfv{label=h}{b1}
          \fmfv{label.angle=-150,label=$\mu$}{t1}
          \fmfpoly{phantom}{c1,l1,c2,l2}
          \fmfv{label.angle=180,label=$e$}{l1}
          \fmfpoly{phantom}{c4,l3,c5,l4}
          \fmfv{label.angle=180,label=$h$}{l3}
        \end{fmfgraph*}
      \end{fmffile}
    \end{minipage}
    \hspace{40px}
    \begin{minipage}{0.35\columnwidth}
      (b)\hfill\mbox{~}
      \vspace{5px}\\
      \begin{fmffile}{item_31b}
        \begin{fmfgraph*}(100,70)
          \fmfstraight
          \fmftopn{t}{8}
          \fmfbottomn{b}{8}
          \fmf{plain,tension=1.0}{t1,t8}
          \fmf{plain,width=3        }{b1,b8}
          \fmf{phantom,tension=0.0001}{t3,c1,c2,b3}
          \fmf{photon,tension=0}{t3,c1}
          \fmf{photon,tension=0}{c2,b3}
          \fmf{plain,left,tension=0}{c1,c2,c1}
          \fmf{phantom,tension=0.0001}{t6,c3,c4,b6}
          \fmf{photon,tension=0}{t6,c3}
          \fmf{photon,tension=0}{c4,b6}
          \fmf{dashes,left,tension=0}{c3,c4,c3}
          \fmffreeze
          \fmfv{label=h}{b1}
          \fmfv{label.angle=-150,label=$\mu$}{t1}
          \fmfpoly{phantom}{c1,l1,c2,l2}
          \fmfv{label.angle=110,label.dist=10,label=$e$}{l1}
          \fmfpoly{phantom}{c3,l3,c4,l4}
          \fmfv{label.angle=100,label.dist=10,label=$h$}{l3}
        \end{fmfgraph*}
      \end{fmffile}
    \end{minipage}
  \end{center}
  \vspace{-10px}
  \caption{Item \#31, the mixed eVP- and hadronic VP contribution,
    comes from the Uehling correction to the hadronic VP correction.
    See Fig.~6(c) in Karshenboim \etal's~\cite{Korzinin:2013:PRD88_125019}.}
    \label{fig:item_31}
\end{figure}
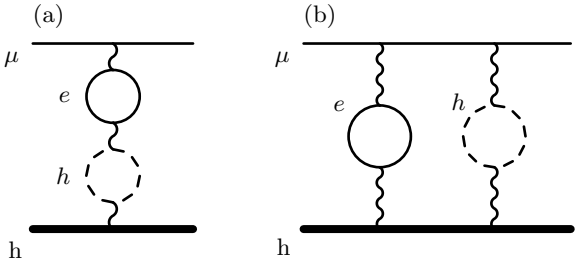

Item \#31 is the mixed $e$VP + hVP (see Fig.\,\ref{fig:item_31}) which has only been calculated by the Karshenboim group \cite{Korzinin:2013:PRD88_125019} (Tab.\,VIII c). We adopt their value as \textit{our choice}.

\begin{figure}
  \begin{center}
      \begin{fmffile}{item_32}
        \begin{fmfgraph*}(70,70)
          \fmfstraight
          \fmfleftn{i}{9}
          \fmfrightn{o}{9}
          \fmf{plain,tension=1.0}{i5,t1,t2,t3,o5}
          \fmf{plain,width=3}{i1,b2,o1}
          \fmf{photon,tension=0}{t2,b2}
          \fmf{phantom}{i7,c1}
          \fmf{phantom}{o7,c2}
          \fmf{photon,tension=0,left=0.3}{t1,c1}
          \fmf{photon,tension=0,left=0.3}{c2,t3}
          \fmf{plain,left,tension=0.5}{c1,c2,c1}
          \fmffreeze
          \fmfv{label.angle=100,label.dist=10,label=$\mu$}{c1}
          \fmfv{label=h}{i1}
          \fmfv{label.angle=-150,label=$\mu$}{i5}
        \end{fmfgraph*}
      \end{fmffile}
      \end{center}
  \caption{Item \#32, muon VP in SE contribution, is automatically included
  in a rescaled electronic $^3$He$^+$ QED value of higher order SE contributions
  (see text).}
  \label{fig:item_32}
\end{figure}
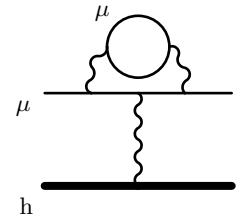

Item \#32, the muon VP in SE correction shown in Fig.\,\ref{fig:item_32} is not included as a separate item in our Tab.\,\ref{tab:LS:QED}. It should already be automatically included in the QED contribution which has been rescaled from the QED of electronic $^3$He$^+$ by a simple mass replacement $m_e\rightarrow m_\mu$ \cite{Karshenboim:PC:2015}. 
This is the case only for QED contributions where the particle in the loop is the same as the bound particle - like in this case, a muon VP correction in a muonic atom. The size of this item \#32 can be estimated from the relationship found by Borie \cite{Borie:1981:HVP}, that the ratio of hadronic to muonic VP is 0.66. With the Karshenboim group's value of item \#30 \cite{Korzinin:2013:PRD88_125019} one would obtain a value for item \#32 of $-0.0004/0.66\mev = -0.0006\mev$. This contribution is contained in our item \#21, together with the dominating item \#12 (see also p.\,10 of Ref.\,\cite{Borie:2014:arxiv_v7}). 

Item \#21 is a higher-order correction to $\mu$SE and $\mu$VP of order $\alpha^2(Z\alpha)^4$ and $\alpha^2(Z\alpha)^6$. This item has only been calculated by Borie \cite{Borie:2014:arxiv_v7} (Tab.\,2, Tab.\,6). On p.\,10 she points out that this contribution includes the ``fourth order electron loops'', which is our item \#12. It also contains our item \#32. We adopt her value as \textit{our choice}.

Item \#14 is the hadronic VP of order $\alpha(Z\alpha)^4$. 
It has been calculated by Borie \cite{Borie:2014:arxiv_v7} (Tab.\,6) and the Martynenko group \cite{Krutov:2014:JETP120_73} (No.\,29, Tab.\,1). 
Borie assigns a 5\% uncertainty to their value. However, in her Ref.\,\cite{Borie:2014:arxiv_v7} there are two different values of item \#14, the first on p.\,5 (0.219\mev) and the second in Tab.\,6 on p.\,16 (0.221\mev). Regarding the given uncertainty this difference is not of interest. In our Tab.\,\ref{tab:LS:QED}, we report the larger value which is further from that of the Martynenko group in order to conservatively reflect the scatter.
Martynenko \etal\ did not assign an uncertainty to their value. However, for \mud \cite{Krutov:2011:PRA84_052514} they estimated an uncertainty of 5\%. As \textit{our choice} we take the average of their values and adopt the uncertainty of 5\% (0.011\mev). 

Item \#17 is the Barker-Glover correction \cite{Barker:1955}. It is a recoil correction of order $(Z\alpha)^4m_r^3/M^2$ and includes the nuclear Darwin-Foldy term that arises due to the Zitterbewegung of the nucleus. As already discussed in App.\,A of \cite{Krauth:2016:mud}, we follow the atomic physics convention \cite{Jentschura:2011:DF}, which is also adopted by CODATA in their report from 2010 \cite{Mohr:2012:CODATA10} and 2014 \cite{Mohr:2016:CODATA14}. This convention implies that item \#17 is considered as a recoil correction to the energy levels and not as a part of the rms charge radius.
This term has been calculated by Borie \cite{Borie:2014:arxiv_v7} (Tab.\,6), the Martynenko group \cite{Krutov:2014:JETP120_73} (No.\,21, Tab.\,1), and Jentschura \cite{Jentschura:2011:PRA84_012505} and \cite{Jentschura:2011:SemiAnalytic} (Eq.\,A.3). As \textit{our choice} we use the number given by Borie and Jentschura as they give one more digit.

Item \#18 is the term called ``recoil, finite size'' by Borie. It is of order $(Z\alpha)^5\langle r \rangle_{(2)}/M$ and is linear in the \textit{first} Zemach moment. It has first been calculated by Friar \cite{Friar:1978:Annals} (see Eq.\,F5 in App.\,F) for hydrogen and has later been given by Borie \cite{Borie:2014:arxiv_v7} for \mud, \muHef, and \muHet. We discard item \#18 because it is considered to be included in the elastic TPE \cite{Pachucki:PC:2015,Yerokhin:2016:RCFS}. It has also been discarded in \mup \cite{Antognini:2013:Annals}, \mud \cite{Krauth:2016:mud}, and \muHef \cite{Diepold:2016:muHe4theo}. For the muonic helium-3 ion, item \#18 in \cite{Borie:2014:arxiv_v7} (Tab.\,6) amounts to 0.4040\mev, which is five times larger than the experimental uncertainty of about 0.08\mev (see Eq.~\ref{eq:uncertainty}), so it is important that the treatment of this contribution is well understood.

Item \#22 and \#23 are relativistic recoil corrections of order $(Z\alpha)^5$ and $(Z\alpha)^6$, respectively. 
Item \#22 has been calculated by Borie \cite{Borie:2014:arxiv_v7} (Tab.\,6), the Martynenko group \cite{Krutov:2014:JETP120_73} (No.\,22, Tab.\,1), and Jentschura \cite{Jentschura:2011:SemiAnalytic} (Eq.\,32). They agree perfectly. 
Item \#23 has only been calculated by the Martynenko group \cite{Krutov:2014:JETP120_73} (No.\,23, Tab.\,1) whose value we adopt as \textit{our choice}.

Item \#24 are higher order radiative recoil corrections of order $\alpha(Z\alpha)^5$ and $(Z^2\alpha)(Z\alpha)^4$. This item has been calculated by Borie \cite{Borie:2014:arxiv_v7} (Tab.\,6) and the Martynenko group \cite{Krutov:2014:JETP120_73} (No.\,25, Tab.\,1). Their values differ by 0.015\mev. As \textit{our choice} we adopt the average.

Item \#28 is the radiative (only eVP) recoil of order $\alpha(Z\alpha)^5$. It consists of three terms which have been calculated by Jentschura and Wundt \cite{Jentschura:2011:SemiAnalytic} (Eq.\,46). We adopt their value as \textit{our choice}. Note that a second value (0.0072\mev) is found in \cite{Jentschura:2011:PRA84_012505}. However, this value is just one of the three terms, namely the seagull term, and is already included in \#28 (see \cite{Jentschura:2011:SemiAnalytic}, Eq.\,46).


The total sum of the QED contributions without explicit nuclear structure dependence is summarized in Tab.\,\ref{tab:LS:QED} and amounts to
\begin{equation}\label{eq:LS:QED}
        \ELSradind = \LSVAL\pm\LSERR\mev.
\end{equation}
Note that Borie, on p.\,15 in Ref.\,\cite{Borie:2014:arxiv_v7} attributes an uncertainty of 0.6\mev to her total sum. 
The origin of this number remains unclear \cite{Borie:PC:2017}. 
Its order of magnitude is neither congruent with the other uncertainties given in Ref.\,\cite{Borie:2014:arxiv_v7} nor with other uncertainties collected in our summary. Thus it will not be taken into account.

\subsection{Nuclear structure contributions}
\label{sec:LS:nuclstruc}
Terms that depend on the nuclear structure are separated into one-photon exchange (OPE) contributions and two-photon exchange (TPE) contributions. 

The OPE terms (also called \textit{radius-dependent contributions}) represent the finite size effect which is by far the largest part of the nuclear structure contributions and are discussed in Sec.\,\ref{sec:LS:Radius}. They are parameterizable with a coefficient times the rms charge radius squared. These contributions are QED interactions with nuclear form factor insertions. 

The TPE terms can be written as a sum of elastic and inelastic terms, where the latter describe the polarizability of the nucleus. These involve contributions from strong interaction and therefore are much more complicated to evaluate, which explains why the dominant uncertainty originates from the TPE part. The TPE contributions are discussed in more detail in Sec.\,\ref{sec:LS:Pol}.

The main nuclear structure corrections to the $n$S states have been given up to order $(Z\alpha)^6$ by Friar \cite{Friar:1978:Annals} (see Eq.\,(43a) therein)

\begin{widetext}
\begin{equation}\label{app:eq:friar}
        \Delta E_{\rm fin. size} = \frac{2\pi Z\alpha}{3}|\Psi(0)|^2\left( \langle r^2\rangle - \frac{Z\alpha m_r}{2} \langle r^3\rangle_{(2)} + (Z\alpha)^2(F_{\rm REL} + m_r^2F_{\rm NREL}) \right),
\end{equation}
\end{widetext}

where $\Psi(0)$ is the muon wave function at the origin, $\langle r^2\rangle$ is the second moment of the charge distribution of the nucleus, i.e.\ the square of the rms charge radius, $\rel^2$. 
$\langle r^3\rangle_{(2)}$ is the Friar moment~\footnote{\label{footnote:friar}$\langle r^3\rangle_{(2)}$ has been called ``third Zemach moment'' in \cite{Friar:1978:Annals}. To avoid confusion with the Zemach radius \rze in the 2S hyperfine structure we adopt the term ``Friar moment'', as recently suggested by Karshenboim \etal~\cite{Karshenboim:2015:PRD91_073003}.}, and $F_{\rm REL}$ and $F_{\rm NREL}$ contain various moments of the nuclear charge distribution (see Eq.\,(43b) and (43c) in Ref.\,\cite{Friar:1978:Annals}). Analytic expressions for some simple model charge distributions are listed in App.\,E of Ref.\,\cite{Friar:1978:Annals}.

As the Schr\"odinger wavefunction at the origin $\Psi(0)$ is nonzero only for S states, it is in leading order only the S states which are affected by the finite size. However, using the Dirac wavefunction a nonzero contribution appears for the \TwoPOne level \cite{Ivanov:2001:LS}. This contribution affects the values for the Lamb shift and the fine structure and is taken into account in the section below.

The Friar moment $\langle r^3\rangle_{(2)}$ has not been included in \mud \cite{Krauth:2016:mud} because of a cancellation \cite{Friar:1997:PRA56_5173,Pachucki:2011:PRL106_193007,Friar:2013:PRC88_034004} with a part of the inelastic nuclear polarizability contributions. The TRIUMF-Hebrew group pointed out \cite{NevoDinur:2016:TPE,Hernandez:2016:POLupdate}, that in the case of \muHet however, a smaller uncertainty might be achieved treating each term separately. This discussion is not finished yet and we will therefore continue with the more conservative treatment as before. See Sec.\,\ref{sec:LS:Pol}.

\begin{landscape}
\begin{table}
\begin{minipage}{\linewidth}
\renewcommand{\baselinestretch}{1.1}
\renewcommand{\arraystretch}{1.5}
\caption[Nuclear structure-independent contributions to the Lamb shift]{
  All known {\bf nuclear structure-independent} contributions to the
  Lamb shift in \muHet. Values are in meV.
  Item numbers ``\#'' in the 1st column follow the nomenclature of
  Refs.~\cite{Antognini:2013:Annals,Krauth:2016:mud}, which in turn follow the supplement 
  of Ref.~\cite{Pohl:2010:Nature_mup1}.
  Items ``\#`` with a dagger $^\dagger$ were labeled ``New'' in Ref.~\cite{Antognini:2013:Annals},
  but we introduced numbers in Ref.\,\cite{Krauth:2016:mud} for definiteness.
  For Borie~\cite{Borie:2014:arxiv_v7} we refer to the most recent 
  arXiv version-7 which contains several corrections
  to the published paper~\cite{Borie:2012:LS_revisited_AoP} 
  (available online 6 Dec.\ 2011).
  For the Martynenko group, numbers \#1 to \#29 refer to rows in Tab.~I of 
  Ref.~\cite{Krutov:2014:JETP120_73}.
  Numbers in parentheses refer to equations in the respective paper.
  }
\label{tab:LS:QED}
\setlength\tabcolsep{1mm}
\setlength{\extrarowheight}{0.2mm}
\centering
    \fontsize{6pt}{6pt}\selectfont
    \hspace*{-6mm}
    \begin{tabular}{ c | l |f{7} l | f{5} l | f{6} l | f{4}@{\qquad}l | f{4}@{ }f{5}  c  c}
      \hline
      \hline
\# & Contribution                                            & \cntl{2}{Borie (B)}
                                                                              & \cntl{2}{Martynenko group (M)}
                                                                                                 & \cntl{2}{Jentschura (J)}
                                                                                                                                        & \cntl{2}{Karshenboim group (K)}
                                                                                                                                             & \cnt{3}{Our choice} \\
   &                                                         & \cntl{2}{\cite{Borie:2014:arxiv_v7}}
                                                                          & \cntl{2}{Krutov \etal~\cite{Krutov:2014:JETP120_73}}  
                                                                                                 & \cntl{2}{Jentschura, Wundt \cite{Jentschura:2011:SemiAnalytic}} 
                                                                                                                                        & \cntl{2}{Karshenboim \etal~\cite{Karshenboim:2012:PRA85_032509}} 
                                                                                                                                             & \cnt{2}{value} & source & Fig. \\
   &                                                         && 
                                                                          && 
                                                                                                  & \cntl{2}{Jentschura \cite{Jentschura:2011:PRA84_012505}} 
                                                                                                                                        & \cntl{2}{Korzinin \etal~\cite{Korzinin:2013:PRD88_125019}} 
                                                                                                                                             &                                 \\
\hline
 1 & NR one-loop electron VP (eVP)                           &                                                                        &                
                                                                          & 1641.8862                                                  & \#1
                                                                                        & 1641.885                                    & \cite{Jentschura:2011:PRA84_012505}  
                                                                                                     &                                &                
                                                                                                                   &                  &                
                                                                                                                                      &    & \\        
 2 & Rel.\ corr.\ (Breit-Pauli)                              & (0.50934)~\footnote{Does not contribute to the sum in Borie's approach.}
                                                                                                                                      & Tab.\,1        
                                                                          &  0.5093
                                                                           & \#7+\#10              
                                                                                        &   0.509344                                  & \cite{Jentschura:2011:SemiAnalytic}(17),   
                                                                                                                                        \cite{Jentschura:2011:PRA84_012505}   
                                                                                                     & (0.509340)                     & \cite{Karshenboim:2012:PRA85_032509} Tab.\,IV 
                                                                                                                   &           &                            
                                                                                                                                      &   & \\              
 3 & Rel.\ one-loop eVP                                       &  1642.412                                                             & Tab. p.\,4    
                                                                          &                                                           &            
                                                                                        &                                             &             
                                                                                                     &                                &             
                                                                                                                   &           &                    
                                                                                                                                      &   & \\      

19 & Rel.\ RC to eVP,   $\alpha(Z\alpha)^4$                   &  -0.0140                                                              & Tab.\,1+6     
                                                                          &                                                           &             
                                                                                        &                                             &             
                                                                                                     &                                &             
                                                                                                                   &           &      &   & \\      

   & Sum of the above                                        & 1642.3980                                                              &  3+19      
                                                                          &  1642.3955                                                &  1+2       
                                                                                        & 1642.3943                                   &  1+2       
                                                                                                     & 1642.3954                      & \cite{Korzinin:2013:PRD88_125019} Tab.\,I   
                                                                                                                   &1642.3962  &~ \pm~0.0018     
                                                                                                                                      & avg & \ref{fig:uehling} \\    
\hline
 4 & Two-loop eVP (K\"all$\acute{\mathrm{e}}$n-Sabry)        &   11.4107                                                              & Tab. p.\,4   
                                                                          &          11.4070                                           & \#2         
                                                                                        &                                             &             
                                                                                                     &                                &             
                                                                                                                   &  11.4089         &\pm~0.0019      & avg.     & \ref{fig:item_4} \\ 
 5 & One-loop eVP in 2-Coulomb lines $\alpha^2(Z\alpha)^2$   &   1.674                                                                & Tab.\,6              
                                                                          &   1.6773                                                  & \#9                
                                                                                        &   1.677290                                  & \cite{Jentschura:2011:SemiAnalytic}(13)  
                                                                                                     &                                &                    
                                                                                                                   & 1.6757&~\pm~0.0017 
                                                                                                                                      & avg.  & \ref{fig:item_5} \\ %
   & Sum of 4 and 5                                      &   13.0847                                                                  &   4+5        
                                                                          &   13.0843                                                 &   4+5        
                                                                                        &                                             &               
                                                                                                     &  13.0843                       & \cite{Korzinin:2013:PRD88_125019} Tab.\,I  
                                                                                                                   & (13.0846)\footnote{Sum of \textit{our choice} of item \#4 and \#5, written down for comparison with the Karshenboim group.}
                                                                                                                          &   
                                                                                                                                      &  & \\ 
\hline
6+7& Third order VP                                          &   0.073(3)                                                             &  p.\,4           
                                                                          &   0.0689                                                  &  \#4+\#12+\#11            
                                                                                        &                                             &                      
                                                                                                     & 0.073(3)                       & \cite{Korzinin:2013:PRD88_125019} Tab.\,I 
                                                                                                                   & 0.0710   & \pm~0.0036        
                                                                                                                                      & avg. \\ 
\hline
   29 & Second-order eVP contribution $\alpha^2(Z\alpha)^4 m $ &                                                                      &             
                                                                         &  0.0018                                                   & \#8+\#13 
                                                                                       &                                             &             
                                                                                                    & 0.00558                        & \cite{Korzinin:2013:PRD88_125019} Tab.\,VIII ``eVP2'' 
                                                                                                                  & 0.0037    & ~\pm~ 0.0019       
                                                                                                                                     &   avg\\          
\hline
   9   & Light-by-light ``1:3'': Wichmann-Kroll              & -0.01969                                                              & p.\,4         
                                                                         &   -0.0197                                                 & \#5 
                                                                                       &                                             &             
                                                                                                    &                                &             
                                                                                                                  &                  &             
                                                                                                                                     &     & \ref{fig:lbl}a \\
  10   & Virtual Delbr\"{u}ck, ``2:2'' LbL                   &                                                                       &             
                                                                         &  \multicolumn{1}{l}{\multirow{2}{1mm}{~\,$\left.\rule{0pt}{3ex}\right\}0.0064$}}
                                                                                                                                     & \multirow{2}{0mm}{\#6} 
                                                                                       &                                             &             
                                                                                                    &                       &
                                                                                                                  &  
                                                                                                                                     & 
                                                                                                                                       & 
                                                                                                                                                       & \ref{fig:lbl}b\\
 9a$^\dagger$ & ``3:1'' LbL                                         &                                                                  &             
                                                                         &                                                           &             
                                                                                       &                                             &             
                                                                                                    &                     &  
                                                                                                                  &           &                    
                                                                                                                                     &      & \ref{fig:lbl}c\\
       & Sum: Total light-by-light scatt.                    & -0.0134(6)                                                               &   p.5+Tab.6          
                                                                          & -0.0133                                                   &  9+10+9a     
                                                                                        &                                             &            
                                                                                                     &  -0.0134(6)                    & \cite{Korzinin:2013:PRD88_125019} Tab.\,I 
                                                                                                                   &   -0.0134 & \pm~0.0006  
                                                                                                                                      & K \\ 
\hline
20 & $\mu$SE and $\mu$VP                                     &  -10.827368                                                             & Tab.\,2+6  
                                                                          &  -10.8286                                                  & \#24       
                                                                                        &                                         &             
                                                                                                     &                                &             
                                                                                                                   &  -10.8280  &\pm~ 0.0006        
                                                                                                                                      & avg. &\ref{fig:onephotonSE}\\ 
11 & Muon SE corr.\ to eVP  $\alpha^2(Z\alpha)^4$            &  (-0.1277)~\footnote{In App.\,C of \cite{Borie:2014:arxiv_v7}, incomplete. Does not contribute to the sum in Borie's approach, see text.}
                                                                                                                                     & Tab.\,16   
                                                                          &  -0.0627 & \#28        
                                                                                        &  -0.06269                                   & \cite{Jentschura:2011:SemiAnalytic}(29) 
                                                                                                     & -0.06269                      & \cite{Korzinin:2013:PRD88_125019} Tab.\,VIII (a)
                                                                                                                   &   -0.06269      &             
                                                                                                                                &J, K & \ref{fig:item_11}\\ 
\hline
12 & eVP loop in self-energy  $\alpha^2(Z\alpha)^4$ \quad 
                                                             & \cnt{1}{ incl. in 21}                                                  &             
                                                                          &  -0.0299                                                  & \#27       
                                                                                        &                                             &             
                                                                                                     & -0.02992                       & \cite{Korzinin:2013:PRD88_125019} Tab.\,VIII (d) 
                                                                                                                   & \cnt{1}{incl. in 21}   &        
                                                                                                                                      & B  & \ref{fig:item_12} \\
 30 & Hadronic VP loop in self-energy  $\alpha^2(Z\alpha)^4 m$  
                                                             &                                                                        &             
                                                                          &                                                           &             
                                                                                        &                                             &             
                                                                                                     & -0.00040(4)                    & \cite{Korzinin:2013:PRD88_125019} Tab.\,VIII (e) 
                                                                                                                   & -0.00040   &\pm~0.00004             
                                                                                                                                      & K  & \ref{fig:item_30} \\          

13 & Mixed eVP + $\mu$VP                                     &  0.00200                                                               & p.\,4        
                                                                          & 0.0022                                                    & \#3         
                                                                                        &                                             &             
                                                                                                     & 0.00383                        & \cite{Korzinin:2013:PRD88_125019} Tab.\,VIII (b) 
                                                                                                                   & 0.0029      & \pm ~0.0009               
                                                                                                                                      &   avg   & \ref{fig:item_13} \\   

 31 &   Mixed eVP + hadronic VP                   &                                                                       &             
                                                                         &                                                           &             
                                                                                       &                                             &             
                                                                                                    & 0.0024(2)                      & \cite{Korzinin:2013:PRD88_125019} Tab.\,VIII (c) 
                                                                                                                   & 0.0024    &\pm ~0.0002               
                                                                                                                                     &  K  & \ref{fig:item_31} \\          

21 & Higher-order corr.\ to  $\mu$SE and $\mu$VP              &  -0.033749                                                        & Tab.\,2+6  
                                                                          &                                                           &             
                                                                                        &                                             &             
                                                                                                     &                                &             
                                                                                                                   &        -0.033749          &             
                                                                                                                                      &   B\\        

   & Sum of 12, 30, 13, 31, and 21                           &  -0.031749                                                              &  13+21     
                                                                          & -0.0277                                                   &      12+13       
                                                                                        &                                             &             
                                                                                                     &  -0.0241(2)         & 12+30+13+31            
                                                                                                                   & -0.0288          &       
                                                                                                                                      &  sum\\        

\hline
14 & Hadronic VP 
                                                             &   0.221(11)                                                   &  Tab.\,6   
                                                                         &  0.2170                                           & \#29        
                                                                                       &                                             &             
                                                                                                    &                                &             
                                                                                                                  &   0.219    & \pm ~ 0.011 
                                                                                                                                     & avg. \\          
\hline
17 & Recoil corr.\ $(Z\alpha)^4m_r^3/M^2$ (Barker-Glover)     & 0.12654                                                              & Tab.\,6        
                                                                         &  0.1265                                                   & \#21 
                                                                                       & 0.12654                                     & \cite{Jentschura:2011:SemiAnalytic}(A.3) \cite{Jentschura:2011:PRA84_012505}(15) 
                                                                                                    &                                &                
                                                                                                                  &  0.12654        &      
                                                                                                                                     & B, J \\           
18 & Recoil, finite size                                   &                 (0.4040(10))~\footnote{Is not included, because it is a part of the TPE, see text.}            &             
                                                                         &                                                           &          
                                                                                       &                                             &             
                                                                                                    &                                &             
                                                                                                                  &                  &           
                                                                                                                                     &  \\       
 22 & Rel.\ RC $(Z\alpha)^5$                                  & -0.55811                                                             & p.9+Tab.6     
                                                                         & -0.5581                                                   & \#22          
                                                                                       & -0.558107                                   & \cite{Jentschura:2011:SemiAnalytic}(32) 
                                                                                                    &                                &               
                                                                                                                  &      -0.558107    &                      
                                                                                                                                     & J \\            
23 & Rel.\ RC $(Z\alpha)^6$                                   &                                                                        &             
                                                                         &  0.0051                                                   & \#23        
                                                                                       &                                             &             
                                                                                                    &                                &             
                                                                                                                  &   0.0051        &             
                                                                                                                                     &  M \\       

24 & Higher order radiative recoil corr.                    &  -0.08102                                                              & p.9+Tab.6                
                                                                         & -0.0656                                                   & \#25 
                                                                                       &                                             &                    
                                                                                                    &                                &                    
                                                                                                                  & -0.0733      & \pm ~0.0077                  
                                                                                                                                     & avg.\\               
28$^\dagger$ & Rad.\ (only eVP) RC $\alpha(Z\alpha)^5$ &                                                                      &             
                                                                         &                                                      &                                       
                                                                                       & 0.004941                           & 
                                                                                                    &                                &             
                                                                                                                  &   0.004941       &                    
                                                                                                                                     &J  \\          

\hline
\hline
&&&&&&&&&&&&&\\[-3ex]
 &  \bf Sum                                            & \cntl{2}{1644.3916~\footnote{Including item \#18 and \#r3' yields 1644.9169\,meV, which is Borie's value from Ref.~\cite{Borie:2014:arxiv_v7} page 15. On that page she attributes an uncertainty of 0.6\mev to that value. This number is far too large to be correct, so we ignore it.}
}

                                                                         & \cntl{2}{$1644.3431$~
}                                                                                
                                                                                       &                                             &             
                                                                                                    &                                &             
                                                                                                                  & \TABLSVAL  & \pm~\TABLSERR 
                                                                                                                                     & \\   [-3ex]          
&&&&&&&&&&&&&\\

 \hline
 \hline

    \end{tabular}

    \end{minipage}
  \end{table}
\end{landscape}

%

\subsubsection{One-photon exchange contributions (finite size effect)}
\label{sec:LS:Radius}

Finite size contributions have been calculated by Borie (\cite{Borie:2014:arxiv_v7} Tab.\,14), the Martynenko group (\cite{Krutov:2014:JETP120_73} Tab.\,1), and the Karshenboim group (\cite{Karshenboim:2012:PRA85_032509} Tab.\,III). All of these contributions are listed in Tab.\,\ref{tab:LS:Radius}, labeled with \#r$i$.

Most of the terms, given in Tab.\,\ref{tab:LS:Radius}, can be parameterized as $c \cdot \rrel$ 
with coefficients $c$ in units of meV\,\insqfm. Borie and Karshenboim \etal\ have provided the contributions in this parameterization, whereas Martynenko \etal\ provide the total value in units of energy. However, the value of their coefficients can be obtained by dividing their numbers by \rrel. The value they used for the charge radius \rel is 1.9660\,fm~\footnote{This value has been introduced by Borie \cite{Borie:2014:arxiv_v7} as an average of several previous measurements \cite{Sick:2008:rad_scatt,Rooij:2011:HeSpectroscopy,CancioPastor:2012:PRL108}.} \cite{Martynenko:PC:2016}. In this way the numbers from Martynenko \etal\ can be compared with the ones from the other groups.

\begin{figure}
  \begin{center}
    \begin{fmffile}{ope}
      \begin{fmfgraph*}(70,70)
        \fmftop{i1,o1}
        \fmfbottom{i2,o2}
        \fmf{plain,tension=1.0}{i1,v1,o1}
        \fmf{plain,width=3}{i2,v2,o2}
        \fmf{photon,tension=0}{v1,v2}
        \fmfv{decor.shape=circle,decor.filled=full,decor.size=10}{v2}
        \fmfv{label=h}{i2}
        \fmfv{label.angle=-150,label=$\mu$}{i1}
      \end{fmfgraph*}
    \end{fmffile}
  \end{center}
  \caption{
      Item \#r1, the leading nuclear finite size correction
      stems from a one-photon interaction with a helion form factor insertion,
      indicated by the thick dot.}
  \label{fig:OPE}
\end{figure}
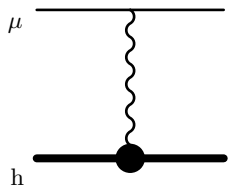

Item \#r1, the leading term of Eq.\,(\ref{app:eq:friar}), is the one-photon exchange with a helion form factor (FF) insertion (see Fig.\,\ref{fig:OPE}). Item \#r1 is of order $(Z\alpha)^4m_r^3$ and accounts for 99\% of the OPE contributions. Borie (\cite{Borie:2014:arxiv_v7} Tab.\,14, $b_a$), the Martynenko group (\cite{Krutov:2014:JETP120_73} No.\,14), and the Karshenboim group (\cite{Karshenboim:2012:PRA85_032509} Tab.\,III, $\Delta_{FNS}^{(0)}$) obtain the same result which we adopt as \textit{our choice}. This contribution is much larger than the following terms, but its absolute precision is worse, which we indicate by introducing an uncertainty. For that we take the value from Borie which is given with one more digit than the values of the other authors and attribute an uncertainty of 0.0005\mev, which may arise from rounding.

Item \#r2 and \#r2' are the radiative correction of order $\alpha(Z\alpha)^5$. The equation used for the calculation of item \#r2 is given in Eq.\,(10) of \cite{Eides:1997:two-photon}. It has been calculated by Borie \cite{Borie:2014:arxiv_v7} (Tab.\,14, $b_b$) and the Martynenko group \cite{Krutov:2014:JETP120_73} (No.\,26, only Eq.\,(92)). Note that the value from the Martynenko group was published with a wrong sign.\\
Very recently the Martynenko group updated their calculation of higher-order finite size corrections \cite{Faustov:2017:rad_fin_size} using more realistic, measured nuclear form factors. The results contain a coefficient (in our work termed item \#r2) which agrees with the old value, and an additional, previously unkown term which cannot be parametrized with $\rh^2$ and therefore is given as a constant. This constant is found in our Tab.\,\ref{tab:LS:Radius} as item \#r2'. In Ref.\,\cite{Faustov:2017:rad_fin_size} the values are given for the 1S state but can easily be transferred to the 2S state via the $1/n^3$ scaling. For the 2S state this results in 
\begin{equation}
\begin{split}
1/8\times(&-0.6109) \mev\\
       =& ~1/8\times(-0.1946\,\rh^2 + 0.1412) \mev\\ 
       =& ~-0.0243\mev/\fm^2 \rh^2 + 0.0177\mev.
\end{split}
\end{equation} 
Borie and Martynenko get the same result for item \#r2, which we adopt as \textit{our choice}. Additionally we adopt the constant term from Martynenko as item \#r2'.

Item \#r3 and \#r3' are the finite size corrections of order $(Z\alpha)^6$. They have first been calculated in Ref.\,\cite{Friar:1978:Annals}. Item \#r3 and \#r3' consider third-order perturbation theory in the finite size potential correction and relativistic corrections of the Schr\"odinger wave functions. There are also corrections in the TPE of the same order $(Z\alpha)^6$, but these are of different origin.
Borie \cite{Borie:2014:arxiv_v7} (Tab.\,14, $b_c$ and Tab.\,6) and the Martynenko group \cite{Krutov:2014:JETP120_73} (Eq.\,(91)) follow the procedure in Ref.\,\cite{Friar:1978:Annals} and then separate their terms into a part with an explicit \rrel dependence (item \#r3) and another one which is usually evaluated with an exponential charge distribution, since a model independent calculation of this term is prohibitively difficult \cite{Borie:2014:arxiv_v7}.
Differences in sorting the single terms have already been noticed in the \mud case \cite{Krauth:2016:mud}, where we mentioned that e.g.~the term $\rr \langle \ln(\mu r)\rangle$ in $F_{\rm REL}$ of Eq.\,\ref{app:eq:friar} is attributed to \#r3 and \#r3' by Martynenko \etal\ and Borie, respectively. The difference in this case amounts to 0.007\mev for \#r3'. Note that in Eq.\,(91) from the Martynenko group \cite{Krutov:2014:JETP120_73}, the charge radius has to be inserted in units of GeV$^{-1}$, with $\rel = 1.966\fm\,\widehat{=}\,9.963\,\mathrm{GeV}^{-1}$.
Item \#r4 is the one-loop $e$VP correction (\textit{Uehling}) of order $\alpha(Z\alpha)^4$. It has been calculated by all three groups, Borie \cite{Borie:2014:arxiv_v7} (Tab.\,14, $b_d$), Martynenko \etal~\cite{Krutov:2014:JETP120_73} (No.\,16, Eq.\,(69)), and Karshenboim \etal~\cite{Karshenboim:2012:PRA85_032509} (Tab.\,III, $\Delta E_{FNS}^{(2)}$).
On p.\,31 of \cite{Borie:2014:arxiv_v7}, Borie notes that she included the correction arising from the K\"all\'en-Sabry potential in her $b_d$. This means that her value already contains item \#r6, which is the two-loop $e$VP correction of order $\alpha^2(Z\alpha)^4$.
Item \#r6 has been given explicitly only by the Martynenko group \cite{Krutov:2014:JETP120_73} (No.\,18, Eq.\,73). The sum of Martynenko \etal's \#r4 and \#r6 differs by 0.016\mev/fm$^2$ from Borie's result. Using a charge radius of 1.9660\,fm this corresponds to roughly 0.06\mev and, hence, causes the largest uncertainty in the radius-dependent OPE part. The origin of this difference is not clear \cite{Borie:PC:2017,Martynenko:PC:2017}. A clarification of this difference is desired but does not limit the extraction of the charge radius.
As \textit{our choice} we take the average of the sum (\#r4+\#r6) of these two groups. The resulting average does also reflect the value for \#r4 provided by Karshenboim \etal~\cite{Karshenboim:2012:PRA85_032509}.

Item \#r5 is the one-loop $e$VP correction (\textit{Uehling}) in second order perturbation theory (SOPT) of order $\alpha(Z\alpha)^4$. It has been calculated by all three groups, Borie \cite{Borie:2014:arxiv_v7} (Tab.\,14, $b_e$), the Martynenko group \cite{Krutov:2014:JETP120_73} (No.\,17, Eq.\,70), and the Karshenboim group \cite{Karshenboim:2012:PRA85_032509} (Tab.\,III, $\Delta E_{FNS}^{(1)}$).
On p.\,31 of \cite{Borie:2014:arxiv_v7}, Borie notes that she included the two-loop corrections to $\epsilon_{VP2}$ in her $b_e$. 
This means that her value already contains item \#r7, which is the two-loop $e$VP in SOPT of order $\alpha^2(Z\alpha)^4$. 
Item \#r7 has only been given explicitly by the Martynenko group \cite{Krutov:2014:JETP120_73} (No.\,19). The sum of Martynenko \etal's \#r5+\#r7 differs by 0.003\mev from Borie's result. As \textit{our choice} we take the average of the sum (\#r5+\#r7) of these two groups.
Again here, {\it our choice} reflects the value for \#r5 provided by Karshenboim \etal~\cite{Karshenboim:2012:PRA85_032509}, too.

Item \#r8 is the finite size correction to the \TwoPOne level of order $(Z\alpha)^6$. It has only been calculated by Borie \cite{Borie:2014:arxiv_v7} (Tab.\,14, $b(2p_{1/2}$). This correction is the smallest in this section and is the only term which affects the \TwoPOne level. In consequence, the effect on the Lamb shift is inverse, i.e.~if the 2P level is lifted ``upwards'', the Lamb shift gets larger. Thus, in contrast to Borie, we include this correction with a positive sign. At the same time this term decreases the fine structure ($\TwoPThree - \TwoPOne$ energy difference) and is hence listed in Tab.\,\ref{tab:fs} as item \#f10 with a negative sign.

The total sum of the QED contributions with an explicit dependence of \rrel is summarized in Tab.\,\ref{tab:LS:Radius} and amounts to
\begin{multline}\label{eq:LS:Radius}
        \ELSrad(\rrel)\\ = -\RADVALERR\mev\insqfm\,\rrel+ 0.1354(33) \mev.
\end{multline}

\subsubsection{Two-photon exchange contributions to the Lamb shift}
\label{sec:LS:Pol}

\begin{figure}
  \begin{center}
    \begin{minipage}{0.3\columnwidth}
      (a)\hfill\mbox{~}
      \vspace{0px}\\
      \begin{fmffile}{tpe_elastic_1}
        \begin{fmfgraph*}(70,60)
          \fmftop{i1,o1}
          \fmfbottom{i2,o2}
          \fmf{plain}{i1,t1,txx,t2,o1}
          \fmf{plain,width=3}{i2,b1,bb,b2,o2}
          \fmfv{decor.shape=circle,decor.filled=full,decor.size=10}{b1}
          \fmfv{decor.shape=circle,decor.filled=full,decor.size=10}{b2}
          \fmf{photon,tension=0}{t1,b1}
          \fmf{photon,tension=0}{b2,t2}
          \fmfv{label=h}{i2}
          \fmfv{label.angle=-150,label=$\mu$}{i1}
        \end{fmfgraph*}
      \end{fmffile}
    \end{minipage}
    \hspace{30px}
    \begin{minipage}{0.3\columnwidth}
      (c)\hfill\mbox{~}
      \vspace{0px}\\
      \begin{fmffile}{tpe_inelastic_1}
        \begin{fmfgraph*}(70,60)
          \fmftop{i1,o1}
          \fmfbottom{i2,o2}
          \fmf{plain}{i1,t1,txx,t2,o1}
          \fmf{plain,width=3}{i2,b1}
          \fmf{plain,width=3}{b2,o2}
          \fmfpoly{smooth,filled=30,pull=1.4,tension=0.2,background=white+blue}{b1,b10,b2,b11}
          \fmffreeze
          \fmfshift{14up}{b10}
          \fmfshift{14down}{b11}
          \fmf{photon,tension=0}{t1,b1}
          \fmf{photon,tension=0}{b2,t2}
          \fmfv{label=h}{i2}
          \fmfv{label.angle=-150,label=$\mu$}{i1}
        \end{fmfgraph*}
      \end{fmffile}
    \end{minipage}
    \vspace{6ex}\\
    \begin{minipage}{0.3\columnwidth}
      (b)\hfill\mbox{~}
      \vspace{0px}\\
      \begin{fmffile}{tpe_elastic_2}
        \begin{fmfgraph*}(70,60)
          \fmftop{i1,o1}
          \fmfbottom{i2,o2}
          \fmf{plain}{i1,t1,txx,t2,o1}
          \fmf{plain,width=3}{i2,b1,bb,b2,o2}
          \fmfv{decor.shape=circle,decor.filled=full,decor.size=10}{b1}
          \fmfv{decor.shape=circle,decor.filled=full,decor.size=10}{b2}
          \fmf{photon,tension=0}{t1,b2}
          \fmf{photon,tension=0}{b1,t2}
          \fmfv{label=h}{i2}
          \fmfv{label.angle=-150,label=$\mu$}{i1}
        \end{fmfgraph*}
      \end{fmffile}
    \end{minipage}
    \hspace{30px}
    \begin{minipage}{0.3\columnwidth}
      (d)\hfill\mbox{~}
      \vspace{0px}\\
      \begin{fmffile}{tpe_inelastic_2}
        \begin{fmfgraph*}(70,60)
          \fmftop{i1,o1}
          \fmfbottom{i2,o2}
          \fmf{plain}{i1,t1,txx,t2,o1}
          \fmf{plain,width=3}{i2,b1}
          \fmf{plain,width=3}{b2,o2}
          \fmfpoly{smooth,filled=30,pull=1.4,tension=0.2,background=white+blue}{b1,b10,b2,b11}
          \fmffreeze
          \fmfshift{14up}{b10}
          \fmfshift{14down}{b11}
          \fmf{photon,tension=0}{t1,b2}
          \fmf{photon,tension=0}{b1,t2}
          \fmfv{label=h}{i2}
          \fmfv{label.angle=-150,label=$\mu$}{i1}
        \end{fmfgraph*}
      \end{fmffile}
    \end{minipage}
  \end{center}
  \caption{\label{fig:tpe}
    (a)+(b) Elastic \ELSFriar{},  and (c)+(d) inelastic \ELSinelast{}
    two-photon exchange (TPE) contribution.
    The thick dots in (a) indicate helion form factor insertions.
    The blob in (c) and (d) represents all possible excitations of the nucleus.
  }
\end{figure}
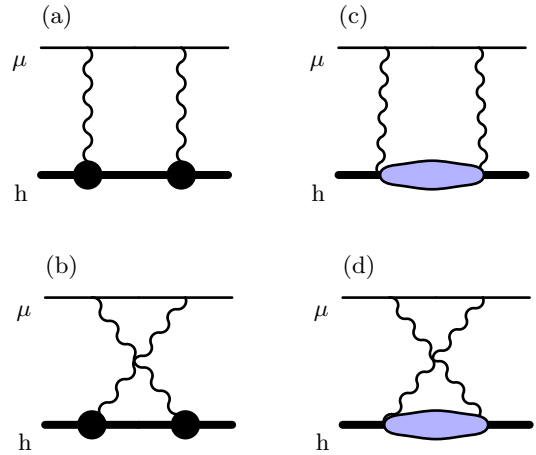

Historically, the two-photon exchange (TPE) contribution to the Lamb shift (LS) in muonic atoms has been considered the sum of the two parts displayed in Fig.\,\ref{fig:tpe}(a,b) and (c,d), respectively:
\begin{equation}
  \label{eq:tpe}
  \ELSTPE = \ELSFriar + \ELSinelast
\end{equation}
with the elastic ``Friar moment'' contribution  \ELSFriar~\footnote{formerly known as ``third Zemach moment'', see footnote \textsuperscript{\ref{footnote:friar}} on p.\,\pageref{footnote:friar} for disambiguation.} and the inelastic part \ELSinelast, frequently termed ``polarizability''.

The elastic part, \ELSFriar ~is shown in Fig.\,\ref{fig:tpe}(a,b). It is sensitive to the shape of the nuclear charge distribution, beyond the leading \rr~dependence discussed in Sec.\,\ref{sec:LS:Radius}. This part is traditionally parameterized as being proportional to the third power of the rms charge radius and it already appeared in Eq.\,(\ref{app:eq:friar}) as the second term proportional to $ \langle r^3\rangle_{(2)} $. The coefficient depends on the assumed radial charge distribution. 

The inelastic part, \ELSinelast ~is shown in Fig.\,\ref{fig:tpe}(c,d). It stems from virtual excitations of the nucleus. The inelastic contributions are notoriously the least well-known theory contributions and limit the extraction of the charge radius from laser spectroscopy of the Lamb shift.

Eq.\,(\ref{eq:tpe}) is valid for the nuclear contributions as well as for the nucleon contributions. This means that elastic and inelastic parts have to be evaluated for both, respectively.


\begin{landscape}
%

\begin{table}
\begin{minipage}{\linewidth}
  \footnotesize
  \setlength\extrarowheight{3pt}
  \centering
  \caption[Nuclear structure-dependent contributions to the Lamb shift]{
    Coefficients of the {\bf nuclear structure-dependent} one-photon exchange (OPE) contributions to the Lamb shift of \muHet.
    The values from the Martynenko group shown here are the published ones divided by $(1.9660\,\mathrm{fm})^2$, which is the radius they used. The numbers \#$i$ from the Martynenko group refer to rows in Tab.\,1 of Ref.\,\cite{Krutov:2014:JETP120_73} and numbers in parenthesis to Eqs.\ therein.
    KS: K\"all\'en-Sabry, VP: vacuum polarization, SOPT: second-order perturbation theory.
    Values are in meV/fm$^2$, except for {\#r2'} and {\#r3'}.
  }
  \label{tab:LS:Radius}
  \fontsize{7pt}{7pt}\selectfont

  \begin{tabular}{l|l |f{5} 
    c     |f{5}   c    |f{4}      c   | f{7}@{ }f{5} c}
    \hline
    \hline
       \#& Contribution  & \cntl{2}{Borie (B)}
                                     & \cntl{2}{Martynenko group (M)}
                                                   & \cntl{2}{Karshenboim group (K)}
                                                                  & \cnt {3}{Our choice} \\
       &               & \cntl{2}{Borie \cite{Borie:2014:arxiv_v7} Tab.14}
                                     & \cntl{2}{Krutov \etal~\cite{Krutov:2014:JETP120_73}}
                                                   & \cntl{2}{Karshenboim \etal~\cite{Karshenboim:2012:PRA85_032509}}
                                                                  &         &            &                    \\
       &               &    &        & \cntl{2}{Faustov \etal~\cite{Faustov:2017:rad_fin_size}}
                                                   &        &                    
                                                                  & \cnt{2}{value}
                                                                                         &  \cnt{1}{source}   \\
    \hline
    r1 & Leading fin.\ size corr., $(Z\alpha)^4$
                       & -102.520                                                        & $b_a$  
                                      & -102.52                                          & \#14, (61)  
                                                   &  -102.52                            & $\Delta E_{FNS}^{(0)}$
                                                                  & -102.520 
                                                                          & \pm~0.0010
                                                                              & B,M,K \\
    r2 & Radiative corr., $\alpha(Z\alpha)^5$
                       & -0.0243~\footnote{Borie uses Eq.\,(10) of \cite{Eides:1997:two-photon} to calculate 
                                                  this term. 
                                                  For further explanations, see text.}
                                                                                         & $b_b$  
                                      & -0.0243~\footnote{The value in Eq.\,92 of \cite{Krutov:2014:JETP120_73} was published with a wrong sign.}
                                                                                         & \#26, (92) 
                                                   &                                     & 
                                                                  & -0.0243  &           & B,M  \\
    r3 & Finite size corr. order $(Z\alpha)^6$
                       &-0.1275                                                          & $b_c$  
                                      & -0.1301
                                                                                         & \#26, (91)  
                                                   &                                     & 
                                                                  &  -0.1288 & \pm0.0013
                                                                                         &  avg. \\
    \hline
    r4 & Uehling corr.\ (+KS), $\alpha(Z\alpha)^4$
                       &                                                                 &       
                                     & -0.3310                                           & \#16, (69)  
                                                   &   - 0.323                           & $\Delta E_{FNS}^{(2)}$
                                                                  &             &  
                                                                                         &  \\
    r6 &Two-loop VP corr., $\alpha^2(Z\alpha)^4$
                       &                                                                 &        
                                     & -0.0026                                           &\#18, (73) 
                                                   &                                     & 
                                                                  &             &       
                                                                                         &  \\
    sum  & r4+r6                        
                       &-0.3176                                                          & $b_d$  
                                     & -0.3336                                           &       
                                                   &                                     &
                                                                  & -0.3256 & \pm 0.0080
                                                                                         &  avg.    \\
    \hline
    r5 & One-loop VP in SOPT, $\alpha(Z\alpha)^4$
                       &                                                                 &       
                                     & -0.5196                                           &\#17, (70)  
                                                   &   -  0.520                          & $\Delta E_{FNS}^{(1)}$
                                                                  &             & 
                                                                                         &   \\
    r7 & Two-loop VP in SOPT, $\alpha^2(Z\alpha)^4$
                       &                                                                 &         
                                     & -0.0063                                           & \#19~\footnote{This term is represented by Fig.\,9(a,b,c,d) from the Martynenko group \cite{Krutov:2014:JETP120_73}. This figure includes equation (76) therein.} 
                                                   &                                     & 
                                                                  &             &
                                                                                         &  \\
     sum& r5+r7                         
                       & -0.5217                                          & $b_e$  
                                     & -0.5259                            &       
                                                   &                      &       
                                                                  &-0.5238 &\pm 0.0021
                                                                             &   avg.   \\
    \hline
    r8 & Corr.\ to the $2P_{1/2}$ level
                       & 0.00409  ~\footnote{The sign is explained in the text.}
                                                                          & $b(2p_{1/2})$  
                                     &                                    &       
                                                   &                      & 
                                                                  & 0.00409 &
                                                                             & B \\
    \hline
       & Sum of coefficients
                       &-103.507(5)~\footnote{The summed coefficient is given in Ref.~\cite{Borie:2014:arxiv_v7} on p.\,15, where Borie indicates the uncertainty of 0.005\,meV.}
                                                                          &       
                                     & -103.5339                          &       
                                                   &   -  103.37          & $\Delta E_{FNS}$
                                                               &-103.5184 &  \pm 0.0098~\footnote{This uncertainty is the one obtained from averaging the above values (0.0084\,meV) and the one given by Borie in her sum of (0.005\,meV) added in quadrature.}
                                                                             &      \\
    \hline
    \lft{11}{~} \\
    \hline
    r2'&  Rad.\ corr.\ $\alpha(Z\alpha)^5$\,[meV]~\footnote{Belongs to \#r2. Not parametrizable with $\rh^2$.}
                      &                                                   &
                                      &  0.0177              & \cite{Faustov:2017:rad_fin_size}
                                                   &                      &
                                                               & 0.0177 &
                                                                              & M    \\
    r3'& Remaining order $(Z\alpha)^6$\,[meV]~\footnote{Belongs to \#r3. Depends on the charge distribution in a non-trivial way, see text.}
                       &  0.121                             & Tab.~6 
                                     &  0.11445              & (91)  
                                                   &                      & 
                                                                  &    0.1177&  
                                                                             \pm0.0033&avg.   \\
    \hline
    \hline
    &&&&&&&&&& \\
       & \bf Sum
                       & \multicolumn{2}{l|}{$-103.507~{\rh}^2$ + 0.121\,meV}                  
                                     & \multicolumn{2}{l|}{$-103.5339~{\rh}^2$ + 0.1322\,meV} 
                                                   &      -   103.37~{\rh}^2           &
                                                                  & \multicolumn{3}{l}{ \bf -103.5184(98)~$\boldsymbol{\rh^2}$ + 0.1354(33)\,meV} \\
    &&&&&&&&&& \\
    \hline
    \hline
  \end{tabular}
\end{minipage}
\end{table}
\end{landscape}

The nuclear parts of \ELSTPE\ are then given as $\delta E^A_{\rm Friar}$ and $\delta E^A_{\rm inelastic}$ for a nucleus with A nucleons, and the nucleon parts as $\delta E^N_{\rm Friar}$ and $\delta E^N_{\rm inelastic}$.

With that, the total (nuclear and nucleon) TPE is given as~\footnote{Compared to the notation of the TRIUMF-Hebrew group \cite{NevoDinur:2016:TPE}, the terms in Eq.\,(\ref{eq:tpe2}) correspond to $\delta^A_{\rm Zem}$, $\delta^N_{\rm Zem}$, $\delta^A_{\rm pol}$, and $\delta^N_{\rm pol}$, respectively.}
\begin{equation}
  \label{eq:tpe2}
  \ELSTPE = \delta E^A_{\rm Friar} + \delta E^N_{\rm Friar} + \delta E^A_{\rm inelastic} + \delta E^N_{\rm inelastic}.
\end{equation}

We refer here to two calculations of the TPE contributions.
The first stems from the TRIUMF-Hebrew group, who perform \textit{ab initio} calculations using two different nuclear potentials. They have published two papers on the TPE in muonic helium-3 ions: Detailed calculations are given in Nevo Dinur \etal~\cite{NevoDinur:2016:TPE}, and updated results are found in Hernandez \etal~\cite{Hernandez:2016:POLupdate}.
The second calculation has been performed by Carlson \etal~\cite{Carlson:2016:tpe}, who obtain the TPE from inelastic structure functions via dispersion relations.

The two calculations are very different, so that comparisons of any but the total value may be inexact \cite{Carlson:2016:tpe}. An attempt to compare the different approaches is given in Tab.\,II of Ref.\,\cite{Carlson:2016:tpe}. Here, we want to refer to this table only and later compare the total values as suggested. Note that we proceed differently to our previous compilation for \mud \cite{Krauth:2016:mud} (Tab.\,3), where we listed and compared 16 individual terms (labeled \#p1...16) which together yield the sum of the four terms of Eq.\,(\ref{eq:tpe2}).
\\\\
The nuclear Friar moment contribution is calculated by the TRIUMF-Hebrew group to be $\delta E^A_{\rm Friar}=10.49(24)\mev$ \cite{NevoDinur:2016:TPE,Hernandez:2016:POLupdate}. 
Previous values have been given by Borie \cite{Borie:2014:arxiv_v7} (10.258(305)\mev) and Krutov \etal\,\cite{Krutov:2014:JETP120_73} (10.50(10)\mev)\footnote{Sum of 10.28(10)\mev and 0.2214(22)\mev, which correspond to line 15 and 20 from Tab.\,1 in Ref.\,\cite{Krutov:2014:JETP120_73}, respectively.} using a Gaussian charge distribution and assuming an rms radius of $1.966(10)$\,fm.
These uncertainties do not include the (rather large) dependence of the calculation on the charge distribution \cite{Krutov:2014:JETP120_73,Sick:2014:HeZemach}. This type of uncertainty is gauged within the ab-initio calculation of \cite{NevoDinur:2016:TPE} by using two different state-of-the-art nuclear potentials. 
We therefore use the more recent value provided by the TRIUMF-Hebrew group. Their value also agrees with a value of 10.87(27)\mev which is obtained in \cite{NevoDinur:2016:TPE} from the third Zemach moment $\langle r^3\rangle_{(2)} = 28.15(70)\,\mathrm{fm}^3$ that was extracted from electrons scattering off $^3$He by Sick\,\cite{Sick:2014:HeZemach}. 
\\ 
The nuclear polarizability contribution from the TRIUMF-Hebrew group is $\delta E^A_{\rm inelastic}=4.16(17)\mev$ \cite{NevoDinur:2016:TPE,Hernandez:2016:POLupdate}. The first calculation of the nuclear polarizability contribution in \muHet has been published in 1961 \cite{Joachain:1961:pol}. The recent value from the TRIUMF-Hebrew group replaces a former one of $4.9\mev$ from Rinker \cite{Rinker:1976:he_pol} which has been used for more than 40 years now.
\\\\
As mentioned before, the total TPE contribution has a nuclear part and a nucleon part. The nucleon Friar moment contribution from the TRIUMF-Hebrew group amounts to $\delta E^N_{\rm Friar}=0.52(3)\mev$. They obtain this value using $\delta E^N_{\rm Friar}(\mup)=0.0247(13)\mev$ from \mup and scale it according to Eq.\,(17) in Ref.\,\cite{NevoDinur:2016:TPE}. This procedure has also been done in \cite{Krauth:2016:mud} for \mud~\footnote{
In Eq.\,(12) of Ref.\,\cite{Krauth:2016:mud}, we used a scaling of the nucleon TPE contribution by the reduced mass ratio to the third power, which is only correct for  $\delta E^N_{\rm inelastic}$. $\delta E^N_{\rm Friar}$ should be scaled with the fourth power \cite{Friar:2013:PRC88_034004,NevoDinur:2016:TPE}. 
This is due to an additional $m_r$ scaling factor compared to the proton polarizability term.
 This mistake has no consequences for \mud yet, as the nuclear uncertainty is much larger, but the correct scaling is relevant for \muHet and \muHef.
}. $\delta E^N_{\rm Friar}(\mup)$ is a sum of the elastic term $(0.0295(13)\mev)$ and the non-pole term $(-0.0048\mev)$ which have been obtained by Carlson \etal\ in Ref.\,\cite{Carlson:2011:PRA84_020102}.\\
The nucleon polarizability contribution from the TRIUMF-Hebrew group amounts to $\delta E^N_{\rm inelastic}=0.28(12)\mev$. It is obtained using the proton polarizability contribution from \mup and scaling it with the number of protons and neutrons~\footnote{
Assuming isospin symmetry, the value of the neutron polarizability contribution used in \cite{NevoDinur:2016:TPE} is the same as the one of the proton, but an additional uncertainty of 20\% is added, motivated by studies of the nucleon polarizabilities \cite{Myers:2014:comptonScatt}.
}, as well as with the wavefunction overlap, 
according to Eq.\,(19) of Ref.\,\cite{NevoDinur:2016:TPE}. Furthermore it is corrected for estimated medium effects and possible nucleon-nucleon interferences. The proton polarizability contribution used here amounts to $0.0093(11)\mev$ and is the sum of an inelastic term ($0.0135\mev$ \cite{Carlson:2014:PRA89_022504}) and the proton subtraction term $\delta^{p}_{\rm subtraction}=-0.0042(10)\mev$ which has been calculated for muonic hydrogen in Ref.\,\cite{BirseMcGovern:2012}.
\\\\
Summing up all nuclear and nucleon contributions evaluated by the TRIUMF-Hebrew group \cite{NevoDinur:2016:TPE,Hernandez:2016:POLupdate} yields a total value of the \ELSTPE ~of \cite{NevoDinur:2016:TPE,Hernandez:2016:POLupdate}
\begin{equation}
\begin{split}\label{eq:LS:TPEpotentials}
  \ELSTPE(&{\rm nuclear~potentials}) \\
            =& ~\delta E^A_{\rm Friar} + \delta E^N_{\rm Friar} + \delta E^A_{\rm inelastic} + \delta E^N_{\rm inelastic}\\ 
            =& ~15.46(39)\mev.~\text{\footnotemark}
\end{split}
\end{equation}
\footnotetext{As explained in the introduction, we use a different sign convention, which explains the minus sign in Refs.\,\cite{NevoDinur:2016:TPE,Hernandez:2016:POLupdate}.}

Recently, Carlson \etal\ \cite{Carlson:2016:tpe} have also calculated the TPE in \muHet. Their result of 
\begin{equation}
  \label{eq:LS:TPEdispersion}
  \ELSTPE(\rm dispersion~relations) = 15.14(49)\mev
\end{equation}
is in agreement with the one from the TRIUMF-Hebrew group. 
As \textit{our choice} we take the average of Eqs.\,(\ref{eq:LS:TPEpotentials}) and (\ref{eq:LS:TPEdispersion}) and remain with
\begin{equation}
  \label{eq:LS:pol}
  \ELSTPE = \POLVALERR\mev.
\end{equation}
As conservative uncertainty we use the larger one (from Eq.\,(\ref{eq:LS:TPEdispersion})) and add in quadrature half the spread. A weighted average of the two values (Eq.\,(\ref{eq:LS:TPEpotentials}) and (\ref{eq:LS:TPEdispersion})) which would reduce the total uncertainty is not adequate as certain contributions are effectively fixed by the same data \cite{Gorchtein:PC:2016}. 

\subsection{Total Lamb shift in \muHet}
\label{sec:LS_total}
Collecting the
radius-independent (mostly) QED contributions 
listed in Tab.~\ref{tab:LS:QED} and 
summarized in Eq.~(\ref{eq:LS:QED}),
the radius-dependent contributions 
listed in Tab.~\ref{tab:LS:Radius} and 
summarized in Eq.~(\ref{eq:LS:Radius}),
and the complete TPE contribution \ELSTPE{}
from Eq.~(\ref{eq:LS:pol}),
we obtain for the $\mathrm{2S\rightarrow2P}$ energy difference in \muHet
\begin{widetext}
\begin{equation}
  \label{eq:LS:full}
  \begin{aligned}
    \Delta E(2S_{1/2}\rightarrow2P_{1/2}) ~ = & ~\, 1644.3466(~\,146)\,\mathrm{meV} \\
    & + ~~\, 0.1354(~~~33)\,\mathrm{meV}
                                          &-& ~ \RADVALERR ~ \rh^2 ~ \mathrm{meV/fm^2}\\
    & + ~ \POLVALFINAL00(5200)\,\mathrm{meV}\\
    = &  ~ \, 1659.78(52)~\mathrm{meV}&-& ~ 103.518(10) ~ \rh^2 ~ \mathrm{meV/fm^2},
  \end{aligned}
\end{equation}
\end{widetext}
where in the last step we have rounded the values to reasonable accuracies.

One should note that the uncertainty of \POLERRFINAL\,meV 
from the nuclear structure corrections \ELSTPE{},
Eq.~(\ref{eq:LS:pol}), 
is about 30 times larger than the combined uncertainty of all 
radius-independent terms summarized in Tab.~\ref{tab:LS:QED},
and 13 times larger than the uncertainty in the coefficient of the $\rh^2$-dependent term (which amounts to 0.038\,meV for $\rh=1.966$\,fm).
A further improvement of the two-photon exchange contributions in 
light muonic atoms is therefore strongly desirable.


\section{2S hyperfine splitting}
\label{sec:HFS}

The 2S hyperfine splitting (HFS) in muonic helium-3 ions has been calculated by Borie \cite{Borie:2014:arxiv_v7} and Martynenko \cite{Martynenko:2008:muheHFS}.
(There is also the more recent paper \cite{Martynenko:2010:2SHFS_muHe} from Martynenko \etal, but it is less detailed and reproduces all numbers from \cite{Martynenko:2008:muheHFS}, with one exception to be discussed for \#h27.)
The values are summarized in Tab.~\ref{tab:hfshelium} and labeled with \#h$i$.

 We also adapted the ordering according to increasing order/complexity of the terms and grouped them thematically as: Fermi energy with anomalous magnetic moment and relativistic corrections discussed in Sec.~\ref{sec:EFermi}, vacuum polarization and vertex corrections in Sec.~\ref{sec:VP}, nuclear structure contributions and corrections listed in Sec.~\ref{sec:nuclstruc}, and the weak interaction contribution in Sec.~\ref{sec:weak}.

\subsection{Fermi energy with muon anomalous magnetic moment and Breit corrections}
\label{sec:EFermi}
\subsubsection{h1 and h4 Fermi energy and muon AMM correction}\label{sec:EFermi2}
Item \#h1 is the Fermi energy $\Delta E_\mathrm{Fermi}$ which defines the main splitting of the 2$S$ hyperfine levels. Borie and the Martynenko group have both calculated the Fermi energy, however, their values disagree by 0.055\mev (see Tab.\,\ref{tab:hfshelium}). For the calculation Borie uses Eq.\,(13) in her Ref.\,\cite{Borie:2014:arxiv_v7}. Martynenko uses Eq.\,(6) in his Ref.\,\cite{Martynenko:2008:muheHFS}. The Fermi energy is calculated using fundamental constants only. Thus we repeated the calculation for both equations, the one from Borie and the one from Martynenko which resulted to be the same: Both equations yield the same result, as they should, which is
\begin{equation}\label{eq:Fermi}
\Delta E_{\rm Fermi} = \frac{8(\alpha^4 Z^3) m_r^3 }{3n^3m_{\mu} m_p}\mu_h=-171.3508\,\mathrm{meV},
\end{equation}
where $m_\mu$ is the muon mass, $m_p$ is the proton mass, $m_r$ is the reduced mass, and $\mu_h$ is the 
helion magnetic moment to nuclear magneton ratio of $\mu_h = -2.127\,625\,308(25)$ \cite{Mohr:2016:CODATA14}. We use the value in Eq.\,(\ref{eq:Fermi}) as \textit{our choice}. This value agrees neither with Borie's value ($-171.3964\mev$) nor with the one from the Martynenko group ($-171.341\mev$).

The value for the Fermi energy corrected for the muon anomalous magnetic moment (AMM) $a_{\mu}$ is then also updated to 
\begin{equation}\label{eq:FermiAMM}
\Delta E_\mathrm{Fermi,AMM}=\Delta E_\mathrm{Fermi}\cdot(1+a_{\mu})=-171.5506\,\mathrm{meV}
\end{equation}
with a correction of $-0.1998$\,meV.

All further corrections from Borie given as coefficients $\epsilon$, are applied to this value analogous to
\begin{equation}
\Delta E_\mathrm{Fermi,AMM}\cdot(1+\epsilon).
\end{equation}
Note, that in Tab.\,\ref{tab:hfshelium}, for the contributions given by Borie, we use her coefficients but apply them to our value of the Fermi Energy given in Eq.\,(\ref{eq:FermiAMM}). The value for the Fermi Energy in Eq.\,(\ref{eq:FermiAMM}) is evaluated to a precision of $0.0001\mev$. If the number of significant digits from Borie's coefficients is too small to yield this precision we attribute a corresponding uncertainty. 
For example item \#h28* has the coefficient $\epsilon_{2\gamma}=0.0013$; here the coefficient is only precise up to a level of 0.00005, which we include as uncertainty. This uncertainty is propagated upon multiplication with the Fermi energy (Eq.\,(\ref{eq:FermiAMM})) and then yields 0.0086\mev.

\subsubsection{h2 Relativistic Breit correction}
Item \#h2 is the relativistic Breit correction of order $(Z\alpha)^6$. It is given congruently by both authors as $\Delta E_\mathrm{F,rel}^{\mathrm{B}}=-0.0775\,$meV and $\Delta E_\mathrm{F,rel}^{\mathrm{M}}=-0.078\,$meV, respectively.
We take the number from Borie as \textit{our choice}, which is given with one more digit and attribute an uncertainty of 0.0001\mev due to the precision in her coefficient.

\subsection{Vacuum polarization and vertex corrections}
\label{sec:VP}
\subsubsection{h8 and h9: Electron vacuum polarization in a one-photon one-loop interaction (h8) and in a one-photon two-loop interaction (h9)}
\label{sec:eVP}

The Feynman diagrams corresponding to \#h8 and \#h9 are analogous to those shown in Figs.~\ref{fig:uehling} and \ref{fig:item_4}, respectively, and constitute the analogs to the Uehling- and K\"all\'en-Sabry contributions in the Lamb shift. \#h8 is of order $\alpha(Z\alpha)^4$, \#h9 is of order $\alpha^2(Z\alpha)^4$.

Borie calculates the main electron VP contribution
("by modification of the magnetic interaction between muon and nucleus"),
which is a one-photon one-loop interaction.
It amounts to a correction $\epsilon_{VP1} = 0.00315$, which results in an energy shift of $-0.5405\mev$ (\#h8).
 She also gives $\epsilon_{VP1}=2.511\cdot10^{-5}$ for  one-photon two-loop interactions, resulting in $-0.0043\mev$ (\#h9).
These terms are evaluated on p.\,21 of her document \cite{Borie:2014:arxiv_v7}, using her Eq.\,(16). 

Martynenko calculates these contributions to be $-0.540\mev$ and $-0.004\mev$, respectively.
These values are found in the table in Ref.\,\cite{Martynenko:2008:muheHFS}.

Martynenko mentions that his value for our item \#h9 consists of his Eqs.\,(15,16).
The numerical result from Eq.\,(15) corresponds to two separate loops (see our Fig.~\ref{fig:item_4}(a)) and is given as $-0.002\,$meV, whereas Eq.~(16) describes the two nested two-loop processes where an additional photon is exchanged within the electron VP loop (see our Fig.~\ref{fig:item_4}(b,c)).
One can conclude that its numerical value is also $-0.002\,$meV.

Both authors give congruent results within their precisions, as \textit{our choice} we write down the numbers by Borie which are given with one more digit. We attribute an uncertainty to item \#h8 due to the precision in Borie's coefficient.\\

\subsubsection{h5 and h7: Electron vacuum polarization in SOPT in one loop (h5) and two loops (h7)}
\label{sec:eVPSopt}
Items \#h5 and \#h7 are the SOPT contributions to items \#h8 and \#h9, respectively.

Borie's value for our item \#h5 is given by the coefficient $\epsilon_{VP2}=0.00506$ and her value for our item \#h7 by $\epsilon_{VP2}=3.928\cdot10^{-5}$. This results in energy shifts of $-0.8680(9)\mev$ and $-0.0067\mev$, respectively (those values are for point nuclei; the finite size correction is taken into account in our \#h25 and \#h26). The uncertainty in item \#h5 originates from the precision of $\epsilon_{VP2}$.

The corresponding values from Martynenko are $-0.869\mev$ (\#h5) and $-0.010\mev$ (\#h7).

Due to slight differences between the two authors, as \textit{our choice} we take the average of items \#h5 and \#h7, respectively. The uncertainty of item \#h5 is the above uncertainty and half the spread between both authors added in quadrature.

 \subsubsection{h13 and h14: Vertex correction ($\hat{=}$ self energy happening at the muon-photon vertex)}
 \label{sec:vertex}
Item \#h13 is the muon self-energy contribution of order $\alpha(Z\alpha)^5$ 
(it is the analogue to a part of item \#20 in the Lamb shift, see Fig.\,\ref{fig:onephotonSE}a). 
It has only been calculated by Borie as
 \begin{equation}
 \epsilon_\mathrm{vertex}=\alpha(Z\alpha)  \left( \ln{2}-\frac{5}{2} \right)=-0.9622\cdot10^{-4}\cdot Z.
  \end{equation}
Its numerical value is thus $0.0330\mev$, however this includes a muon VP contribution of $-0.0069\mev$ (\#h12, see Sec.~\ref{sec:mVP}). For our item \#h13, we use the value from Borie as \textit{our choice}. We therefore should not include \#h12, which is discussed later.

 Borie also cites a higher order correction of Brodsky and Erickson \cite{Brodsky:1966:radiative:hyperfine} which results in a correction of $-0.211\cdot10^{-4}\hat{=}-0.0036\mev$ (\#h14). 
Very probably the sign of the energy shift is not correct because the coefficient is negative, but the Fermi energy of helium-3 also has a negative sign, thus the energy shift should be positive.
(The analogous contributions in muonic hydrogen and deuterium are negative, which is a further hint to a wrong sign since the helium-3 Fermi energy is negative, contrary to hydrogen and deuterium.)

\subsubsection{h12: Muon VP and muon VP SOPT}\label{sec:mVP}
Item \#h12 is the one-loop muon vacuum polarization. Borie on p.\,19 (below the equation of $\epsilon_\mathrm{vertex}$) of Ref.\,\cite{Borie:2014:arxiv_v7} gives the coefficient as $0.3994\cdot10^{-4}\cdot Z$. In combination with the Fermi energy this yields $-0.0069\mev$.
Martynenko obtains a value of $-0.007\mev$ which is congruent to Borie's value.
However, Borie's value of this contribution is already included in our item \#h13, which has been discussed in the previous section. Hence, we do not include it separately in `our choice'.

 \subsubsection{h18 Hadronic vacuum polarization}
\label{sec:hVP}
Item \#h18 is the hadronic vacuum polarization. Borie gives this contribution as $ \epsilon_\mathrm{hVP}=0.2666\cdot10^{-4}\cdot Z$, which amounts to $-0.0091\mev$ on p.\,19 of her paper. 
This contribution is analogous to our Fig.~\ref{fig:uehling}, but with a hadronic loop in the photon line.
Since Martynenko does not provide a value for hadronic VP in muonic helium-3 ions, we use Borie's value as `our choice'.

\subsection{Nuclear structure and finite size corrections}
\label{sec:nuclstruc}
Analogously to Sec.~\ref{sec:LS:nuclstruc}, we categorize the nuclear structure contributions to the 2S HFS as one-photon exchange (OPE) and two-photon exchange (TPE) processes, respectively.
We list first the by far dominant contribution to nuclear structure: the Zemach term, which is an elastic TPE process. 
The following subsections describe the known elastic TPE corrections in the 2S HFS.
So far, to our knowledge there are yet no calculations with respect to the {\it inelastic} TPE contribution to the 2S HFS. Thus we only give a simplified estimate with a large uncertainty. 
Later the section is concluded with the one-photon exchange (OPE) corrections to nuclear structure in the 2S HFS.

\subsubsection{h20 Zemach term and h23, h23b*, h28* nuclear recoil}\label{sec:Zemachterm}

Item \#h20 is the elastic TPE and the main finite size correction to the 2S HFS. This correction arises due to the extension of the magnetization density (Bohr-Weisskopf effect) and is also called the Zemach term \cite{Zemach:1956}. The Zemach term is usually parameterized as \cite{FriarSick:2004:Zemach}
\begin{equation}\label{eq:zemach_term}
  \Delta E_{\rm Zemach}^{\rm HFS} = - \Delta E_{\rm Fermi, AMM}~ 2(Z\alpha)m_r~ \rze
\end{equation}
with $m_r$ being the reduced mass and \rze the Zemach radius of the nucleus \cite{Sick:2014:HeZemach}
\begin{equation}
\rze = -\frac{4}{\pi}\int_0^\infty [G_E(q)G_M(q)-1]\,\frac{dq}{q^2}.
\end{equation}
Here, $G_E(q)$ and $G_M(q)$ are the electric and magnetic form factors of the nucleus, respectively.

The corresponding coefficient to the Fermi energy in Eq.\,(\ref{eq:zemach_term}) is given by Borie on p.\,23 of \cite{Borie:2014:arxiv_v7} as 
\begin{equation}
\epsilon_{\rm Zem} = -2(Z\alpha)m_r~\rze = -0.01506\mathrm{\,fm^{-1}}~\rze. 
\end{equation}
With our Fermi energy from Eq.\,(\ref{eq:FermiAMM}), item \#h20 is 
 \begin{equation}\label{eq:zemach_term2}
\Delta E_{\rm Zemach}^{\rm HFS} = 2.5836~\rze\mev/\mathrm{fm} = 6.5312(413)\mev,
\end{equation}
where, in the second step, we inserted the most recent Zemach radius from Sick \cite{Sick:2014:HeZemach} ($\rze = 2.528(16)\,$fm).

Note that Borie's published value of $\Delta E_{\rm Zemach}^{\rm HFS}$ differs from the one given here, because she uses a different Zemach radius of $\rze = 2.562\,$fm, assuming a Gaussian charge distribution.

Martynenko, in his Ref.\,\cite{Martynenko:2008:muheHFS}, gives a value of $\Delta E_{\rm str}^{\rm HFS} = 6.047\mev$. This value contains a recoil contribution and is thus not directly comparable with our item \#h20. However, this value has been updated \cite{Martynenko:PC:2016} and is now available as two separate values of $\Delta E_{\rm str}^{\rm HFS} = 6.4435\mev = (6.4085 + 0.0350_{\rm recoil})\mev$. 
The first can be compared to Eq.\,(\ref{eq:zemach_term2}).
The second is the recoil correction and listed in our table as item \#h23. Martynenko notes \cite{Martynenko:2008:muheHFS} that changing from a Gaussian to a dipole parameterization results in a change of the final number of 2\%. 

Regarding our item \#h20, we do not consider the respective value from Martynenko because it is model-dependent and therefore carries a large uncertainty. This uncertainty can be avoided using the model-independent Zemach radius from Sick and the coefficient given by Borie as stated above.

A new contribution which hasn't been calculated for \mup and \mud is our item \#h23b*. It is an additional recoil contribution 
which amounts to 0.038\mev. It has only been calculated by Martynenko and we adopt his value as \textit{our choice}. In order to account for the precision given by Martynenko, we write 0.0380(5)\mev.

Another contribution which has not been calculated for \mup and \mud is item \#h28*. It is a two-photon recoil correction, calculated by Borie in 1980 \cite{Borie:1980:mu3HeLS}, who followed the procedure of Grotch and Yennie \cite{Grotch:1969:EPM}. This contribution is not listed in Borie's recent Ref.\,\cite{Borie:2014:arxiv_v7}, but should be included \cite{Borie:PC:2014}. It is given by $\epsilon_{2\gamma}= 0.0013$ and therefore results in -0.2230(86)\mev, using our Fermi energy from Eq.\,(\ref{eq:FermiAMM}). The attributed uncertainty originates from the number of significant digits in $\epsilon_{2\gamma}$ (the value of the coefficient is considered to be accurate only to $\pm0.00005$). Regarding the contributions given by Martynenko, no overlap is found, which is why we list this item separately.

\subsubsection{h24 electron VP contribution to two-photon exchange}
\label{sec:h24}
Item \#h24, the electron VP contribution to the 2S HFS elastic two-photon exchange in muonic helium-3 ions is only calculated by Martynenko \cite{Martynenko:2008:muheHFS}.
The corresponding Feynman diagrams are shown in Fig.~4 of his helium 2S HFS paper \cite{Martynenko:2008:muheHFS}. These are analogous to our Fig.~\ref{fig:tpe}, but with a VP loop in one of the exchange photons.
A numerical value of the contribution is given in his Eq.\,(38) of 0.095\,meV and thus enters \textit{our choice}, where we write 0.0950(5)\mev and therefore account for the precision given by Martynenko.

\subsubsection{h15, h16, h17 radiative corrections to the elastic two-photon exchange}
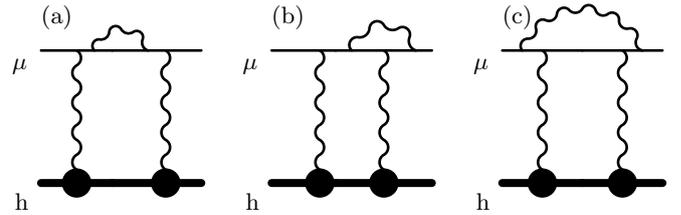
\begin{figure}
  \begin{center}
    \begin{minipage}{0.22\columnwidth}
      (a)\hfill\mbox{~}
      \vspace{7px}\\
      \begin{fmffile}{item_h17ba}
        \begin{fmfgraph*}(60,50)
          \fmfstraight
          \fmftopn{t}{10}
          \fmfbottomn{b}{10}
          \fmf{plain}{t1,t5,t10}
          \fmf{plain,width=3}{b1,b5,b10}
          \fmf{photon,tension=0.1,left=0.8}{t4,t7}
          \fmf{photon}{t3,b3}
          \fmf{photon}{t8,b8}
          \fmfv{decor.shape=circle,decor.filled=full,decor.size=10}{b3}
          \fmfv{decor.shape=circle,decor.filled=full,decor.size=10}{b8}
          \fmfv{label=h}{b1}
          \fmfv{label.angle=-150,label=$\mu$}{t1}
        \end{fmfgraph*}
      \end{fmffile}
    \end{minipage}
    \hspace{25px}
    \begin{minipage}{0.22\columnwidth}
      (b)\hfill\mbox{~}
      \vspace{7px}\\
      \begin{fmffile}{item_h17bc}
        \begin{fmfgraph*}(60,50)
          \fmfstraight
          \fmftopn{t}{11}
          \fmfbottomn{b}{11}
          \fmf{plain}{t1,t5,t11}
          \fmf{plain,width=3}{b1,b5,b11}
          \fmf{photon,tension=0.1,left=0.8}{t6,t10}
          \fmf{photon}{t4,b4}
          \fmf{photon}{t8,b8}
          \fmfv{decor.shape=circle,decor.filled=full,decor.size=10}{b4}
          \fmfv{decor.shape=circle,decor.filled=full,decor.size=10}{b8}
          \fmfv{label=h}{b1}
          \fmfv{label.angle=-150,label=$\mu$}{t1}
        \end{fmfgraph*}
      \end{fmffile}
    \end{minipage}
    \hspace{25px}
    \begin{minipage}{0.22\columnwidth}
      (c)\hfill\mbox{~}
      \vspace{7px}\\
      \begin{fmffile}{item_h17bb}
        \begin{fmfgraph*}(60,50)
          \fmfstraight
          \fmftopn{t}{9}
          \fmfbottomn{b}{5}
          \fmf{plain}{t1,t5,t9}
          \fmf{plain,width=3}{b1,b3,b5}
          \fmf{photon,tension=0.1,left=0.7}{t2,t8}
          \fmf{photon}{t3,b2}
          \fmf{photon}{t7,b4}
          \fmfv{decor.shape=circle,decor.filled=full,decor.size=10}{b2}
          \fmfv{decor.shape=circle,decor.filled=full,decor.size=10}{b4}
          \fmfv{label=h}{b1}
          \fmfv{label.angle=-150,label=$\mu$}{t1}
       \end{fmfgraph*}
      \end{fmffile}
    \end{minipage}
  \end{center}
  \caption{ 
    (a) Item \#h15, $\mu$SE contribution to the elastic two-photon exchange;
    (b) item \#h16 the vertex correction to the elastic two-photon exchange, which results in two terms (the vertex correction can take place either at one or the other photon); and
    (c) item \#h17, spanning photon contribution to the elastic two-photon exchange, also referred to as jellyfish diagram.}
  \label{fig:h15h16h17}
\end{figure}

Items \#h15, \#h16, and \#h17 are radiative corrections to the elastic two-photon exchange in the 2S hyperfine structure and represented in Fig.\,\ref{fig:h15h16h17}. They are partially given in Martynenko's Ref.\,\cite{Martynenko:2008:muheHFS}, but have been updated \cite{Martynenko:PC:2017} and result to be $-0.0101\mev$ (\#h15), $0.0333\mev$ (\#h16), and $0.0074\mev$ (\#h17). These numbers include recoil corrections and are based on Eqs.(24)-(27) from the Martynenko group \cite{Faustov:2014:radrec} and use a dipole parameterization of the helion form factor, as well as $\rh=1.966\,$fm.
For the moment, we will adapt these preliminary numbers including recoil considerations into \textit{our choice}.

\subsubsection{h22 inelastic two-photon exchange in the hyperfine structure} 
In contrast to the Lamb shift, no calculations are available for the inelastic two-photon exchange (polarizability contribution) in the 2S HFS. We give an estimate of this value by calculating the ratio between the polarizability contribution and the Zemach term in the 1S ground state of (electronic) $\mathrm{^3He^+}$ and assume the ratio to be similar for the 2S state in \muHet. 

The 1S Zemach term for electronic $\mathrm{^3He^+}$ is found by using Eq.\,(\ref{eq:zemach_term}), but with the muon mass replaced by the electron mass and $n=1$.
Using the Zemach radius \rze from Friar and Payne \cite{Friar:2005:PRC72} a value of 1717\,kHz is obtained. In order to obtain the total sum (polarizability + Zemach) of 1442\,kHz \cite{Friar:2005:PRC72}, a polarizability term of order $-300$\,kHz is missing. The ratio is then roughly $-1/6$. The Zemach term for muonic helium-3 ions (our item h20), obtained above, yields $\Delta E_{\rm Zem}\approx 6.5\mev$. The estimate for the polarizability contribution consequently follows with $\Delta E^{\rm HFS}_{\rm pol.}\approx-1.0\pm1.0\mev$, which includes a conservative 100\% uncertainty.

\subsubsection{h25 and h26 finite size correction to electron VP}
\label{sec:eVPfinitesize}
Borie gives the electron VP contributions \#h8 and \#h5 (eVP processes in OPE, see Sec.~\ref{sec:VP}) which are based on a point nucleus.
Additionally, she provides modified contributions which include the finite size effect on electron VP.
These are $\epsilon_{VP1}'=0.00295$ and $\epsilon_{VP2}'=0.00486$, respectively.
The difference between those values and \#h8 and \#h5 constitute finite size corrections.
Multiplied with the Fermi energy (including the AMM), these yield 0.0343(9)\,meV each and we attribute them to \#h25 and \#h26, analogous to the previous CREMA summaries. The uncertainty originates from the precision in Borie's coefficients.
Note that these are OPE processes.

\subsubsection{h27 and h27b nuclear structure correction in leading order  and SOPT}
This correction is only given by Martynenko. The two terms are found in  Fig.~5(a) and (b) of Ref.\,\cite{Martynenko:2008:muheHFS}, for leading and second order, respectively.
This correction is also an OPE process.
Care has to be taken here because 
this contribution is given as 0.272\,meV in \cite{Martynenko:2008:muheHFS}, but as 0.245\,meV in a 2010 follow up paper \cite{Martynenko:2010:2SHFS_muHe} (however, this is the only term that changed between \cite{Martynenko:2008:muheHFS} and \cite{Martynenko:2010:2SHFS_muHe}).
As compared to muonic deuterium, Martynenko only gives the sum (h27 $+$ h27b) and not the single contributions.
In \cite{Martynenko:2008:muheHFS}  the formulas he uses to calculate h27 and h27b are explicitly given as
\begin{equation}
\Delta E^{\mathrm{HFS}}_\mathrm{1\gamma,str}
=-\frac{4}{3}(Z\alpha)^2 m_r^2r_M^2 \cdot E_\mathrm{Fermi}\cdot\frac{1-n^2}{4n^2}
\label{eq:1gammaStr}
\end{equation}
\begin{multline}
\Delta E^{\mathrm{HFS}}_\mathrm{str,SOPT}(2S)\\
=\frac{4}{3}(Z\alpha)^2m_1^2r_E^2 \cdot E_\mathrm{Fermi}(2S)\cdot 
(\ln(Z\alpha) - \ln 2), \label{eq:strSOPT}
\end{multline}
where $m_r$ is the reduced mass of the muon, $m_1$ is the muon mass, and $r_E$ and $r_M$ are the charge and magnetic radii, respectively.
Martynenko states to use $r_E\approx r_M=1.844\pm0.045$\,fm which is known to be outdated.

However, inserting Martynenko's Fermi energy, the radius he used, and fundamental constants into Eqs.\,(\ref{eq:1gammaStr}) and (\ref{eq:strSOPT}) yields a sum of 0.2251$\pm$0.0001\,meV which is neither congruent with \cite{Martynenko:2008:muheHFS} nor \cite{Martynenko:2010:2SHFS_muHe}.

Using Sick's 2014 values \cite{Sick:2014:HeZemach} for the charge and magnetic radii yields 0.2577$\pm$0.0001\,meV.

In the course of some private communications with Martynenko, he provided us his most current value of 0.2421\,meV for the sum of h27+h27b, and we use this preliminarily as \textit{our choice}.\vspace{20px} 


\subsection{h19 weak interaction}
\label{sec:weak}
The contribution of the weak interaction to the 2S HFS of helium-3 is only given 
by Borie. She cites Eides \cite{Eides:2012:Weak} and provides $\epsilon_\mathrm{weak}=1.5\cdot 10^{-5}\hat{=}-0.0026\mev$, which we adopt as \textit{our choice}.\vspace{20px}

\subsection{Total 2S HFS contribution}
In total, the 2S HFS contributions are given by
\begin{widetext}
\begin{equation}
  \label{eq:hfs:total1}
  \begin{aligned}
    \Delta E^{\rm HFS}(\TwoSOneFOne - \TwoSOneFCero) =&~ -172.7457(89) \mev +2.5836\mev/\mathrm{fm}~r_Z &+&~ \Delta E_{\rm pol.}^{\rm HFS}\\  
                            =&~ \hspace{50px}-166.2145(423) &-&~ 1.0(1.0)\mev\\      
                            =&~ -167.2(1.0)\mev.
  \end{aligned}
\end{equation}
\end{widetext}
Here, in the first line, we separated out the Zemach contribution and the estimate of the polarizability contribution. In the second line, the Zemach radius $r_Z = 2.528(16)\,$fm \cite{Sick:2014:HeZemach} is inserted and the estimated value of $\Delta E_{\rm pol.}^{\rm HFS}$ is shown. The polarizability is the dominant source of uncertainty in the hyperfine structure and prevents a precise determination of the helion Zemach radius from the measured transitions in the muonic helium-3 ion \cite{CREMA:mu3he}. A calculation of the polarizability contribution is therefore highly desirable. 
Until then a precise measurement of the 1S or 2S HFS in muonic helium-3 ions can be used to experimentally determine a value of the polarizability contribution $\Delta E_{\rm pol.}^{\rm HFS}$. In essence, the measurement of the 2S HFS by the CREMA collaboration can be used to give the total TPE contribution to the HFS, $\Delta E^\mathrm{HFS}_\mathrm{TPE} = 2.5836\mev/\mathrm{fm}~r_Z + \Delta E_\mathrm{pol.}^{\rm HFS}$ with an expected uncertainty of 0.1\mev.

%
\begin{landscape}
\begin{table}
\begin{minipage}{\linewidth}
  \footnotesize
  \setlength\extrarowheight{2pt}
  \centering
  \caption{All contributions to the {\bf 2S hyperfine splitting (HFS)}.
    The item numbers h$i$ in the first column follow
    the entries in Tab.\,3 of Ref.\,\cite{Antognini:2013:Annals}.
    However, the terms are now sorted by increasing complexity, analogous to their order in the text.
    For Martynenko, numbers \#1 to \#13 refer to rows in 
    Tab.\,I of his Ref.\,\cite{Martynenko:2008:muheHFS}, 
    whereas numbers in parentheses refer to equations therein.
    Borie~\cite{Borie:2014:arxiv_v7} gives the values as coefficients $\epsilon$ 
    to be multiplied with the sum of (h1+h4) of 'our choice' values. We list the resulting values 
    in meV.
    AMM: anomalous magnetic moment,
    PT: perturbation theory,
    VP: vacuum polarization,
    SOPT: second order perturbation theory,
    TOPT: third order perturbation theory.
    All values are in meV.
    Values in brackets do not contribute to the total sum.
  }
    \fontsize{6pt}{6pt}\selectfont
    \label{tab:hfshelium}
    \begin{tabular}{l|l|f{3} l l| f{3} l l| f{2} l c}
      \hline
      \hline
      & \cntl{1}{Contribution} 
                    & \cntl{3}{Borie (B)} 
                                       & \cntl{3}{Martynenko group (M)} 
                                                         & \cnt{3}{Our choice}\\
      & \cntl{1}{~} 
                    & \cntl{3}{Borie \cite{Borie:2014:arxiv_v7}} 
                                       & \cntl{3}{Martynenko \cite{Martynenko:2008:muheHFS}} 
                                                         & \cnt{2}{value} & \cnt{1}{source}\\
      \hline
      h1& Fermi splitting, $(Z\alpha)^4$          
                    &   (  -171.3964)     &&     p.\,19         
                                       & -171.341 &&\#1, (6)
                                                         &   -171.3508~\footnote{calculated in this work and given in Eq.~(\ref{eq:Fermi}).}     
                                                                        & 
                                                                            & \\
      h4& $\mu$AMM corr., $\alpha(Z\alpha)^4$  
                    &     ( -0.1999)      &&
                                       & -0.200 &&\#2, (7)
                                                         &  -0.1998  & & \\
      sum& (h1+h4)                 
                    & -171.5963 &&p.\,19
                                       &  (-171.541) &&               
                                                         & 
                                                                   & 
                                                                      & \\
      h2& Breit corr., $(Z\alpha)^6$           
                    & -0.0775 &$\pm$ 0.0001&p.\,19
                                       & -0.078&&\#3, (8)
                                                         & -0.0775 &$\pm$ 0.0001    & B\\
      \hline
          h8& One-loop eVP in OPE, $\alpha(Z\alpha)^4$ ($\epsilon_\mathrm{VP1}$)
                    & -0.5404  &$\pm$ 0.0009&p.\,21         
                                       &-0.540 &&\#4, (12)
                                                         & -0.5404 &$\pm$ 0.0009  & B\\
      h9& Two-loop eVP in OPE, $\alpha^2(Z\alpha)^4$ ($\epsilon_\mathrm{VP1}$)
                    & -0.0043  &&p.\,21         
                                       & -0.004 &&\#5, (15,16)
                                                         &-0.0043   & 
                                                                      & B\\
       h5& One-loop eVP in OPE, SOPT, $\alpha(Z\alpha)^4$ ($\epsilon_\mathrm{VP2}$)
                    & -0.8680 &$\pm$ 0.0009&p.\,21         
                                       &-0.869 &&\#7, (24) 
                                                         & -0.8685  & $\pm$ 0.0010 & avg.\\
      h7& Two-loop eVP in OPE, SOPT, $\alpha^2(Z\alpha)^4$ ($\epsilon_\mathrm{VP2}$)
                    & -0.0067  &&p.\,21         
                                       &-0.010&   &          \#8, (29,30)
                                                         & -0.0084 & $\pm$ 0.0017 & avg.\\
       h13& Vertex, $\alpha(Z\alpha)^5$          
                    &0.0330&&p.\,19         
                                       & &&         
                                                         & 0.0330&
                                                                      & B\\
      h14& Higher order corr.\ of (h13), part with ln($\alpha$)
                    & 0.0036~\footnote{The sign from Borie is wrong and has been corrected here, see Sec.~\ref{sec:vertex}.} &&p.\,19         
                                       & && 
                                                         &0.0036 &  & B\\   

      h12& one-loop $\mu$VP in 1$\gamma$ int., $\alpha^6$
                    & (-0.0069)& incl.\ in h13 &p.\,19 \& p.\,21 
                                      & -0.007  &&\#6, (12)
                                                         & \cnt{2}{incl.\ in h13}     & B\\
      h18& Hadronic VP, $\alpha^6$               
                    &  -0.0091  &&p.\,19         
                                      &    && 
                                                         &   -0.0091 &  & B\\
      \hline
      h20& Fin.\ size (Zemach) corr.\ to $\Delta E_\mathrm{Fermi}$, $(Z\alpha)^5$
                    & 6.5312~\footnote{Calculated by combining Borie's coefficient with Sick's $r_Z$.} 
                            & (=2.5836~$r_Z$/fm )
                                                         &p.\,23         
                                      &  6.4085
                                               &($\pm$ 0.1)~\footnote{This uncertainty reflects the change in this contribution when moving from dipole parameterization to a Gaussian one.}
                                                 & priv.comm.
                                                         & 2.5836
                                                                 & $r_Z$/fm 
                                                                      &  B \\
      h23& Recoil of order $(Z\alpha)(m_1/m_2$)ln($m_1/m_2)E_F$
                    &
                        &&      
                                      & 0.0350
                                         & 
                                           &priv.comm. 
                                                         &    0.0350 & 
                                                                      &M \\
      
   h23b*&Recoil of order $(Z\alpha)^2(m_1/m_2)E_F$ 
                    & 
                       &&      
                               & 0.038
                                 & 
                                   &  \#13, (48)
                                                         & 0.0380 &$\pm$ 0.0005
                                                                      & M\\

   h28*& Two-photon recoil 
                    &  -0.2230
                       &$\pm 0.0086$& \cite{Borie:1980:mu3HeLS}     
                               &
                                 & 
                                   & 
                                                         & -0.2230 &$\pm$ 0.0086
                                                                      & B\\

      h24& eVP in two-photon-exchange, $\alpha^6$
                    &          &&            
                                      & 0.095 & 
                                                &\#10, (38)
                                                         & 0.0950 & $\pm$ 0.0005
                                                                      & M\\                                                                      
      h15& muon self energy contribution in TPE, w/recoil
                    &          &&             
                                       & -0.0101 &&priv.comm.
                                                         & -0.0101 &  & M\\
      h16& vertex correction contribution in TPE, w/recoil
                    &          &&             
                                       &  0.0333 &&priv.comm.
                                                         & 0.0333 &  & M\\
      h17& jelly fish correction contribution in TPE, w/recoil
                    &          &&             
                                       & 0.0074  &&priv.comm.
                                                         & 0.0074 &  & M\\
                                                         
                                                                  \hdashline
      h22a& Helion polarizability, $(Z\alpha)^5$
                    &   & 
                              &                                        &          &&            
                                                         &   & 
                                                                      &  \\
      h22b& Helion internal polarizability, $(Z\alpha)^5$
                    &  & 
                              & 
                                       &          && 
                                                         & &                                                                       & \\
      sum& (h22a+h22b) 
                    &   & 
                              &
                                       &   & 
                                                   & 
                                                         & (-1.0   &$\pm$ 1.0)~\footnote{Is a preliminary estimate, see text. It is therefore listed separately in the sum below.} &  \\
   \hdashline

      h25& eVP corr.\ to fin.\ size in OPE (sim.\ to $\epsilon_\mathrm{VP2}$)
                    & 0.0343  &$\pm$ 0.0009& p.\,21      
                                       &        &&
                                                         & 0.0343 
      ~\footnote{Difference of two terms in Borie~\cite{Borie:2014:arxiv_v7}, see also Sec.~\ref{sec:eVPfinitesize}.}
                                                                   &$\pm$ 0.0009  & B\\
      h26& eVP corr.\ to fin.\ size in OPE (sim.\ to $\epsilon_\mathrm{VP1}$)
                    & 0.0343  &$\pm$ 0.0009& p.\,21      
                                       &        &&
                                                         & 0.0343  &$\pm$ 0.0009  & B\\
      h27+h27b& Nucl.\ struct.\ corr. in SOPT, $\alpha(Z\alpha)^5$
                    &           &&            
                                       & 0.2421 &&priv.comm.
                                                         & 0.2421  &  & M\\
                                                            \hline
      h19& Weak interact.\ contr.               
                    & -0.0026
                              &$\pm$ 0.0001&p.\,21         
                                       &      &&        
                                                         & -0.0026       &$\pm$ 0.0001  & B\\
      \hline
      \hline
      &&&&&&&&&&\\
			      & \bf Sum 
                    & -166.6988~\footnote{Borie's sum given in this table differs from her published one of -166.3745\,meV \cite{Borie:2014:arxiv_v7}. This is because we used an updated value of the Fermi energy (see Sec.\,\ref{sec:EFermi2}), a different value for the Zemach radius $r_Z$ (see Sec.\,\ref{sec:Zemachterm}), and included item \#h28* which has not been considered in Ref.\,\cite{Borie:2014:arxiv_v7}.} & 
                               & 
                                       &-165.1998~\footnote{Martynenko's sum given in this table is different from the (superseded) published one of -166.615\,meV \cite{Martynenko:2008:muheHFS} because several terms have been changed and added upon private communication.}    & 
                                                    &            
                                                         & \bf -172.\bf7457&$\pm$ \bf 0.0089          &\\
			      & 
                    & & 
                               &
                                       &    & 
                                                    &            
                                                         &  \bf 2.\bf5836& \bf \,$\boldsymbol{r_Z/\mathrm{fm}}$          &\\
			      & 
                    & & 
                               &
                                       &    & 
                                                    &            
                                                         & \bf -1.\bf0&$\pm$ \bf 1.0          &\\
      &&&&&&&&&&\\
      \hline
      \hline
    \end{tabular}
\end{minipage}
\end{table}
\end{landscape}

\clearpage


\begin{landscape}
\begin{table}
\begin{minipage}{\linewidth}
  \footnotesize
  \setlength\extrarowheight{3pt}
  \fontsize{8pt}{8pt}\selectfont
  \centering
  \caption{
    Contributions to the {\bf 2P fine structure}. Items \# with an asterisk * 
    denote new contributions in this compilation.
    The items \#f7a, \#f7d, and \#f7e
    originate from the same graphs as the Lamb shift items \#11, \#12, and \#30,
    respectively.
    VP: vacuum polarization, AMM: anomalous magnetic moment, KS: K\"all\'en-Sabry. 
    All values are in meV.}
  \label{tab:fs}
  \begin{tabular}{l|l  |f{5}         |f{5}  l        |f{7} l    |f{3} l l}
    \hline
    \hline
      \#  & Contribution & \cntl{1}{Borie (B)}
                                     & \cntl{2}{Martynenko group (M)}
                                                    & \cntl{2}{Karshenboim group (K)}
                                                                 & \cnt{3}{Our choice} \\
        &              & \cntl{1}{Borie \cite{Borie:2014:arxiv_v7} Tab.\,7} 
                                     & \cntl{2}{Elekina \etal~\cite{Elekina:2010:2Pmu3He} Tab.\,1} 
                                                    &  \cntl{2}{Karshenboim \etal~\cite{Karshenboim:2012:PRA85_032509}} 
                                                                 &     &
                                                                                  &     \\
          &              & \cntl{1}{~} 
                                     & \cntl{2}{~} 
                                                    &  \cntl{2}{Korzinin \etal~\cite{Korzinin:2013:PRD88_125019}} 
                                                                 &     &
                                                                                  &     \\
    \hline
    f1  & Dirac
                       &  144.4157    &             & &            &&     &         &     \\
    f2  & Recoil        
                       &   -0.1898    &             & &            &&     &         &     \\
    f3  & Contrib.\ of order $(Z\alpha)^4$
                       &              &  144.18648  & l.\,1
                                                      &            &&     &         &     \\
    f4a & Contrib.\ of order $(Z\alpha)^6$
                       &              &    0.01994  & l.\,3
                                                      &            &&     &         &     \\
    f4b & Contrib.\ of order $(Z\alpha)^6\,m^2/M$
                       &              &   -0.00060  & l.\,4
                                                      &            &&     &         &     \\
    sum &  (f1+f2) or (f3+f4)
                       &  144.2259    &  144.20582  & &            &
                                                                 & 144.2159    & $\pm\,0.0100$ & avg. \\
    \hline
    f5a & eVP corr.\ (Uehling), $\alpha(Z\alpha)^4$  
                       &             &  0.12925     & l.\,5 
                                                      &            &&     &         &   \\
    f5b & eVP corr.\ SOPT, $\alpha(Z\alpha)^4$  
                       &             &  0.14056     & l.\,7
                                                      &            &&     &         &   \\
    f13*& eVP corr.\ SOPT, $\alpha^2(Z\alpha)^4$  
                       &             &  0.00028     & l.\,9
                                                      &            &&     &         &   \\
    sum &  f5+f13*
                       &  0.2696     &  0.27009     & &  0.26920                      & \cite{Karshenboim:2012:PRA85_032509} Tab.IV
                                                                 &   0.2696    & $\pm\,0.0004$ & avg. \\
    \hline
    f6a & two-loop $e$VP corr.\ (KS), $\alpha^2(Z\alpha)^4$
                       &             &  0.00098     & l.\,10+11 
                                                      &            &&  0.0010     &         & M   \\
    f6b & two-loop $e$VP in SOPT, $\alpha^2(Z\alpha)^4$
                       &  0.0021     &  0.00234     & l.\,12+13
                                                      & 0.00242                    & \cite{Korzinin:2013:PRD88_125019} Tab.IX ``eVP2''
                                                                 &  0.0024      &          & K   \\
    f7a & $\alpha^2(Z\alpha)^4 m$, like \#11
                       &             &              & & 0.000606                  & \cite{Korzinin:2013:PRD88_125019} Tab.IX (a)
                                                                 &  0.0006      &          & K   \\
    f7d & $\alpha^2(Z\alpha)^4 m$, like \#12
                       &             &              & & 0.00164                   & \cite{Korzinin:2013:PRD88_125019} Tab.IX (d)
                                                                 &  0.0016      &          & K   \\
    f7e & $\alpha^2(Z\alpha)^4 m$, like \#30$^*$
                       &             &              & & 0.000019(2)               & \cite{Korzinin:2013:PRD88_125019} Tab.IX (e)
                                                                 &  0.0000      &          & K   \\
    f11*& $\alpha(Z\alpha)^6$
                       &             & -0.00055     & l.\,8
                                                      &          && -0.0006     &         & M   \\
    f12*& one-loop $\mu$VP, $\alpha(Z\alpha)^4$
                       &             &  0.00001     & l.\,6
                                                      &          &&  0.0000      &         & M   \\
    \hline
    f8  & AMM (second order)
                       &  0.3232     &              & &          &&              &         &     \\
    f9  & AMM (higher orders)
                       &  0.0012     &              & &          &&              &         &     \\            
    sum & Total AMM (f8+f9)
                       &  0.3244    &  0.32446      & l.\,2
                                                      &          && 0.3244 
                                                                        &       &avg. \\
    \hline
    f10 & Finite size, $(Z\alpha)^6$~\footnote{This is item \#r8,
 evaluated for a helion radius of 1.966(10)\,fm \cite{Borie:2014:arxiv_v7}, see text. The uncertainty is propagated from the charge radius, but is negligible.}
                       & -0.0158    &               & &          && -0.0158 
                                                                        & $\pm\,0.0002$    & B \\
    \hline
    \hline
    &&&&&&\\
       & \bf Sum       &  144.8062  &  144.80315    & &          &
                                                                 &  \TABFSVAL & \bf $\pm$\,\TABFSERR &  \\
    &&&&&&\\
    \hline
    \hline
  \end{tabular}
\end{minipage}
\end{table}
\end{landscape}

%
\section{2P levels}
\label{sec:2Plevels}
%

\subsection{2P fine structure}
\label{sec:fs}

Fine structure (FS) contributions have been calculated by Borie \cite{Borie:2014:arxiv_v7} (Tab.\,7), the Martynenko group \cite{Elekina:2010:2Pmu3He} (Tab.\,1), and the Karshenboim group \cite{Karshenboim:2012:PRA85_032509} (Tab.\,4) and \cite{Korzinin:2013:PRD88_125019} (Tab.\,9). All of these contributions are listed in Tab.\,\ref{tab:fs} and labeled with \#f$i$. 

The leading fine structure contribution of order $(Z\alpha)^4$ has been calculated by Borie using the Dirac wavefunctions (same as in Lamb shift). Her result (our item \#f1) has to be corrected by a recoil term (item \#f2) in order to be compared with the result from the Martynenko group. They use a nonrelativistic approach (our item \#f3) and then add relativistic corrections (our item \#f4a+b). Their total results differ by 0.02\mev. We take the average as \textit{our choice} and remain with an uncertainty of 0.01\mev. This is by far the dominant uncertainty in the 2P fine structure.

Item \#f5a and \#f5b are the one-loop $e$VP of order $\alpha(Z\alpha)^4$ in leading order and SOPT. Item \#f13* is the one-loop $e$VP contribution of order $\alpha^2(Z\alpha)^4$ in SOPT. All three items are given individually by the Martynenko group \cite{Elekina:2010:2Pmu3He} in lines 5, 7, and 9 of their Tab.\,1.
In Tab.\,7 of \cite{Borie:2014:arxiv_v7}, Borie's term ``Uehling(VP)'' presumably contains all these three items. Karshenboim \etal~\cite{Karshenboim:2012:PRA85_032509} (Tab.\,4) also calculate the sum of these items. All agree within 0.0009\mev and we take the average as \textit{our choice} which coincides with Borie's value.

Item \#f6a and \#f6b are the two-loop $e$VP (\textit{K\"all\'en-Sabry}) contribution of order $\alpha^2(Z\alpha)^4$ in leading order and SOPT. These terms have been calculated by Martynenko \etal~\cite{Elekina:2010:2Pmu3He} (Tab.\,1, line 10+11 and 12+13, respectively). Borie \cite{Borie:2014:arxiv_v7} and the Karshenboim group \cite{Korzinin:2013:PRD88_125019} (Tab.\,IX) only calculated our item \#f6b. We therefore adopt the value provided by the Martynenko group for item \#f6a and the Karshenboim group's value of \#f6b as they included some higher order terms as well.

Items \#f7a, \#f7d, and \#f7e are of order $\alpha^2(Z\alpha)^4$ and have been calculated with high accuracy by the Karshenboim group \cite{Korzinin:2013:PRD88_125019} (Tab.\,IX). They correspond to the same Feynman diagrams as the Lamb shift items \#11, \#12, and \#30, shown in Figs.\,\ref{fig:item_11}, \ref{fig:item_12}, and \ref{fig:item_30}, respectively. We adopt the values from the Karshenboim group as \textit{our choice}.

Item \#f11* is a contribution of order $\alpha(Z\alpha)^6$ which has been calculated by Martynenko \etal~\cite{Elekina:2010:2Pmu3He} (Tab.\,1, line 8).
Item \#f12* is the one-loop $\mu$VP of order $\alpha(Z\alpha)^4$ which has been calculated by the Martynenko group as well \cite{Elekina:2010:2Pmu3He} (Tab.\,1, line 6). We adopt both of these values as \textit{our choice}. 

The sum of items \#f8 and \#f9 is the muon anomalous magnetic moment (AMM) contribution of order $(Z\alpha)^4$. These items are labeled by Borie \cite{Borie:2014:arxiv_v7} as ``second order'' and ``higher orders'', respectively. Martynenko \etal~\cite{Elekina:2010:2Pmu3He} (Tab.\,1, line 2) provide the sum of these. Both groups agree very well. As \textit{our choice} we adopt the average.

Item \#f10 is the finite size correction to the \TwoPOne level of order $(Z\alpha)^6$ which has only been calculated by Borie \cite{Borie:2014:arxiv_v7}. It is the same correction which appears in the radius dependent part of the Lamb shift as \#r8, with opposite sign and evaluated with a helion charge radius of 1.966(10)\,fm \cite{Borie:2014:arxiv_v7}. We adopt Borie's value as \textit{our choice} and add the uncertainty which we obtain from the given charge radius.

The total sum of the FS contributions is summarized in Tab.\,\ref{tab:fs} and amounts to
\begin{equation}\label{eq:fs}
        \EFS = \FSVAL\mev \pm \FSERR \mev.
\end{equation}
It will enter the calculation of the 2P hyperfine structure in the following section. Note, that the uncertainty originates only from differences in the treatment of Dirac term (sum of items \#f1 to \#f4). 

\subsection{2P hyperfine structure}
\label{sec:hfs}
The 2P hyperfine splitting is described by the Breit Hamiltonian. Off-diagonal terms appear in the matrix representation of this Hamiltonian in the basis of \TwoPOneFOne, \TwoPOneFCero, \TwoPThreeFTwo, and \TwoPThreeFOne. These terms lead to a mixing of energy levels with same quantum number $F$ (see Fig.\,\ref{fig:energy_level}). This has first been calculated by Brodsky and Parsons \cite{Brodsky:1967:zeemanspectrum} for hydrogen and later has also been evaluated for muonic hydrogen by Pachucki \cite{Pachucki:1996:LSmup}. 
In previous publications \cite{Antognini:2013:Annals,Krauth:2016:mud}, we also discussed the mixing of hyperfine states. 

The traditional way \cite{Brodsky:1967:zeemanspectrum,Pachucki:1996:LSmup} is to calculate the FS (without perturbations from the HFS $F$ state mixing) and then include the so obtained FS in the evaluation of the Breit matrix. The centroids of the diagonal elements are now the virtual levels \TwoPOne and \TwoPThree. \textit{Afterwards} the mixing is included (via diagonalization) which means that the actual centroid is not at the position of the virtual levels anymore.



The 2P hyperfine structure has been calculated by Borie \cite{Borie:2014:arxiv_v7} (Tab.\,9) and Martynenko \etal~\cite{Elekina:2010:2Pmu3He} (Tab.\,2). We also calculated the splittings following Pachucki \cite{Pachucki:1996:LSmup}, who did the evaluation for \mup. The values which are listed in our Tab.\,\ref{tab:2Phfs} are not the published values, but the values which result when including our FS value from Sec.\,\ref{sec:fs}.

\begin{table*}
  \setlength\extrarowheight{7pt}
  \centering
  \caption[2P levels from fine- and hyperfine splitting]{
    {\bf 2P levels from fine- and hyperfine splitting}.
    All values are in meV relative to the 2P$_{1/2}$ level.
    The columns labeled with Borie and Martynenko include their HFS calculations, 
    but our value of the fine structure (2P$_{3/2} - $2P$_{1/2}$ 
    energy splitting) $\EFS=\FSVALERR\,$meV from Eq.\,(\ref{eq:fs}).
    The column 'following \cite{Pachucki:1996:LSmup}' is calculated in this work following the 
    treatment of Pachucki for \muHet, also including our value of the fine structure.
    Uncertainties arise from differences between the published values and from the uncertainty in the fine structure value 
    \EFS.
    }
  \label{tab:2Phfs}
  \begin{tabular}{l f{6} f{9} f{8} f{14}}
    \hline
    \hline
                   & \lft{1}{Borie \cite{Borie:2014:arxiv_v7}} 
                              & \lft{1}{Martynenko \cite{Elekina:2010:2Pmu3He}}
                                          & \lft{1}{following \cite{Pachucki:1996:LSmup}}
                                                       & \lft{1}{Our choice}    \\[0.5ex]
    \hline
    \TwoPOneFOne  & -14.7877  & -14.8080  &  -14.7990  &  -14.7979(102) \\
    \TwoPOneFCero &  43.8458  &  43.9049  &   43.8797  &   43.8754(296) \\
    \hline
    \TwoPThreeFTwo& 135.7580  & 135.7552  &  135.7527  &  135.7554 (27)(101)_{\rm FS} \\
    \TwoPThreeFOne& 160.0410  & 160.0459  &  160.0494  &  160.0452 (42)(101)_{\rm FS} \\
    \hline
    \hline
  \end{tabular}
\end{table*}

Borie in her Tab.\,9 lists the energies of the four 2P hyperfine levels relative to the \TwoPOne fine structure state where she already included the $F$ state mixing. We reproduced her results using the Eqs.\ given in her Tab.\,9 and then inserted our \EFS from our Eq.\,(\ref{eq:fs}). The result is listed in the second column of Tab.\,\ref{tab:2Phfs}. Borie mentions, she used the shielded helion magnetic moment, whereas the (unshielded) magnetic moment should be used. The change, however, appears only on the seventh digit and is therefore negligible.

In their Tab.\,2, Martynenko \etal\ provide the total splittings of the hyperfine structure levels, and at the end of their Sec.\,3, they list the term $\Delta = 0.173\mev$ originating from the mentioned $F$ state mixing. In order to include this term, the numbers in their Tab.\,2 first have to be divided according to the weight given by the number of $m_F$ states. $\Delta$ has then to be added to the two $F=1$ states. Furthermore, for the \TwoPThree states, we add our \EFS. The result is listed in the third column of our Tab.\,\ref{tab:2Phfs}.

Additionally, following Pachucki \cite{Pachucki:1996:LSmup}, we repeat his calculations in \mup for \muHet. The off-diagonal elements are given by Eq.\,(85) of \cite{Pachucki:1996:LSmup}
\begin{multline}
  \langle  \,  \TwoPOneFOne \, |\, V\, |\, \TwoPThreeFOne \,\rangle\\
        = \frac{1}{3}(Z\alpha)^4 \frac{m_r^3}{m_\mu m_h}(1+\kappa) 
          \left(1+\frac{m_\mu}{m_h}\frac{1+2\kappa}{1+\kappa}\right)\left(-\frac{\sqrt{2}}{48}\right),
\end{multline}
where we included the correct $Z$ scaling. $m_r$ is the reduced mass of the muonic helium-3 ion, $m_\mu$ ($m_h$) is the mass of the muon (helion), and $\kappa = -4.18415$~\footnote{The helion anomalous magnetic moment is obtained using the respective equation on p.\,17 of Borie's Ref.\,\cite{Borie:2014:arxiv_v7}, where this magnitude is denoted as $\kappa_2$.} is the 
helion anomalous magnetic moment.
The diagonal terms are given by Eq.\,(86) therein
\begin{multline}
  E_{\rm HFS}(\TwoPOne)\\ =  \frac{1}{3}(Z\alpha)^4 \frac{m_r^3}{m_\mu m_h}(1+\kappa) 
                           \left(\frac{1}{3}+\frac{a_\mu}{6}+\frac{1}{12}\frac{m_\mu}{m_h}\frac{1+2\kappa}{1+\kappa}\right) 
                           \label{eq:2Pone}
\end{multline}
\begin{multline}
  E_{\rm HFS}(\TwoPThree)\\ =  \frac{1}{3}(Z\alpha)^4 \frac{m_r^3}{m_\mu m_h}(1+\kappa) 
                             \left(\frac{2}{15}-\frac{a_\mu}{30}+\frac{1}{12}\frac{m_\mu}{m_h}\frac{1+2\kappa}{1+\kappa}\right) 
                             \label{eq:2Pthree}
\end{multline}
with the anomalous magnetic moment of the muon $a_\mu = 1.165\,920\,89(63)\times10^{-3}$ \cite{Mohr:2016:CODATA14}.

Furthermore, Pachucki adds corrections due to vacuum polarization in his Eq.\,(89) and (90). With correct $Z$ scaling these are
\begin{align}
  \delta E_{\rm HFS}(\TwoPOne)   =& \frac{1}{3}(Z\alpha)^4\frac{m_r^3}{m_\mu m_h}(1+\kappa) \cdot0.00022 \\
  \delta E_{\rm HFS}(\TwoPThree) =& \frac{1}{3}(Z\alpha)^4\frac{m_r^3}{m_\mu m_h}(1+\kappa) \cdot 0.00008.
\end{align}
They have to be added to Eqs.\,(\ref{eq:2Pone}) and (\ref{eq:2Pthree}), respectively.
Diagonalizing the matrix given in Eq.\,(91) of Ref.\,\cite{Pachucki:1996:LSmup} with entries determined by the above equations yields the values given as \textit{our choice} in Tab.\,\ref{tab:2Phfs}.
The diagonalization yields an $F$ mixing of $\Delta = 0.1724\mev$.
In the same manner as for the sections above, \textit{our choice} in Tab.\,\ref{tab:2Phfs} takes into account the spread of values from the different authors and additionally the uncertainty of our value of the fine structure which we obtained in Sec.\,\ref{sec:fs}. It is astonishing that the splitting of the \TwoPOne states differs by as much as 0.04\mev between Borie and Martynenko. These states do not overlap with the nucleus, so it should be possible to determine them to much better precision. A precise calculation of these splittings is therefore highly welcome.

\section{Summary}
We have compiled all available contributions necessary to extract a charge radius of the helion from the Lamb shift measurement in muonic helium-3 ions, performed by the CREMA collaboration. 

The total of the Lamb shift contributions are summarized in Eq.\,\ref{eq:LS:full}.\\
The nuclear structure-independent contributions of the Lamb shift, given in Tab.\,\ref{tab:LS:QED}, show good agreement within the four (groups of) authors. The uncertainty is dominated by the hadronic VP (\#14) and higher order radiative recoil corrections (\#24). The total uncertainty in Tab.\,\ref{tab:LS:QED}, however, is in the order of 0.01\mev and therefore sufficiently good (see also Eq.~\ref{eq:uncertainty}).\\
The nuclear structure-dependent part of the Lamb shift completely dominates the theoretical uncertainties. The one-photon exchange (finite size) contributions, where the coefficients are given in Tab.\,\ref{tab:LS:Radius}, have an uncertainty which corresponds to 0.04\mev, which already is above the ``ideal'' precision, mentioned in the introduction. This uncertainty is dominated by a disagreement in the terms \#r4 and \#r6. 
The much larger uncertainty, however, stems from the two-photon exchange contributions (TPE), given in Eq.\,(\ref{eq:LS:pol}). Recently, two groups have published new calculations on the TPE with a precision of about 3\% ($\sim0.5\mev$). In terms of the helion charge radius this uncertainty corresponds to about 
\begin{equation}
\sigma_{\rm theory}(\rh)\approx \pm 0.0013\fm. 
\end{equation}
The expected experimental uncertainty will be about an order of magnitude smaller.
Thus, improving the theoretical uncertainty directly improves the extraction of the charge radius.

Isotope shift measurements generally benefit from cancellations of theory contributions that limit the absolute charge radii \cite{CancioPastor:2012:PRL108,Jentschura:2011:IsoShift}. For the present case of the muonic helium isotope shift it will be useful to exploit possible correlations between the nuclear and nucleon structure contributions, which dominate the total uncertainty of the muonic radii. The correlations could lead to a reduction of the uncertainty of the muonic isotope shift determination and shed light on the $4\,\sigma$ discrepancy in the electronic isotope shift measurements, see Fig.\,\ref{fig:iso_shift}. A further investigation of these correlations is therefore desired.

The total of the 2S HFS contributions are given in Tab.\,\ref{tab:hfshelium} and summarized in Eq.\,\ref{eq:hfs:total1}.
The uncertainty in the 2S HFS is completely dominated by the polarizability contribution, where no calculation exists. We have given a very rough estimate. The second largest uncertainty in the 2S HFS originates from the Zemach radius term (Bohr-Weisskopf effect).
The upcoming results of the CREMA experiment will be able to extract a value for the TPE in the 2S hyperfine splitting (sum of polarizability and Zemach radius contribution) from measured data. In this case the uncertainty will be limited by the experimental uncertainty. 

For the 2P levels, we collect all fine structure terms from the various authors (Tab.\,\ref{tab:fs}) which are then used to calculate the hyperfine structure by means of the Breit matrix. The results are compared with two other groups (Tab.\,\ref{tab:2Phfs}). Here, the largest uncertainty originates from the leading order contributions (\#f1 to \#f4) in the fine structure (which is still sufficiently good) and from differing published values of the \TwoPThree splitting. A clarification of this difference would be very welcome.
\\\\
Note added in proof: After this manuscript was accepted for publication, a paper by Karshenboim \etal\,\cite{Karshenboim:2016:mu3he}  about the Lamb shift theory in muonic helium and tritium was published. They discuss the 2S-2P Lamb shift and the 2P fine- and hyperfine structure. The 2S hyperfine structure is not treated therein. The comparison of their values with ours has to be done carefully because Karshenboim \etal\ treat the mixing of the hyperfine levels (Brodsky Parsons contribution) differently. In their work the mixing is added as a perturbation to the fine structure. The traditional way, however, is to use the unperturbed fine structure and add the mixing as a perturbation to the hyperfine levels, which is what we do. Comparing the values one therefore has to subtract/add the Brodsky Parsons term printed in bold italic in \cite{Karshenboim:2016:mu3he}. Furthermore Karshenboim \etal\ neglect some known higher order terms and increase the uncertainty due to estimates of non-listed higher order contributions.
The comparison with the values in Ref. \cite{Karshenboim:2016:mu3he} yields the following (the numbers shown here are \textit{adapted} to the traditional treatment of the Brodsky Parsons contribution): For the radius-independent QED Lamb shift without TPE, Karshenboim \etal\ obtain a value of 1644.35(2)\mev which is in very good agreement with ours (Eq.\,\ref{eq:LS:QED}). In order to compare the radius-dependent (finite size) part we use a helion charge radius of 1.966\fm \cite{Borie:2014:arxiv_v7}. The value of Karshenboim \etal\ is then $-399.69(23)^\mathrm{theo}\mev$ which differs by 0.33(23)\mev ($1.4\sigma$) from our value of $-400.02(4)^\mathrm{theo}\mev$. This difference is the largest between our values and the ones from Karshenboim \etal. For the 2P fine structure, Karshenboim \etal\ obtain a value of $144.800(5)\mev-0.004\,\rh^2\mev/\fm^2$ which differs by 0.0142\mev ($1.3\sigma$) from ours. Regarding the 2P$_{1/2}$ hyperfine structure, the value from Karshenboim \etal\ of $-58.7150(7)\mev$ differs by 0.0417\mev ($1.3\sigma$) and has by far the smaller uncertainty. In our case the uncertainty arises from the huge difference between Borie and Martynenko. The 2P$_{3/2}$ splitting of $-24.2925(7)\mev$ agrees very well with our value.\\  
However, all these differences are considerably smaller than the uncertainty of the two-photon contribution which we assumed to be 0.52\mev while Karshenboim \etal\ increase it to 0.86\mev. The final result for the charge radius will therefore not be changed significantly.

\section{Acknowledgments}
\label{sec:ack}

We are grateful to 
E.~Borie and A.P.~Martynenko for insightful comments and for providing us with previously unpublished results.
We thank M.~Gorchtein and N.~Nevo Dinur for helpful discussions about the two-photon exchange in muonic helium-3 ions and the treatment of the Friar moment contribution.
We acknowledge valuable contributions in general from
S.~Bacca, 
N.~Barnea,
M.~Birse,
E.~Borie, 
C.E.~Carlson,
M.~Eides,
J.L.~Friar,
M.~Gorchtein,
F.~Hagelstein,
C.~Ji,
S.~Karshenboim,
A.P.~Martynenko,
J.~McGovern,
N.~Nevo Dinur,
K.~Pachucki, and
M.~Vanderhaeghen
and are thankful for 
their valuable remarks and insightful discussions. 
%
We proactively thank a future generation of motivated theorists for all future critical compilations of theory terms in light muonic atoms/ions.

The authors acknowledge support from the European Research Council
(ERC) through StG. \#279765 and CoG. \#725039, the Excellence Cluster PRISMA of the Unversity of Mainz, and the Swiss National Science 
Foundation SNF, Projects 200021L\_138175 and 200021\_165854.

\section{Author contribution statement}
B.F.~and J.J.K.~set up the tables and wrote the manuscript. Both contributed equally to the paper. The paper was written under the supervision of and includes many comments and suggestions from A.A., F.K., and R.P., whereas M.D.~participated in the discussion. 
All authors discussed the paper and participated in the review.
%
%

%
\bibliographystyle{mysty1}
\bibliography{ref}

\begin{thebibliography}{10}
\newcommand{\enquote}[1]{`#1'}
\expandafter\ifx\csname urlstyle\endcsname\relax
  \providecommand{\doi}[1]{doi:\discretionary{}{}{}#1}\else
  \providecommand{\doi}{doi:\discretionary{}{}{}\begingroup
  \urlstyle{rm}\Url}\fi

\bibitem{Pohl:2010:Nature_mup1}
R.~Pohl, A.~Antognini, F.~Nez, et~al.
\newblock \emph{Nature} \textbf{466}, 213 (2010).

\bibitem{Antognini:2013:Science_mup2}
A.~Antognini, F.~Nez, K.~Schuhmann, et~al.
\newblock \emph{Science} \textbf{339}, 417 (2013).

\bibitem{Antognini:2013:Annals}
A.~Antognini, F.~Kottmann, F.~Biraben, et~al.
\newblock \emph{Ann.~Phys.} \textbf{331}, 127 (2013).
\newblock [arXiv:1208.2637].

\bibitem{Pohl:2016:mud}
R.~Pohl, F.~Nez, L.~M.~P. Fernandes, et~al.
\newblock \emph{Science} \textbf{353}, 669 (2016).

\bibitem{Krauth:2016:mud}
J.~J. Krauth, M.~Diepold, B.~Franke, et~al.
\newblock \emph{Ann.~Phys.} \textbf{366}, 168  (2016).
\newblock [arXiv:1506.01298].

\bibitem{Mohr:2016:CODATA14}
P.~J. Mohr, D.~B. Newell, and B.~N. Taylor.
\newblock \emph{Rev.~Mod.~Phys.} \textbf{88}, 035009 (2016).

\bibitem{Pohl:2013:ARNPS}
R.~Pohl, R.~Gilman, G.~A. Miller, et~al.
\newblock \emph{Ann.~Rev.~Nucl.~Part.~Sci.} \textbf{63}, 175 (2013).
\newblock [arXiv 1301.0905].

\bibitem{Carlson:2015:Puzzle}
C.~E. Carlson.
\newblock \emph{Prog.~Part.~Nucl.~Phys.} \textbf{82}, 59  (2015).

\bibitem{Hill:2017:PRP}
{Hill, Richard J.}
\newblock \emph{EPJ Web Conf.} \textbf{137}, 01023 (2017).
\newblock [arXiv:1702.01189].

\bibitem{Parthey:2010:PRL_IsoShift}
C.~G. Parthey, A.~Matveev, J.~Alnis, et~al.
\newblock \emph{Phys.~Rev.~Lett.} \textbf{{104}}, {233001} ({2010}).

\bibitem{Jentschura:2011:IsoShift}
U.~D. Jentschura, A.~Matveev, C.~G. Parthey, et~al.
\newblock \emph{Phys.~Rev.~A} \textbf{83}, 042505 (2011).

\bibitem{Antognini:2011:Conf:PSAS2010}
A.~Antognini, F.~Nez, F.~D. Amaro, et~al.
\newblock \emph{Can.~J.~Phys.} \textbf{89}, 47 (2010).

\bibitem{Machleidt:2011:nuclforces}
R.~Machleidt and D.~Entem.
\newblock \emph{Physics Reports} \textbf{503}, 1  (2011).

\bibitem{NevoDinur:2016:TPE}
N.~Nevo~Dinur, C.~Ji, S.~Bacca, et~al.
\newblock \emph{Phys.~Lett.~B} \textbf{755}, 380  (2016).

\bibitem{Antognini:2016:PRP}
{Antognini, A.}, {Schuhmann, K.}, {Amaro, F. D.}, et~al.
\newblock \emph{EPJ Web of Conferences} \textbf{113}, 01006 (2016).

\bibitem{Miller:2013:pol}
G.~A. Miller.
\newblock \emph{Phys.~Lett.~B} \textbf{718}, 1078 (2013).

\bibitem{Jentschura:2015:virtPart}
U.~D. Jentschura.
\newblock \emph{Phys. Rev. A} \textbf{92}, 012123 (2015).

\bibitem{Tucker-Smith:2011}
D.~Tucker-Smith and I.~Yavin.
\newblock \emph{Phys. Rev. D} \textbf{83}, 101702 (2011).

\bibitem{Batell:2011:PV_muonic_forces}
B.~Batell, D.~{McKeen}, and M.~Pospelov.
\newblock \emph{Phys.~Rev.~Lett.} \textbf{107}, 011803 (2011).

\bibitem{Karshenboim:2014:darkForces}
S.~G. Karshenboim, D.~McKeen, and M.~Pospelov.
\newblock \emph{Phys. Rev. D} \textbf{90}, 073004 (2014).

\bibitem{Carlson:2015:BSM}
C.~E. Carlson and M.~Freid.
\newblock \emph{Phys. Rev. D} \textbf{92}, 095024 (2015).

\bibitem{Shiner:1995:heliumSpec}
D.~Shiner, R.~Dixson, and V.~Vedantham.
\newblock \emph{Phys.~Rev.~Lett.} \textbf{74}, 3553 (1995).

\bibitem{Rooij:2011:HeSpectroscopy}
R.~{van Rooij}, J.~S. Borbely, J.~Simonet, et~al.
\newblock \emph{Science} \textbf{333}, 196 (2011).

\bibitem{CancioPastor:2012:PRL108}
P.~Cancio~Pastor, L.~Consolino, G.~Giusfredi, et~al.
\newblock \emph{Phys. Rev. Lett.} \textbf{108}, 143001 (2012).

\bibitem{Patkos:2016:HeIso}
V.~Patk\'o\ifmmode\check{s}\else\v{s}\fi{}, V.~A. Yerokhin, and K.~Pachucki.
\newblock \emph{Phys. Rev. A} \textbf{94}, 052508 (2016).
\newblock [arXiv:1610.04060].

\bibitem{Patkos:2017:HeIsoII}
V.~Patk\'o\ifmmode~\check{s}\else \v{s}\fi{}, V.~A. Yerokhin, and K.~Pachucki.
\newblock \emph{Phys. Rev. A} \textbf{95}, 012508 (2017).
\newblock [arXiv:1612.06142].

\bibitem{Diepold:2016:muHe4theo}
M.~{Diepold}, J.~J. {Krauth}, B.~{Franke}, et~al. (2016).
\newblock [arXiv:1606.05231].

\bibitem{Vutha:2012:H2S2P}
A.~C. Vutha, N.~Bezginov, I.~Ferchichi, et~al.
\newblock \emph{Bull. Am. Phys. Soc.} \textbf{57(5)}, Q1.138 (2012).

\bibitem{Beyer:2013:AdP_2S4P}
A.~Beyer, J.~Alnis, K.~Khabarova, et~al.
\newblock \emph{Ann. d. Phys. (Berlin)} \textbf{525}, 671 (2013).

\bibitem{Peters:2013:AdP}
E.~Peters, D.~C. Yost, A.~Matveev, et~al.
\newblock \emph{Ann. d. Phys. (Berlin)} \textbf{525}, L29 (2013).

\bibitem{Herrmann:2009:He1S2S}
M.~Herrmann, M.~Haas, U.~Jentschura, et~al.
\newblock \emph{Phys.~Rev.~A} \textbf{79}, 052505 (2009).

\bibitem{Kandula:2010:XUV_comb_metrology}
D.~Z. Kandula, C.~Gohle, T.~J. Pinkert, et~al.
\newblock \emph{Phys.~Rev.~Lett.} \textbf{105}, 063001 (2010).

\bibitem{Mihovilovic:2013:ISR_exp_MAMI}
M.~Mihovilovic and H.~Merkel.
\newblock \emph{{AIP}\ Conf.\ Proc.} \textbf{1563}, 187 (2013).

\bibitem{Gasparian:2014:PRad}
A.~Gasparian.
\newblock \emph{EPJ~Web~of~Conf.} \textbf{73}, 07006 (2014).

\bibitem{Gilman:2013:MUSE}
R.~Gilman.
\newblock \emph{{AIP}\ Conf.\ Proc.} \textbf{1563}, 167 (2013).

\bibitem{Sick:2014:HeZemach}
I.~Sick.
\newblock \emph{Phys.~Rev.~C} \textbf{90}, 064002 (2014).

\bibitem{Angeli:2013:radii}
I.~Angeli and K.~Marinova.
\newblock \emph{At.~Data and Nucl.~Data Tables} \textbf{99}, 69  (2013).

\bibitem{Carboni:1978:LS_mu4he}
G.~Carboni, G.~Gorini, E.~Iacopini, et~al.
\newblock \emph{Phys.~Lett.~B} \textbf{73}, 229  (1978).

\bibitem{Hauser:1992:LS_search}
P.~Hauser, H.~P. von Arb, A.~Biancchetti, et~al.
\newblock \emph{Phys. Rev. A} \textbf{46}, 2363 (1992).

\bibitem{Hernandez:2016:POLupdate}
O.~J. Hernandez, N.~Nevo~Dinur, C.~Ji, et~al.
\newblock \emph{Hyp.~Interact.} \textbf{237}, 158 (2016).
\newblock [arXiv:1604.06496].

\bibitem{Carlson:2016:tpe}
C.~E. Carlson, M.~Gorchtein, and M.~Vanderhaeghen.
\newblock \emph{Phys. Rev. A} \textbf{95}, 012506 (2017).

\bibitem{Brodsky:1967:zeemanspectrum}
S.~J. Brodsky and R.~G. Parsons.
\newblock \emph{Phys.~Rev.} \textbf{163}, 134 (1967).

\bibitem{Borie:2012:LS_revisited_AoP}
E.~Borie.
\newblock \emph{Ann.~Phys.} \textbf{327}, 733 (2012).

\bibitem{Borie:2014:arxiv_v7}
E.~Borie.
\newblock \enquote{{Lamb} shift in light muonic atoms -- {Revisited}}.
\newblock arXiv:1103.1772-v7 [physic.atom-ph] (2014).

\bibitem{Krutov:2014:JETP120_73}
A.~Krutov, A.~Martynenko, G.~Martynenko, et~al.
\newblock \emph{J. Exp. Theo. Phys.} \textbf{120}, 73 (2015).

\bibitem{Martynenko:2010:2SHFS_muHe}
A.~P. Martynenko and E.~N. Elekina.
\newblock \emph{Phys.~At.~Nucl.} \textbf{73}, 2074 (2010).

\bibitem{Martynenko:2008:muheHFS}
A.~P. Martynenko.
\newblock \emph{J. Exp. Theo. Phys.} \textbf{106}, 690 (2008).

\bibitem{Faustov:2014:radrec}
R.~Faustov, A.~Martynenko, G.~Martynenko, et~al.
\newblock \emph{Phys.~Lett.~B} \textbf{733}, 354  (2014).

\bibitem{Elekina:2010:2Pmu3He}
E.~N. Elekina and A.~P. Martynenko.
\newblock \emph{Phys.~At.~Nucl.} \textbf{73}, 1828 (2010).

\bibitem{Jentschura:2011:SemiAnalytic}
U.~D. Jentschura and B.~J. Wundt.
\newblock \emph{Eur.~Phys.~J.~D} \textbf{65}, 357 (2011).

\bibitem{Jentschura:2011:PRA84_012505}
U.~D. Jentschura.
\newblock \emph{Phys.~Rev.~A} \textbf{84}, 012505 (2011).

\bibitem{Korzinin:2013:PRD88_125019}
E.~Y. Korzinin, V.~G. Ivanov, and S.~G. Karshenboim.
\newblock \emph{Phys.~Rev.~D} \textbf{88}, 125019 (2013).

\bibitem{Karshenboim:2012:PRA85_032509}
S.~G. Karshenboim, V.~G. Ivanov, and E.~Y. Korzinin.
\newblock \emph{Phys.~Rev.~A} \textbf{85}, 032509 (2012).

\bibitem{KallenSabry:1955}
G.~K{\"a}ll{\'e}n and A.~Sabry.
\newblock \emph{Dan.~Mat.~Fys.~Medd.} \textbf{29}, 1 (1955).

\bibitem{Jentschura:2011:AnnPhys1}
U.~D. Jentschura.
\newblock \emph{Ann.~Phys.} \textbf{326}, 500 (2011).

\bibitem{Pachucki:1996:LSmup}
K.~Pachucki.
\newblock \emph{Phys.~Rev.~A} \textbf{53}, 2092 (1996).

\bibitem{Karshenboim:PC:2015}
S.~G. Karshenboim.
\newblock private communication (2015).

\bibitem{Borie:1981:HVP}
E.~Borie.
\newblock \emph{Z.~Phys.~A} \textbf{302}, 187 (1981).

\bibitem{Krutov:2011:PRA84_052514}
A.~A. Krutov and A.~P. Martynenko.
\newblock \emph{Phys.~Rev.~A} \textbf{84}, 052514 (2011).

\bibitem{Barker:1955}
W.~Barker and F.~Glover.
\newblock \emph{Phys.~Rev.} \textbf{99}, 317 (1955).

\bibitem{Jentschura:2011:DF}
U.~D. Jentschura.
\newblock \emph{Eur.~Phys.~J.~D} \textbf{61}, 7 (2011).

\bibitem{Mohr:2012:CODATA10}
P.~J. Mohr, B.~N. Taylor, and D.~B. Newell.
\newblock \emph{Rev.~Mod.~Phys.} \textbf{84}, 1527 (2012).

\bibitem{Friar:1978:Annals}
J.~L. Friar.
\newblock \emph{Ann.~Phys.} \textbf{122}, 151 (1979).

\bibitem{Pachucki:PC:2015}
K.~Pachucki.
\newblock private communication (2015).

\bibitem{Yerokhin:2016:RCFS}
V.~A. Yerokhin.
\newblock \enquote{{Nuclear recoil in the Lamb shift of hydrogen-like atoms}}.
\newblock {ECT* Workshop on the Proton Radius Puzzle} (2016).

\bibitem{Borie:PC:2017}
E.~Borie.
\newblock private communication (2017).

\bibitem{Karshenboim:2015:PRD91_073003}
S.~G. Karshenboim, E.~Y. Korzinin, V.~A. Shelyuto, et~al.
\newblock \emph{Phys.~Rev.~D} \textbf{91}, 073003 (2015).

\bibitem{Ivanov:2001:LS}
V.~G. Ivanov and S.~G. Karshenboim.
\newblock \emph{Lamb Shift in Light Hydrogen-Like Atoms} pages 637--650.
\newblock Springer Berlin Heidelberg, Berlin, Heidelberg.
\newblock ISBN 978-3-540-45395-6.
\newblock \doi{10.1007/3-540-45395-4_44} (2001).

\bibitem{Friar:1997:PRA56_5173}
J.~Friar and G.~Payne.
\newblock \emph{Phys.~Rev.~A} \textbf{56}, 5173 (1997).

\bibitem{Pachucki:2011:PRL106_193007}
K.~Pachucki.
\newblock \emph{Phys.~Rev.~Lett.} \textbf{106}, 193007 (2011).

\bibitem{Friar:2013:PRC88_034004}
J.~L. Friar.
\newblock \emph{Phys.~Rev.~C} \textbf{88}, 034003 (2013).

\bibitem{Sick:2008:rad_scatt}
I.~Sick.
\newblock \emph{Precise Radii of Light Nuclei from Electron Scattering} pages
  57--77.
\newblock Springer Berlin Heidelberg, Berlin, Heidelberg.
\newblock ISBN 978-3-540-75479-4.
\newblock \doi{10.1007/978-3-540-75479-4_4} (2008).

\bibitem{Martynenko:PC:2016}
A.~P. Martynenko.
\newblock private communication (2016).

\bibitem{Eides:1997:two-photon}
M.~I. Eides and H.~Grotch.
\newblock \emph{Phys.~Rev.~A} \textbf{56}, R2507 (1997).

\bibitem{Faustov:2017:rad_fin_size}
R.~N. Faustov, A.~P. Martynenko, F.~A. Martynenko, et~al.
\newblock {[arXiv:1706.01060 (hep-ph)]}.

\bibitem{Martynenko:PC:2017}
A.~P. Martynenko.
\newblock private communication (2017).

\bibitem{Joachain:1961:pol}
C.~Joachain.
\newblock \emph{Nuclear Physics} \textbf{25}, 317  (1961).

\bibitem{Rinker:1976:he_pol}
G.~A. Rinker.
\newblock \emph{Phys.~Rev.~A} \textbf{14}, 18 (1976).

\bibitem{Carlson:2011:PRA84_020102}
C.~E. Carlson and M.~Vanderhaeghen.
\newblock \emph{Phys.~Rev.~A} \textbf{84}, 020102(R) (2011).
\newblock [arXiv:1101.5965 (hep-ph)].

\bibitem{Myers:2014:comptonScatt}
L.~S. Myers, J.~R.~M. Annand, J.~Brudvik, et~al.
\newblock \emph{Phys. Rev. Lett.} \textbf{113}, 262506 (2014).

\bibitem{Carlson:2014:PRA89_022504}
C.~E. Carlson, M.~Gorchtein, and M.~Vanderhaeghen.
\newblock \emph{Phys.~Rev.~A} \textbf{89}, 022504 (2014).

\bibitem{BirseMcGovern:2012}
M.~C. Birse and J.~A. {McGovern}.
\newblock \emph{Eur.~Phys.~J.~A} \textbf{48}, 120 (2012).
\newblock [arXiv:1206.3030 (hep-ph)].

\bibitem{Gorchtein:PC:2016}
M.~Gorchtein.
\newblock private communication (2016).

\bibitem{Brodsky:1966:radiative:hyperfine}
S.~J. Brodsky and G.~W. Erickson.
\newblock \emph{Phys.~Rev.} \textbf{148}, 26 (1966).

\bibitem{Zemach:1956}
A.~C. Zemach.
\newblock \emph{Phys. Rev.} \textbf{104}, 1771 (1956).

\bibitem{FriarSick:2004:Zemach}
J.~L. Friar and I.~Sick.
\newblock \emph{Phys.~Lett.~B} \textbf{579}, 285 (2004).

\bibitem{Borie:1980:mu3HeLS}
E.~Borie.
\newblock \emph{Z.~Phys.~A} \textbf{297}, 17 (1980).

\bibitem{Grotch:1969:EPM}
H.~Grotch and D.~R. Yennie.
\newblock \emph{Rev. Mod. Phys.} \textbf{41}, 350 (1969).

\bibitem{Borie:PC:2014}
E.~Borie.
\newblock private communication (2015).

\bibitem{Friar:2005:PRC72}
J.~L. Friar and G.~L. Payne.
\newblock \emph{Phys. Rev. C} \textbf{72}, 014002 (2005).

\bibitem{Eides:2012:Weak}
M.~I. Eides.
\newblock \emph{Phys.~Rev.~A} \textbf{85}, 034503 (2012).

\bibitem{CREMA:mu3he}
{CREMA Collaboration}.
\newblock To be published.

\bibitem{Karshenboim:2016:mu3he}
S.~G. Karshenboim, E.~Y. Korzinin, V.~A. Shelyuto, et~al.
\newblock \emph{Phys. Rev. A} \textbf{96}, 022505 (2017).

\end{thebibliography}

\end{document}